\documentclass[trackchanges,resetfootnote, twocolumn]{aastex7}

\usepackage{graphicx}%
\usepackage{multirow}%
\usepackage{amsmath,amssymb,amsfonts}%
\usepackage{amsthm}%
\usepackage{mathrsfs}%
\usepackage{xcolor}%
\usepackage{xspace}
\usepackage[T1]{fontenc}
\usepackage{lineno}
\usepackage{comment}

\newcommand{\Me}{\ensuremath{M_{\oplus}}\xspace} 

\newcommand{\Rj}{\ensuremath{R_{\rm{Jup}}}\xspace}
\newcommand{\Mj}{\ensuremath{M_{\rm{Jup}}}\xspace}

\newcommand{\Lsun}{L_\odot}
\newcommand{\Teff}{\ensuremath{T_\mathrm{eff}}\xspace}
\newcommand{\logg}{\ensuremath{\log{g}}\xspace}
\newcommand{\lbollsun}{\ensuremath{\log(L_\mathrm{bol}/\Lsun)}}

\newcommand{\logco}{\ensuremath{\mathrm{log(^{12}CO / ^{13}CO)}\xspace}}
\newcommand{\logcoo}{\ensuremath{\mathrm{log(C^{16}O / C^{18}O)}\xspace}}
\newcommand{\loghdo}{\ensuremath{\mathrm{log(H_2O / HDO)}\xspace}}

\newcommand{\kms}{km~s$^{-1}$\xspace}

\newcommand{\ko}[1]{\textcolor{blue}{\bf#1}}

\newcommand{\caltech}{Department of Astronomy, California Institute of Technology, Pasadena, CA 91125, USA}
\newcommand{\gps}{Division of Geological \& Planetary Sciences, California Institute of Technology, Pasadena, CA 91125, USA}
\newcommand{\ucsc}{Department of Astronomy \& Astrophysics, University of California, Santa Cruz, CA95064, USA}

\newcommand{\uclagps}{Department of Earth, Planetary, and Space Sciences, University of California, Los Angeles, CA 90095, USA}
\newcommand{\jpl}{Jet Propulsion Laboratory, California Institute of Technology, 4800 Oak Grove Dr.,Pasadena, CA 91109, USA}
\newcommand{\ucsd}{Department of Astronomy \& Astrophysics,  University of California, San Diego, La Jolla, CA 92093, USA}

\newcommand{\ames}{NASA Ames Research Center, MS 245-6, Moffett Field, CA 94035, USA}

\newcommand{\ifahilo}{Institute for Astronomy, University of Hawaii at Hilo, 640 N Aohoku Pl, Hilo, HI 96720, USA}

\newcommand{\carnegiew}{Earth and Planets Laboratory, Carnegie Institution for Science, Washington, DC, 20015}

\newcommand{\jhu}{Department of Physics \& Astronomy, Johns Hopkins University, Baltimore, MD 21218, USA}
\newcommand{\nexsci}{NASA Exoplanet Science Institute, IPAC, Caltech, Pasadena, CA 91125, USA}
\newcommand{\stsci}{Space Telescope Science Institute, Baltimore, MD 21218, USA}
\newcommand{\arizona}{Department of Astronomy/Steward Observatory, University of Arizona, Tucson, AZ 85721, USA}
\newcommand{\herzberg}{NRC Herzberg Astronomy and Astrophysics, Victoria, BC, V9E 2E7, Canada}
\newcommand{\victoria}{Department of Physics and Astronomy, University of Victoria, Victoria, BC, V8P 5C2, Canada}
\newcommand{\umich}{Department of Astronomy, University of Michigan, Ann Arbor, MI 48109, USA}
\newcommand{\ipac}{IPAC, California Institute of Technology, Pasadena, CA 91125, USA}
\newcommand{\ciera}{Center for Interdisciplinary Exploration and Research in Astrophysics, Evanston, IL 60201, USA}

\begin{document}

\title{The Compositions of the HR 8799 Planets Reflect Accretion of Both Solids and Metal-enriched Gas}

\author[0000-0002-6618-1137]{Jerry W. Xuan}
\altaffiliation{Shared first authorship}
\affiliation{\uclagps}
\affiliation{\caltech}
\affiliation{51 Pegasi b Fellow}
\email[show]{jerryxuan@g.ucla.edu}

\author[0000-0003-2233-4821]{Jean-Baptiste Ruffio}
\altaffiliation{Shared first authorship}
\affiliation{\ucsd}
\email{jruffio@ucsd.edu}

\author[0000-0003-1728-8269]{Yayaati Chachan}
\affiliation{\ucsc}
\email{ychachan@ucsc.edu}

\author[0000-0003-3290-6758]{Kazumasa Ohno}
\affiliation{Division of Science, National Astronomical Observatory of Japan, 2-12-1 Osawa, Mitaka, Tokyo 181-8588, Japan}
\email{ohno.k.ab.715@gmail.com}

\author[0000-0002-3239-5989]{Aurora Kesseli}
\affiliation{\nexsci}
\email{aurorak@ipac.caltech.edu}

\author[0000-0001-5061-0462]{Ruth Murray-Clay}
\affiliation{\ucsc}
\email{rmc@ucsc.edu}

\author[0000-0002-1228-9820]{Eve J. Lee}
\affiliation{\ucsd}
\email{evelee@ucsd.edu}

\author[0000-0002-8837-0035]{Julianne I. Moses}
\affiliation{Space Science Institute, Boulder, CO 80301, USA}
\email{jmoses@SpaceScience.org}

\author[0000-0001-6396-8439]{William O. Balmer}
\affiliation{\jhu}
\email{wbalmer@stsci.edu}

\author[0000-0003-3708-241X]{Aneesh Baburaj}
\affiliation{\ucsd}
\affiliation{\ciera}
\email{ababuraj@ucsd.edu}

\author{Geoffrey A. Blake}
\affiliation{\gps}
\email{gab@caltech.edu}

\author[0000-0002-6773-459X]{Doug Johnstone}
\affiliation{\herzberg}
\affiliation{\victoria}
\email{Douglas.Johnstone@nrc-cnrc.gc.ca}

\author[0000-0003-0097-4414]{Yapeng Zhang}
\affiliation{\caltech}
\affiliation{51 Pegasi b Fellow}
\email{yapzhang@caltech.edu}

\author[0000-0002-5375-4725]{Heather A. Knutson}
\affiliation{\gps}
\email{hknutso2@caltech.edu}

\author{Dimitri Mawet}
\affiliation{\caltech}
\affiliation{\jpl}
\email{dmawet@astro.caltech.edu}

\author[0000-0002-5627-5471]{Charles Beichman}
\affiliation{\nexsci}
\affiliation{\jpl}
\email{chas@ipac.caltech.edu}

\author[0000-0003-0786-2140]{Klaus Hodapp}
\affiliation{\ifahilo}
\email{hodapp@ifa.hawaii.edu}

\author[0000-0002-3191-8151]{Marshall D. Perrin}
\affiliation{\stsci}
\email{mperrin@stsci.edu}

\author[0000-0002-9936-6285]{Quinn Konopacky}
\affiliation{\ucsd}
\email{qkonopacky@ucsd.edu}

\author[]{Michael Meyer}
\affiliation{\umich}
\email{mrmeyer@umich.edu}

\author[]{Geoffrey Bryden}
\affiliation{\jpl}
\email{geoffrey.bryden@jpl.nasa.gov}

\author[0000-0002-8963-8056]{Thomas P. Greene}
\affiliation{\ames}
\affiliation{\ipac}
\email{tgreene@ipac.caltech.edu}

\author[0000-0002-0834-6140]{Jarron Leisenring}
\affiliation{\arizona}
\email{jarronl@arizona.edu}

\author[0000-0001-7591-2731]{Marie Ygouf}
\affiliation{\jpl}
\email{marie.ygouf@jpl.nasa.gov}

\author[0000-0001-5578-1498]{Björn Benneke}
\affiliation{\uclagps}
\email{bbenneke@epss.ucla.edu}

\author[0000-0001-9164-7966]{Julie Inglis}
\affiliation{\gps}
\email{jinglis@caltech.edu}

\author[0000-0003-0354-0187]{Nicole L. Wallack}
\affiliation{\carnegiew}
\email{nwallack@carnegiescience.edu}

\begin{abstract}
With four giant planets ($m\sim5-10~\Mj$, $\Teff\sim900-1200$ K) orbiting between 15-70 au, HR 8799 provides an unparalleled testbed for studying giant planet formation and probing compositional trends across the protoplanetary disk. We present new JWST/NIRSpec IFU observations ($2.85-5.3~\mu$m, $R\approx2700$) that now include the spectrum of HR 8799 b, and higher signal-to-noise ratio spectra for HR 8799 c, d, and e compared to that in Ruffio \& Xuan et al. We detect CO, CH$_4$, H$_2$O, H$_2$S, CO$_2$, and for planet b, NH$_3$. We combine the NIRSpec spectra with $1-5 \mu$m photometry to perform atmospheric retrievals that account for disequilibrium chemistry and clouds, and allow C/H, O/H, N/H, and S/H to scale independently. While the four planets are similarly enriched in carbon and oxygen, with C/H and O/H between $3-5\times$ stellar, we observe a tentative trend of increasing S/H -- a tracer of refractory solids -- from $2-5 \times$ stellar with increasing orbital distance. From HR 8799 b's NH$_3$ abundance, we estimate $\rm N/H=21.2^{+16.2}_{-8.8}\times$ stellar, suggesting the outer planet accreted significant amounts of N-rich gas. Overall, the elemental abundance patterns we observe are consistent with a picture where planet b formed between the CO snowline and the more-distant N$_2$ snowline, while the inner planets accreted $3 \times$ stellar CO-enriched disk gas within the CO snowline. The excess volatile mass from pebble drift and evaporation implies an integrated pebble flux of $750 \pm 200$ \Me. The increase in the planets' S/H with orbital distance implies more solid accretion further out, which is quantitatively compatible with expectations from both pebble and planetesimal accretion ($2 \times$ Minimum Mass Solar Nebula) paradigms.

\end{abstract}

\section{Introduction}

HR 8799 was the first multi-planet system discovered via direct imaging and remains the highest multiplicity imaged system to date. With four giant planets \citep{Marois2008science, Marois2010}, debris disks both interior and exterior to the planetary orbits \citep[e.g.][]{Su2009, Booth2016, Faramaz2021, Boccaletti2024}, and an age of $\approx40$ Myr \citep{Zuckerman2011, Bell2015, Faramaz2021}, the system represents a prime laboratory for studying giant planet formation. The existence of multiple giant planets with similar masses, and that likely formed and remained in their present orbital order, makes the HR 8799 system uniquely powerful for tracing potential chemical trends in the planets' atmospheres.

From orbital stability analyses, as well as substellar evolutionary models, the planet masses are estimated to be between $5-10~\Mj$, where the outermost and faintest planet b has the lowest mass \citep[e.g.][]{Marois2010, Gozdziewski2014, wang_dynamical_2018, Zurlo2022}. Due to the relatively high planet masses and compact configuration, previous studies find that these planets have orbital periods consistent with a mean-motion resonance (MMR) chain of 8:4:2:1, and they need to be in such a MMR to maintain long-term stability \citep[e.g.][]{Fabrycky2010, Konopacky2016, wang_dynamical_2018, Godziewski2018, Godziewski2020, Zurlo2022}. A resonant configuration for the HR 8799 planets would suggest a history of convergent migration within a gas-rich circumstellar disk, and hydrodynamical simulations show that inward planetary migration can reproduce the current orbital architecture \citep{Zurlo2022}. Recent work shows that both inward and outward migration might have played a role in shaping the observed debris disk morphology today \citep{Poblete2025}. While the planets are likely to have migrated during their evolution, it remains unclear at what orbital distances the planetary cores formed, and where and when in the history of the circumstellar disk the planets accreted the bulk of their mass. 

There has also been a wealth of spectroscopic studies on the HR 8799 planets (for an excellent summary, see Sec 2 of \citealt{Nasedkin2024}). We highlight a few key findings here. In the near-infrared, these planets have redder colors compared to field brown dwarfs \citep[e.g.][]{skemer_Directly_2014, Bonnefoy2016} and their atmospheres are found to be cloudy and in chemical disequilibrium  \citep[e.g.][]{Janson2010, Bowler2010, barman2015, molliere_Retrieving_2020, Wang2022, Nasedkin2024}. For a long time, CO and H$_2$O were the only molecules convincingly detected in these planets \citep{konopacky_detection_2013, ruffio_Deep_2021}, with the detection of CH$_4$ remaining ambiguous until JWST/NIRSpec observations clearly resolved the $3.3~\mu$m CH$_4$ feature \citep{RuffioXuan2026}. Recently, JWST/NIRCam photometry also revealed the presence of CO$_2$ in these planets \citep{Balmer2025b}, which is suggestive of high atmospheric metallicity. This was confirmed in \citet{RuffioXuan2026}, who also presented new detections of H$_2$S, $^{13}$CO, and C$^{18}$O in these planets. 

Earlier atmospheric retrieval analyses have found evidence of high metallicities for HR 8799 e and c, though with large uncertainties \citep{molliere_Retrieving_2020, Wang2023}. 
More recently, \citet{Nasedkin2024} carried out extensive atmospheric analyses of all four HR 8799 planets using VLTI/GRAVITY observations and archival low-resolution spectroscopy and photometry. Their retrievals found very metal-enriched atmospheres with $\rm [C/H]\approx1-2$, disequilibrium  abundances of CO and CH$_4$, and cloudy atmospheres for all four planets. However, they caution that discrepancies between some of the archival low-resolution spectra could impact the exact abundance values they measure, and predicted that future work with JWST spectroscopy in the $3-5~\mu$m range would be sensitive to differences between some of their models that perform similarly well in the near-infrared. 

In a precursor paper to this work, \citet{RuffioXuan2026} performed atmospheric retrievals using JWST/NIRSpec moderate-resolution ($R\sim2700$) spectroscopy from $3-5~\mu$m and archival photometry from $1-5~\mu$m. They also found super-stellar metallicities for HR 8799 c, d, and e, but revise the values downward from \citet{Nasedkin2024} to $\rm [C/H] < 1.0$. Planet b was not included in \citet{RuffioXuan2026} since it was outside the NIRSpec IFU's field of view in that first observation of the system with JWST/NIRSpec. A novel advance reported in \citet{RuffioXuan2026} was the first measurement of S/H in these planets via H$_2$S, demonstrating that HR 8799 c and d are enriched in sulfur at comparable levels seen for carbon and oxygen, within measurement uncertainties. Their measured super-stellar S/H values show, for the first time, that the HR 8799 planets likely accreted a significant amount of solids from the circumstellar disk. It also highlighted the apparent similarity between the atmospheric abundances of the HR 8799~cde planets and the solar system gas giants, Jupiter and Saturn, suggesting possible similarities in their formation pathways.

Several recent studies have highlighted the value of measuring refractory abundances (e.g. Fe, Si, Mg) in exoplanet atmospheres as tracers of solid accretion \citep{Schneider2021b, Lothringer2021, Turrini2021, Pacetti2022, Chachan2023}. For widely-separated planets like those in HR 8799, sulfur can also be a powerful tracer for solids since it is only present in the solid phase of circumstellar disks at orbital distances $\gtrsim1$ AU \citep{Kama2019}. Measuring refractory abundances allows one to break some of the degeneracies from measurements of carbon and oxygen abundances alone, which are volatile elements that can be in either solid or gas phase at different locations in the disk. Indeed, most directly imaged planets (including HR 8799 bcde) and brown dwarfs have measured C/O ratios close to the solar and stellar value \citep[e.g.][]{ruffio_Deep_2021, Hoch2023, Xuan2024b, Zhang2024, Hsu2024_pds, Snellen2025}, which, by itself, provides limited insight on their formation history. This reveals the limitations of only measuring carbon and oxygen abundances from species such as CO and H$_2$O, which have been pointed out by more recent theoretical studies as well \citep[e.g.][]{mordasini_Imprint_2016, molliere_Interpreting_2022a, Crossfield2023, Chachan2023}. 

Besides C, O, and refractory elements such as S, other elements could also provide complementary constraints on planet formation. One element of particular interest is nitrogen, which is a highly volatile species. Nitrogen is almost entirely in the form of N$_2$ in protoplanetary disks \citep{oberg_bergin2021}, where the N$_2$ snowline is located even farther out than the CO snowline; CO condenses at $\sim30$~K while N$_2$ condenses at $\sim25$~K. Therefore, unless a planet forms at large distances beyond the N$_2$ snowline ($\sim$80~au for HR~8799), it should accrete nitrogen from disk gas. This makes N/H an excellent complement to S/H, which reflects the solid accretion history. Several studies have shown that the nitrogen abundance ratios (e.g. N/H, N/S) can provide important constraints on the formation location of a giant planet \citep{Piso+16,Cridland2020, Turrini2021, Ohno2023}. 
In the Solar System, observations with the Galileo probe and the Juno spacecraft have revealed a super-solar N/H for Jupiter \citep{Wong2004,Li+17,Li+20}, which motivated several studies to suggest a frigid environment at Jupiter's birth location \citep{Owen+99,Oberg2019,Bosman+19,OhnoUeda21} or the accretion of nitrogen-enriched disk gas due to pebble drift \citep[e.g.,][]{Mousis+19,Aguichine+22,Nakazawa&Okuzumi25}. 
In exoplanets and brown dwarfs, N/H can mainly be measured from NH$_3$ or HCN, since the bulk nitrogen reservoir of N$_2$ has negligible opacity in the near- and mid-IR.

In this paper, we present the JWST/NIRSpec spectrum of HR 8799 b, the outermost planet in the system, and even higher quality spectra for HR 8799 c, d, and e than those presented in \citet{RuffioXuan2026}. HR 8799 b has an orbital semi-major axis of $\approx70$ AU, an estimated mass of $\approx6\Mj$ \citep{Zurlo2022, Thompson2023}, and $\Teff\approx900-1000~$K \citep{Nasedkin2024}. We also provide updated abundance measurements of HR 8799 c, d, and e based on this new dataset. This paper is organized as follows. In Section~\ref{sec:obs}, we describe new observations of the HR 8799 system with the JWST/NIRSpec IFU, and the data reduction methods. In Section~\ref{sec:spec_analysis}, we describe the modeling framework, including the atmospheric retrieval setup. We discuss the choice of comparison stellar abundances in Section~\ref{sec:stellar_ab}, before presenting the results in Section~\ref{sec:results}. We discuss lessons learned, and the planet formation implications of our results in Section~\ref{sec:discuss}.

\section{Observations and data reduction}\label{sec:obs}
We observed the HR 8799 system with the JWST/NIRSpec IFU in moderate resolution spectroscopy mode ($R\sim2,700$; filter: F290LP; grating: G395H). The observations are part of Cycle 1 GTO program 1188 (PI: Klaus Hodapp), and were obtained on July 10 and 11 (UT) 2024 with a total integration time of 3.7 hr.\footnote{All the JWST data used in this paper can be found in MAST: \dataset[10.17909/8e4h-kd52]{http://dx.doi.org/10.17909/8e4h-kd52}. We make the reduced NIRSpec spectra available in Zenodo at: \url{https://doi.org/10.5281/zenodo.19355669}.} The observations were split into two observatory rolls, each with $\approx112$ min of integration time. All four planets were included in the IFU in this observation. The same GTO program also observed planets c, d, and e in an earlier observation on July 29 2023, and the results for those planets were published in \citet{RuffioXuan2026}. The star, and therefore planet b, was not included inside the IFU's field of view in the first epoch data from 2023. This is because the potential consequences of including the HR 8799 host star, which strongly saturates the NIRSpec detectors, in the center of the field of view were not well understood that early in the JWST mission. Indeed, the analysis presented in \citet{Ruffio2024} only demonstrated NIRSpec IFU high-contrast capability for a companion at significantly larger separation ($\sim1.6^{\prime\prime}$) compared to the HR~8799 inner planets; HR~8799~e is at $\sim0.4^{\prime\prime}$.  

In this paper, we analyze the second epoch observation, from 2024, for all four planets. While HR 8799 b is the main addition of this paper, we also analyze the 2nd epoch observations of HR 8799 c, d, and e since the new dataset has higher S/N than the 1st epoch dataset published in \citet{RuffioXuan2026}. The higher S/N is due to the longer integration time (3.7 hr vs 2.2 hr), but even more significantly the use of 4 groups per integration (compared to 2 groups in the 1st epoch) which significantly improves the detector noise. The S/N improvement between the two observations is a factor $\times3$ for HR~8799~c and $\times2$ for HR~8799~e for example. We note that in both the 1st and 2nd epoch NIRSpec observations, HR~8799~e landed on top of a stellar diffraction spike, so higher SNR could be achieved on this planet if the observations were scheduled differently.

The planet detection and spectral extraction are performed using the same methods already described in \citet{RuffioXuan2026}, which uses the framework introduced in \citet{Ruffio2024} and the python package \texttt{BREADS}\footnote{\url{https://github.com/jruffio/breads}} \citep{breads}. We briefly summarize the main steps below. 

\subsection{Data calibration and pre-processing}

First, initial reductions are performed using stage 1 of the JWST Science Calibration Pipeline \citep{Bushouse2023}. In the resulting \texttt{*\_rate.fits} files, the correlated 1/f read noise and charge transfer from the saturated stellar core are fitted, and subsequently subtracted, in each column of the detector using a combination of a spline and a Lorentzian profile centered on the host star position.
Flux calibrated detector images (\texttt{*\_cal.fits}) are then generated using stage 2 of the JWST pipeline. A wavelength-dependent centroid of the star is also recalibrated by fitting a STPSF (formerly WebbPSF) model directly to the detector images (i.e., point clouds) without reconstructing a spectral cube. (The wavelength-dependent centroid is a workaround for residual systematics in the NIRSpec IFU instrument model currently, not any true astrophysical position shift.) Then, a continuum normalized spectrum of the star is estimated directly from the speckle field using a spline model to fit the speckle continuum.

\subsection{Planet detection}
\begin{figure*}
    \centering
    \includegraphics[width=\linewidth]{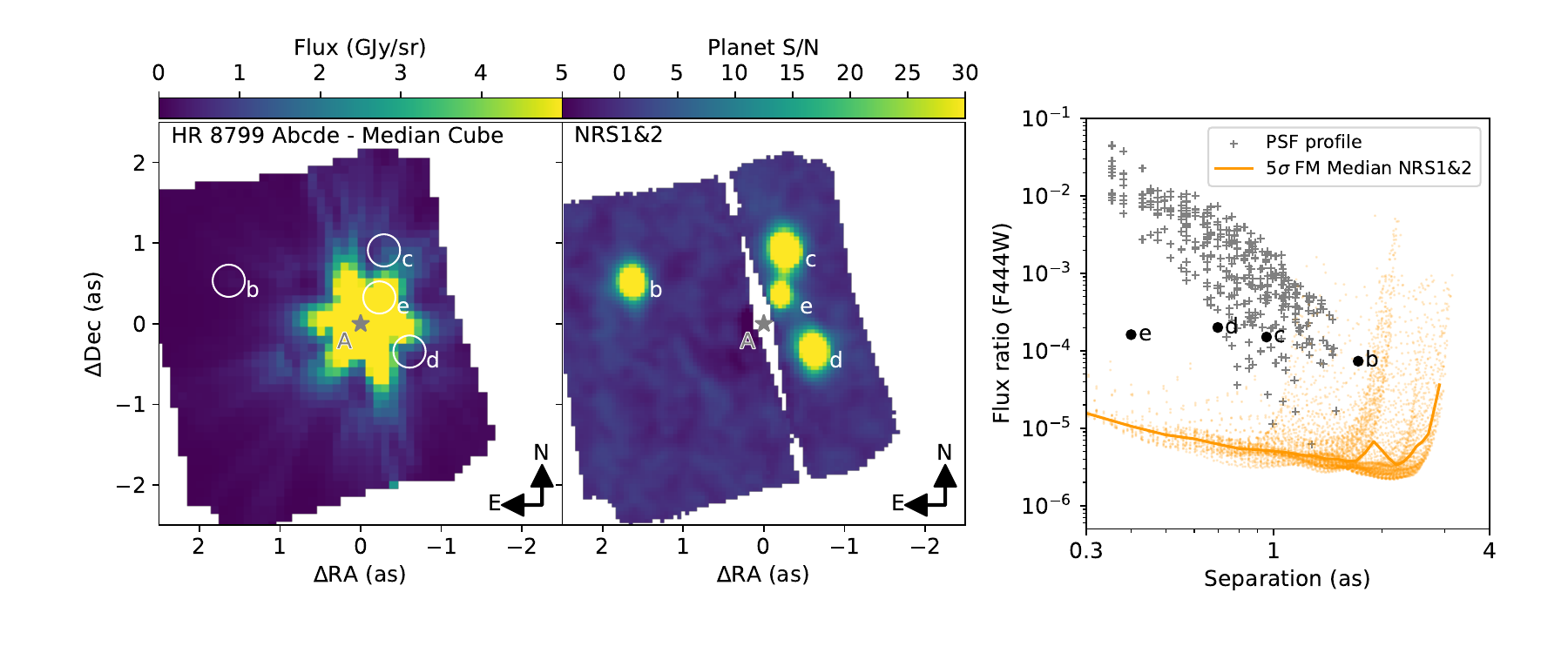}
    \caption{\textbf{\boldmath Detection of the four planets orbiting the star HR~8799 with the moderate resolution mode (R$\sim$2,700) of JWST/NIRSpec IFU between $3-5\,\mu$m.} 
\textbf{(Left)} Median spectral cube using the standard JWST calibration pipeline and combining the two observatory roll angles. This image does not include any PSF subtraction so the planets are hidden behind the starlight.
\textbf{(Middle)} Signal-to-noise ratio map for planet detection. HR~8799~b, c, d, and e are detected with an S/N of 134, 211, 157,and 79 respectively. The S/N calculation is systematics limited and S/N values have been normalized to yield a standard deviation of unity in the absence of planetary signal. The white vertical gaps arise from masking out the IFU slices containing the saturated stellar PSF core. \textbf{(Right)} Planet $5\sigma$ detection limits for the fully combined dataset with each dot representing a spatial pixel in the field of view. The flux ratio is defined in the F444W filter. We use the estimated planet-to-star flux ratios and reference stellar fluxes from ref. \cite{Balmer2025b}. The drop in sensitivity at the largest separations is due to the dithering pattern of the observation and edge effects. }
    \label{fig:snrmap}
\end{figure*}

The four HR 8799 planets are $\sim$1 to 50 times fainter than the speckles at their separations. The planets are therefore detected by fitting a joint model of the planet and the starlight directly to the detector images. A STPSF model combined with a $T_{\mathrm{eff}}=1200\,\mathrm{K}$ BTSettl atmospheric model \citep{Allard2003IAUS..211..325A} is used to model the planet signal. The starlight is modeled using the continuum normalized spectrum of the star combined with a 60-node spline to modulate the speckle continuum on each row of detector pixels.
An estimate of the planet S/N is then obtained at each position in the field of view. This means that a single BTSettl model is used across the entire field of view despite the differences in effective temperature between the planets. However, this approximation does not affect our ability to strongly detect all four planets as shown in Figure~\ref{fig:snrmap}. Similarly to \citet{RuffioXuan2026}, we normalize the S/N maps to ensure that the S/N values have a unit standard deviation when no planet is present. The S/N values are therefore divided by a normalization factor of 1.9 for NRS1 and 1.7 for NRS2.
The flux uncertainties are scaled accordingly to compute the $5\sigma$ detection threshold curves shown in Figure~\ref{fig:snrmap}.

\subsection{Spectral extraction}
For the spectral extraction of the planets, we effectively high-pass filter the dataset to remove the starlight. This step is performed by fitting the starlight-only model everywhere on the detector using the same spline prescription introduced in \citet{RuffioXuan2026}. In this case, spline nodes are placed non-uniformly across the spectral range of each detector to better account for the variable curvature of the slicer traces on the detector. Then, the high-pass filtered planetary spectra are extracted by fitting a STPSF model at each wavelength to the combined sequence of dithers for each observatory roll angle. The flux uncertainties and covariance matrix for each planet spectrum is estimated from the speckle residuals around the planets at similar projected separations.
We refer the reader to \citet{RuffioXuan2026} for more details.

\section{Atmospheric characterization}\label{sec:spec_analysis}
To study the spectra of the HR 8799 planets, we use an atmospheric retrieval framework built around the radiative-transfer code \texttt{petitRADTRANS} version 3.2.0 \citep{molliere_petitRADTRANS_2019, molliere_Retrieving_2020, Nasedkin2024}. This framework has been used extensively on directly imaged companions \citep{Xuan2024b}, and validated on benchmark brown dwarfs and low-mass stars \citep[e.g.][]{Xuan2022, Xuan2024, GWang2025}. 

For the HR 8799 planets, we fit both their NIRSpec spectra and archival photometry from $1-4.8~\mu$m. The photometry includes several medium-band filters from JWST/NIRCam and a single VLT/NACO point which together cover $3$–$4.8~\mu$m \citep{Balmer2025b, Currie2014_hr8799}, complemented by VLT/SPHERE $J$, $H$, $K$ band photometry from $1$–$2.2~\mu$m \citep{Zurlo2016}. The photometry data provide constraints on the spectral continuum of the planet, which has been subtracted in the NIRSpec spectrum during the spectral extraction. We add the log likelihoods from the JWST/NIRSpec spectra and photometry in these retrievals.  

To sample the posteriors, we use the open-source nested sampling package \texttt{pymultinest} \citep{Buchner2014}, which is based on  \texttt{MultiNest} \citep{Feroz2009, Feroz2019}. We adopt 1000 live points and stop sampling when the estimated contribution of the remaining prior volume to the total evidence is $<1\%$. 

\subsection{NIRSpec forward model}
The NIRSpec forward model is described in detail in \citet{RuffioXuan2026}, which we summarize here. For each model spectrum from \texttt{petitRADTRANS}, we apply the following steps. First, we apply a radial velocity (RV) shift to the model to align it with the data. Second, we apply instrumental broadening to the model and resample the model to the data wavelengths. We directly fit for the instrumental resolution as a function of wavelength ($\lambda$) using a linear relation $R_\lambda = r_0 + r\lambda$, where $r$ and $r_0$ are fitted parameters. In each iteration of the fit, the model spectrum is convolved with a variable Gaussian kernel whose standard deviation is calculated from $R_\lambda$ \citep[see][]{Xuan2024d}.

Then, we apply the same spline-based high-pass filtering on the model that we applied on the data, and compute the residuals between the forward model and the data. We account for the covariance matrix in computing the log likelihood as follows

\begin{equation}
    \ln \mathcal{L} = -\frac{1}{2} \left( R^\intercal C^{-1} R + n \ln(2\pi) + \ln\det(C) \right)
\end{equation}

where $R$ is the residual array, $C$ is the covariance matrix, $n$ is the number of data points, and ${\rm det}(C)$ is the determinant of the covariance matrix. Here $C$ has been multiplied by a factor of $e_{\rm mult}$, an error inflation term that we fit for. 

\subsection{Atmospheric retrieval setup}
The physical parameters we fit for include the planet mass and radius, temperature profile, chemical abundances, and cloud structure. We use Gaussian mass priors derived from the stability analysis in \citet{Zurlo2022}. Specifically, we adopt $e=7.6\pm0.9~\Mj$, $d=9.2\pm0.7~\Mj$, $c=7.7\pm0.7~\Mj$, and $b=5.8\pm0.4~\Mj$. The fitted parameters and priors are listed in Table~\ref{tab:param_prior}. The retrieval setup we use for HR 8799 b is similar to that described in \citet{RuffioXuan2026}. Here, we describe the main components and highlight any modifications. The framework described from Section~\ref{sec:opacities} to Section~\ref{sec:pt} is used for HR 8799 b. We implement some simplifications in the retrieval model for HR 8799 c, d, e in Section~\ref{sec:cde_setup}, mainly due to the fact that NH$_3$ was not detected in these three planets. 

\subsubsection{Opacities}\label{sec:opacities}
We include line opacities from CO and its two minor isotopologues $^{13}$CO and C$^{18}$O \citep{Rothman2010}, H$_2$O \citep{Polyansky2018} and HDO \citep{Voronin2010}, CO$_2$ \citep{Rothman2010}, CH$_4$ \citep{hargreaves_Accurate_2020}, NH$_3$ \citep{Coles2019}, HCN \citep{Barber2014}, H$_2$S \citep{Azzam2016}, Na \citep{Allard2019}, K (line profiles by N. Allard, \citealt{molliere_petitRADTRANS_2019}), and FeH \citep{Bernath2020}. For continuum opacities, we include the collision induced absorption (CIA) from H$_2$-H$_2$ and H$_2$-He. In our models, the reference for solar elemental abundances is \citet{asplund_Chemical_2009}.

\begin{deluxetable*}{ll|ll}
\tablecaption{Fitted Parameters and Priors for HR 8799 b Retrieval\label{tab:param_prior}}
\tabletypesize{\small}
\tablehead{
\colhead{Parameter} & \colhead{Prior} & \colhead{Parameter} & \colhead{Prior}
}
\startdata
Mass ($\Mj$)                 & $\mathcal{N}(\mu_{\rm M, dyn}, \sigma_{\rm M, dyn})^{\rm (a)}$ 
  & $T_{\rm ref}$ [$P=10^{2}$] (K) & $\mathcal{U}(2000, 4000)$ \\
Radius ($\Rj$)               & $\mathcal{U}(0.6, 2.0)$ 
  & $\left(d\ln{T}/d\ln{P}\right)_{1}$ [$10^2$] & $\mathcal{N}(0.15, 0.01)$ \\
$[{\rm C/H}]$                & $\mathcal{U}(0.0,1.2)$ 
  & $\left(d\ln{T}/d\ln{P}\right)_{2}$ [$10^1$] & $\mathcal{N}(0.18, 0.04)$ \\
$[{\rm O/H}]$                & $\mathcal{U}(0.0,1.2)$ 
  & $\left(d\ln{T}/d\ln{P}\right)_{3}$ [$10^0$] & $\mathcal{N}(0.21, 0.05)$ \\
$[{\rm N/H}]$                & $\mathcal{U}(-0.5,2.0)$ 
  & $\left(d\ln{T}/d\ln{P}\right)_{4}$ [$10^{-1}$] & $\mathcal{N}(0.16, 0.06)$ \\
\logco                       & $\mathcal{U}(0, 8)$ 
  & $\left(d\ln{T}/d\ln{P}\right)_{5}$ [$10^{-2}$] & $\mathcal{N}(0.08, 0.025)$ \\
\logcoo                      & $\mathcal{U}(0, 8)$ 
  & $\left(d\ln{T}/d\ln{P}\right)_{6}$ [$10^{-3}$] & $\mathcal{N}(0.06, 0.02)$ \\
\loghdo                      & $\mathcal{U}(0, 8)$ 
  & $\left(d\ln{T}/d\ln{P}\right)_{7}$ [$10^{-4}$] & $\mathcal{U}(-0.05, 0.10)$ \\
log(CO$_2$) mass–mixing ratio & $\mathcal{U}(-10, -2)$ 
  & $\left(d\ln{T}/d\ln{P}\right)_{8}$ [$10^{-5}$] & $\mathcal{U}(-0.05, 0.10)$ \\
log(HCN) mass–mixing ratio   & $\mathcal{U}(-10, -2)$ 
  & $\left(d\ln{T}/d\ln{P}\right)_{9}$ [$10^{-6}$] & $\mathcal{U}(-0.05, 0.10)$ \\
log(H$_2$S scale factor)     & $\mathcal{U}(-3, 2)$ 
  & $\left(d\ln{T}/d\ln{P}\right)_{10}$ [$10^{-7}$] & $\mathcal{U}(-0.05, 0.10)$ \\
log($P_{\rm quench, C}$/bar)    & $\mathcal{U}(-5, 2)$ 
  & RV (\kms) & $\mathcal{U}(-50 , 50)$ \\
log($P_{\rm quench, diff}$/bar) & $\mathcal{U}(0, 2)$ 
  & Error multiple$^{\rm (b)}$ & $\mathcal{U}(1, 5)$ \\
$\log(r_{\rm cloud} / \rm {cm})$  & $\mathcal{U}(-7, 1)$ 
  & $r$ & $\mathcal{U}(300, 1400)$ \\
$\log(P_{\rm cloud}/{\rm bar})$ & $\mathcal{U}(-6, 1.5)$ 
  & $r_0$ & $\mathcal{U}(-800, 1200)$ \\
$\sigma_{\rm g}$ & $\mathcal{U}(1.05, 3)$  &  \\
${\rm log}(X_{\rm cloud})$ & $\mathcal{U}(-8, 0)$ & \\
\enddata
\tablecomments{
$\mathcal{U}$ denotes a uniform distribution with bounds in parentheses while $\mathcal{N}$ denotes a Gaussian distribution with the mean and standard deviation in parentheses. The cloud parameters and P-T profile parameters are described in Section~\ref{sec:clouds} and Section~\ref{sec:pt}. Logarithmic values are all base 10. The equivalent table for planets c, d, and e are in Appendix~\ref{app:cde_table}. \\
$^{\rm (a)}$ Mass prior from the stability analysis in \citet{Zurlo2022}, which is $5.8\pm0.4~\Mj$ for HR 8799 b. \\
$^{\rm (b)}$ Error inflation term, which is multiplied to the covariance matrix. 
}
\end{deluxetable*}

\subsubsection{Chemistry}\label{sec:chemistry}
The default equilibrium chemistry grid in \texttt{petitRADTRANS} is parametrized by C/O and [C/H], where the latter acts as a global metallicity scale \citep{molliere_Retrieving_2020}. In other words, it implicitly assumes [C/H]=[S/H]=[N/H], etc. For HR 8799 b, we wish to measure [N/H] and [S/H] separately from [C/H]. To do so, we construct a new chemical grid where the abundances are parameterized by four parameters, [C/H], [O/H], [N/H], and a scale factor for the H$_2$S abundance. Ideally, the chemical grid would include a dimension for [S/H] as well; however, in practice, we find that a grid with four elemental abundances (and the necessary P and T ranges) requires $10$ GB or more memory space, which was impracticable from a computational standpoint.

We use the \texttt{easyCHEM} package \citep{Lei2025} to compute chemical abundances for different species, assuming equilibrium chemistry and including condensation. Our chemical grid spans $[60\,\mathrm{K},\,3500\,\mathrm{K}]$ in temperature ($T$), $[10^{-7}\,\mathrm{bar},\,10^{2}\,\mathrm{bar}]$ in pressure ($P$), and $[0~\mathrm{dex},\,1.2~\mathrm{dex}]$ for [C/H] and [O/H] with 0.06 dex spacing. For [N/H], the grid covers -0.5 to 2.0 dex with 0.1 dex spacing. We adopt a larger range for [N/H] because it is less well-constrained compared to C and O. The grid preferentially covers super-solar values because preliminary retrievals show HR 8799 b's atmosphere to be enriched in C, O, S, and N. We validated the results from the custom grid against the default \texttt{petitRADTRANS} grid for the same [C/H], [O/H], and [N/H] values, and found excellent agreement. 

Out of the box, our grid assumes [C/H]=[S/H]. Therefore, to measure [S/H] from the data, we fit a $f_{\rm H_2S}$ parameter, which adjusts the H$_2$S abundance up or down from the equilibrium value set by [C/H], following \citet{RuffioXuan2026}. From tests with \texttt{easyCHEM}, we find that the H$_2$S abundance tracks well with [S/H], such that the approximation 
\begin{equation}\label{eq:scale_fac}
    {\rm [S/H]} = f_{\rm H_2S} + {\rm [C/H]}
\end{equation}
holds at the $<5\%$ level for the parameter space covered by the grid. On the other hand, the NH$_3$ abundance scales sub-linearly with [N/H] because the dominant nitrogen carrier is N$_2$ in the deep atmosphere of HR 8799 b (see Section~\ref{sec:nitrogen} for a discussion), so fitting a scale factor for the NH$_3$ abundance would not be valid. 

To account for the effects of vertical transport-induced disequilibrium chemistry, we fit several additional parameters. First, we allow carbon quenching by fitting a log($P_{\rm quench, C}$) parameter, which sets the abundances of CO, H$_2$O, and CH$_4$ at $P<P_{\rm quench, C}$ to be equal to their values at $P=P_{\rm quench, C}$ \citep{Zahnle_methane_2014}. To account for NH$_3$ and N$_2$ quenching, we also fit a log($P_{\rm quench, diff}$) parameter 
\begin{equation}\label{eq:quench_diff}
    {\rm log(P_{quench, N}) = log(P_{quench, C}) + log(P_{quench, diff}) }
\end{equation}
such that the abundances of NH$_3$ and N$_2$ at $P<P_{\rm quench, N}$ is equal to the values at $P=P_{\rm quench, N}$. Here, we use the fact that the quench pressure for nitrogen disequilibrium chemistry should always be deeper than that of carbon \citep[e.g.][]{Moses2016, Mukherjee2022}. 

Our carbon disequilibrium does not handle CO$_2$, which \citet{Beiler2024b} showed to be inaccurately captured by a quenching timescale approximation. From chemical kinetics models, the CO$_2$ abundance is nearly constant over pressure for most of the observable atmosphere \citep{Beiler2024b, Wogan2025}. Therefore, in our disequilibrium chemistry models, we treat CO$_2$ separately by retrieving a constant-with-pressure abundance. The disequilibrium  abundance of HCN is not well-described by a single quench pressure, as CH$_4$-CO quenching and N$_2$-NH$_3$ quenching both impact the abundance of HCN. For simplicity, we also treat HCN separately with a constant-over-pressure abundance. We tested retrievals where we quench HCN at $P_{\rm quench, C}$, but found that these were not favored over fitting a constant-with-pressure HCN abundance.

In addition to the disequilibrium chemistry models, we run several `free retrievals' where the abundance of each molecule is assumed to be constant over pressure. We include the following molecules in the baseline free retrieval: CO, $^{13}$CO, C$^{18}$O, CO$_2$, CH$_4$, H$_2$O, HDO, NH$_3$, HCN, and H$_2$S. For each planet, we perform leave-one-out experiments where we remove one molecule and re-run the free retrievals in order to validate molecule and isotopologue detections (see Section~\ref{sec:ccf_molecule}).

\subsubsection{Clouds}\label{sec:clouds}

The EddySed model \citep{ackerman_Precipitating_2001} in \texttt{petitRADTRANS} has been widely adopted \citep[e.g.][]{molliere_Retrieving_2020, Xuan2022, Zhang2023, RuffioXuan2026} in retrieval studies. In this model, the cloud base pressure $P_\mathrm{base}$ is determined by the intersection of the P-T profile and the cloud condensation curve, while the eddy diffusion coefficient and sedimentation efficiency $f_{\rm sed}$ together set the mean cloud particle size \citep{ackerman_Precipitating_2001}. 

Recent studies have pointed to limitations of the EddySed cloud model \citep[e.g.][]{luna_Empirically_2021, Molliere2025}, in particular the lack of flexibility in setting the cloud base pressure and cloud particle sizes. Motivated by this, we adopt a more flexible cloud model in this paper, similar to that in \citet{Nasedkin2025}. In our modified cloud model, we still use optical constants for different cloud compositions. For the HR 8799 planets, we choose MgSiO$_3$ and Fe clouds following \citet{Nasedkin2024}. Then, we freely retrieve the mean cloud particle radius ($r_c$) and the cloud base pressure ($P_{\mathrm{base}}$). The particle size distribution is assumed to be lognormal, and we retrieve for its width $\sigma_g$. In this model, the cloud mass fraction is assumed to be constant with pressure above the cloud base, which corresponds to the expectation for well-mixed atmosphere \citep{Gao2018}. Finally, $X_{\rm cloud}$ is the cloud mass fraction at the cloud base. Each cloud species has a different $r_c$, $P_{\mathrm {base}}$, and $X_{\rm cloud}$.

\subsubsection{Thermal structure}\label{sec:pt}
In this paper, we use the P-T parametrization from \citet{Zhang2023}, which fits for temperature gradients $(d\ln{T}/d\ln{P})$ in different pressure layers, and a single reference temperature, $T_{\rm ref}$. To match the pressure range of the emission contribution function, we adopt 10 pressure layers spaced logarithmically between $10^{2}$ and $10^{-7}$ bars, and fit the reference temperature at $10^2$ bars. Given $T_{\rm ref}$ and the quadratically interpolated $(d\ln{T}/d\ln{P})$ values, we construct the full P-T profile over a grid of 100 pressure layers for the radiative transfer. Following \citet{Zhang2025}, we apply Gaussian priors on the temperature gradients between $10^2-10^{-3}$ bars, which are derived from a set of self-consistent Sonora Diamondback models \citep{Morley2024}. For the temperature gradients outside this pressure range, we apply wide uniform priors (see Table~\ref{tab:param_prior}). 

\begin{deluxetable*}{lccccccc}\label{tab:star_abunds}
\tabletypesize{\footnotesize}
\tablecaption{Elemental Abundances of Stars in the Columba and Carina Associations}
\tablehead{
\colhead{Star} & \colhead{[C/H]} & \colhead{[N/H]} & \colhead{[O/H]} & \colhead{[S/H]} & Association & Abundance reference
}
\startdata
HD 21997             & $0.00 \pm 0.10$ & \nodata            & $-0.03 \pm 0.12$ & $-0.09 \pm 0.10$ & Columba (BF) & 1 \\
HD 49855             & $-0.11 \pm 0.02$ & $-0.32 \pm 0.03$ & $-0.06 \pm 0.03$ & $-0.01 \pm 0.05$ & Carina (BF) & 2 \\
HD 55279             & $0.11 \pm 0.02$ & $-0.13 \pm 0.03$ & $-0.05 \pm 0.02$ & $-0.01 \pm 0.05$ & Carina (BF) & 2 \\
HD 269620            & $-0.14 \pm 0.03$ & $-0.03 \pm 0.02$ & $0.02 \pm 0.04$  & $-0.03 \pm 0.04$ & Columba (CM) & 2 \\
CD-63 336            & $-0.14 \pm 0.03$ & $-0.18 \pm 0.03$ & $-0.07 \pm 0.05$ & $-0.16 \pm 0.04$ & Carina (CM) & 2 \\
HR 8799 (Baburaj+25) & $0.13 \pm 0.04$ & \nodata            & $0.10 \pm 0.07$  & $-0.22 \pm 0.09$ & Columba/Carina & 3\\
\hline
Weighted mean $\pm$ scatter   & $-0.03 \pm 0.11$ & $-0.13 \pm 0.10$ & $-0.05 \pm 0.06$ & $-0.06 \pm 0.08$ & \\
\enddata
\tablecomments{The values in the final row are the weighted mean of all stars and the standard deviation of the different values. The association information is taken from MOCA database (Gagné et al., in prep; \citealt{Gagne2018, Gagne2024}) for all stars except HR 8799, whose membership in Columba or Carina is described in \citet{Faramaz2021}. BF refers to a bona fide member while CM refers to a candidate member in the association \citet{MOCAPY}.}
\tablerefs{(1) \citet{Borthakur2025}, (2) \citet{Apogee2022}, (3) \citet{Baburaj2025}
}
\end{deluxetable*}

\subsubsection{Model simplifications for HR 8799 c, d, e}\label{sec:cde_setup}
For planets c, d, and e, we do not detect NH$_3$ at the $>3\sigma$ level (see Section~\ref{sec:ccf_molecule}), likely due to their hotter $\Teff$ compared to HR 8799 b. For this reason, we simplify the retrieval in two ways. First, we do not fit a log($P_{\rm quench, diff}$) parameter to account for NH$_3$ quenching. Secondly, we use the default chemical grid of \texttt{petitRADTRANS} for these three planets, which does not independently vary N/H. We still fit for the $f_{\rm H_2S}$ parameter to allow S to scale independently of C and O, and report only C/H, O/H, and S/H for HR 8799 c, d, and e. The fitted parameters and priors for these three planets are provided in Table~\ref{tab:param_prior_cde}.

\section{Stellar abundances}\label{sec:stellar_ab}

\subsection{Challenges related to the composition of HR~8799~A}
Atmospheric abundance measurements for exoplanets must be compared to the stellar values, which are typically a proxy for the natal circumstellar disk abundances at the planet formation epoch \citep{Reggiani2024}. However, this connection between the composition of the star and the disk is complicated for HR 8799 A as it is a $\lambda$ Boo type chemically peculiar star. $\lambda$ Boo stars are A-type stars with significant surface depletion of iron-peak elements such as Fe, Ni, and Mn. While $\lambda$ Boo stars have higher abundances in lighter elements such as C, O, N, and S compared to iron-peak elements, these lighter elements also show scatter from star to star, and are not strictly solar \citep{Kamp2001}. The origin of $\lambda$ Boo stars is still a matter of debate, but possibilities include the accretion of gas from a circumstellar object or interactions with a diffuse ISM cloud \citep{Jura2015, Alacoria2022}, the accretion of volatile-rich comets \citep{Gray2002}, or embedded planets that deplete the dust \citep{Kama2015, Jermyn2018}. 

Recently, \citet{Baburaj2025} measured the abundances of several elements in HR 8799 A using optical high-resolution spectra from the Automated Planet Finder \citep{Vogt2014} at Lick Observatory. They find that HR 8799 A's photosphere shows a slightly super-solar carbon abundance ([C/H]=$+0.13\pm0.04$ dex), while the sulfur abundance is sub-solar ([S/H]=$-0.22\pm0.09$ dex).
In \citet{RuffioXuan2026}, we therefore opted to use the abundance of HR~8799 from \citet{Baburaj2025} to normalize the abundances of planets c, d, and e. This resulted in a nearly uniform abundance pattern of C, O, and S for planet c with C/S$=0.9_{-0.2}^{+0.3}\times$ stellar. In that previous paper, sulfur was only marginally detected at $<2\sigma$ in the atmospheres of HR~8799~de, but the inferred abundances were also consistent with a uniform enrichment of C, O, and S.
However, the 0.35 dex difference between [C/H] and [S/H] for the star in \citet{Baburaj2025} ($>3\sigma$ based on quoted uncertainties) has a sizable effect in interpreting the retrieved planetary C and S abundances; it influences the planet C/S ratio by a factor of $\approx2.25$. We note that [C/H] and [O/H] were measured from a spectral fitting method, while [S/H] was measured from an equivalent widths method.
Furthermore, the $\lambda$ Boo nature of HR 8799 A suggests that there could be physical processes that alter the star's photospheric abundances away from the planet-forming disk abundances in unpredictable ways. All of these factors combined motivated us to look for more independent and reliable proxies of the stellar composition of HR~8799~A in this work, which we discuss in the next subsection. 

\begin{figure*}[t]
\centering
\includegraphics[trim={0cm 0cm 0cm 0cm},clip,width=1\linewidth]{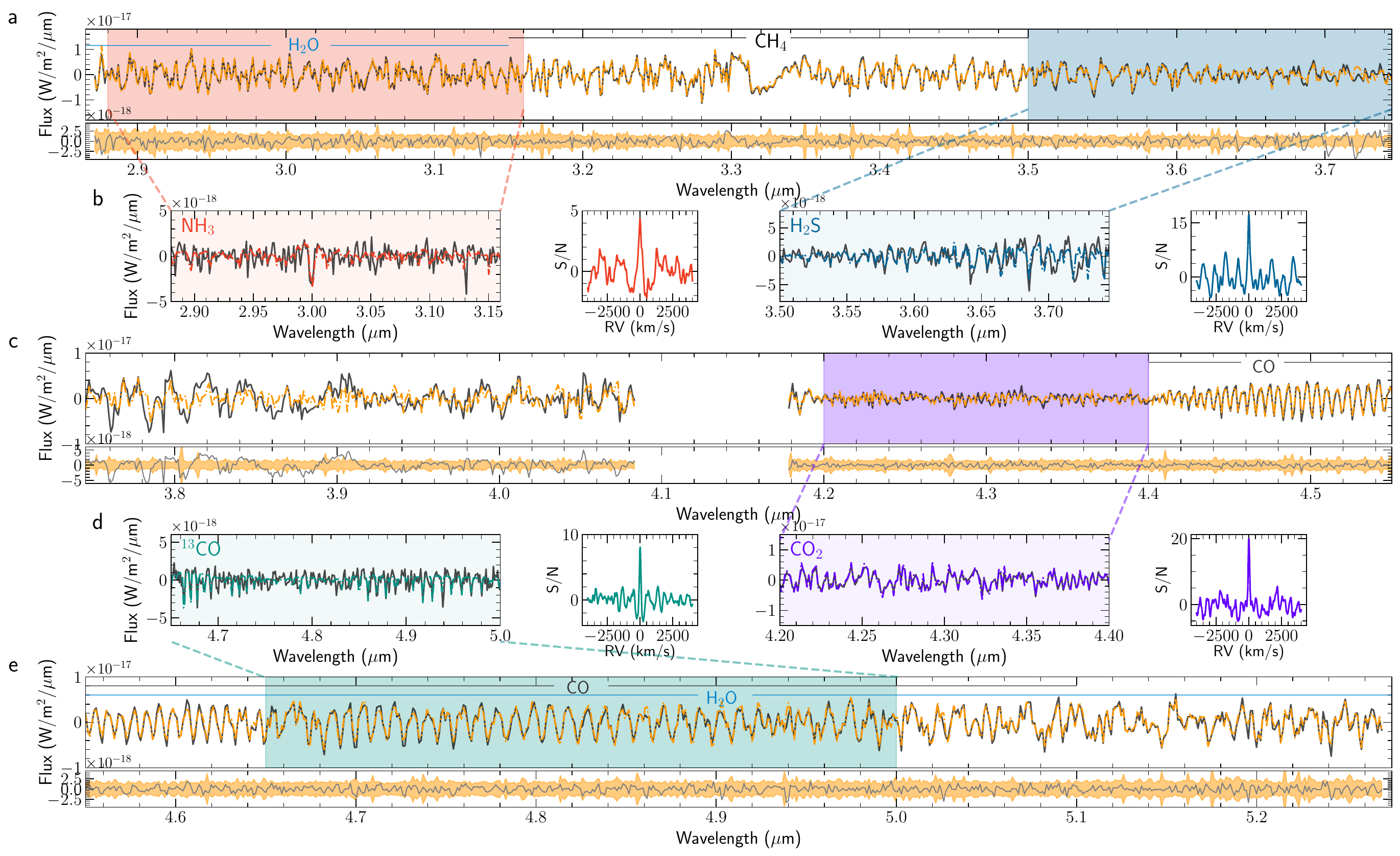}
\caption{\textbf{JWST/NIRSpec spectrum of HR 8799 b.} Panels a, c, e show the observed spectrum ($R\sim2,700$) in black and the best-fit model in orange. In the sub-panels below, the corresponding residuals after subtracting the best-fit \texttt{petitRADTRANS} model are plotted as gray lines and the $2.5\sigma$ uncertainties are shown as orange contours. The factor of 2.5 comes from the best-fit error scaling factor, and is largely driven by the enhanced residuals from $3.75-3.9~\mu$m. Panels b and d show data residuals after fitting an atmospheric model without a given species (NH$_3$, H$_2$S, $^{13}$CO, CO$_2$) in black, and the corresponding molecular templates in color. The similarity between the data residuals and molecular templates indicate that the highlighted species contribute significantly to the planet's spectra. On the right insets, we plot the cross-correlation functions (CCF) between the data residuals and models in the left insets. The CCF provides an estimate of the detection S/N for each molecule or isotopologue.
}\label{fig:spec}
\end{figure*}

\begin{figure}
    \centering
    \includegraphics[width=\linewidth]{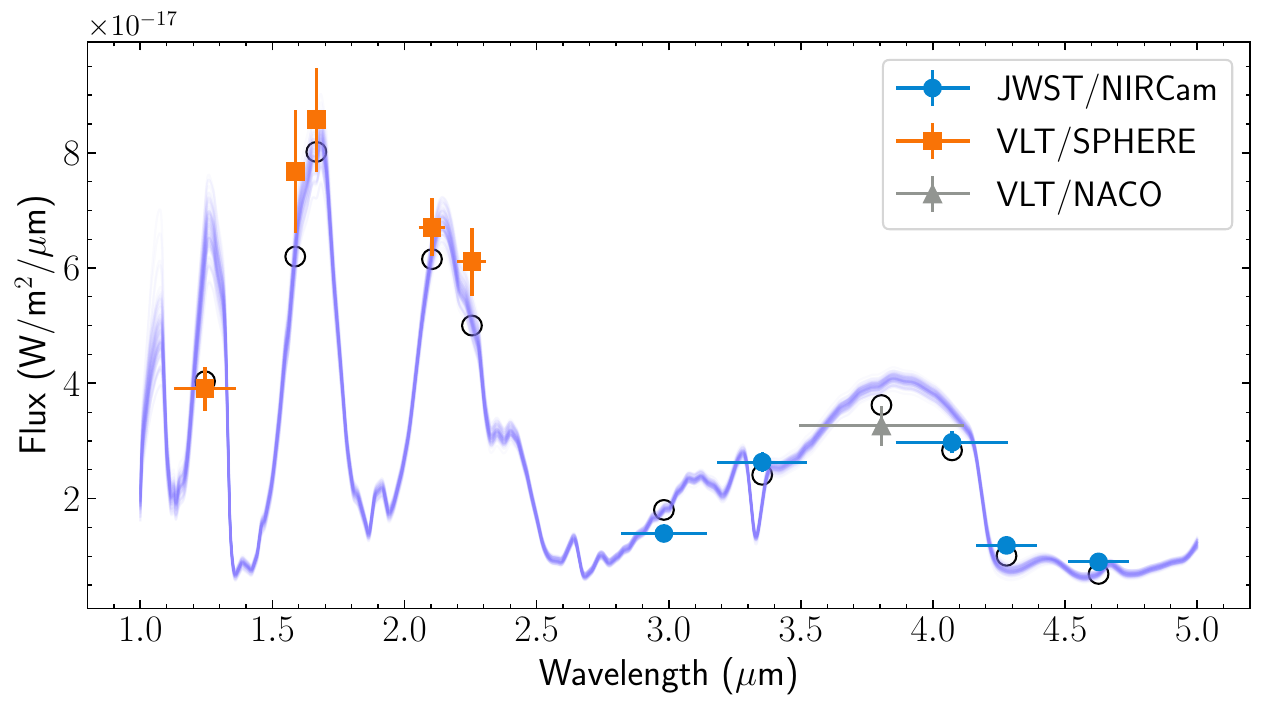}
    \caption{The colored points show photometry data of HR 8799 b from VLT/NACO, VLT/SPHERE, and JWST/NIRCam \citep{Currie2014_hr8799, Zurlo2016, Balmer2025b} used in the retrievals. We show the best-fit photometry model in open circles. Random draws of the model spectrum at $R = 100$ are overlaid in light purple.}
    \label{fig:phot}
\end{figure}

\subsection{A revised estimate of the stellar abundances}
Following the guidelines in \citet{Reggiani2024}, we turn to stars that likely formed within the same molecular cloud as HR 8799 A. HR 8799 A is most likely a member of the Columba or Carina associations, though the star is somewhat detached from these associations in its kinematics. \citet{Faramaz2021} carried out an extensive analysis of potential memberships of HR 8799 A, and concluded that several young stellar groups including Columba and Carina likely all formed nearly contemporaneously in separate star formation bursts within the same molecular cloud. 

Therefore, we compiled C, O, N, and S abundance measurements for other stellar members in Columba and Carina which do not show any chemical peculiarity (see Table~\ref{tab:star_abunds}). We computed the weighted average and scatter of the measurements as a proxy for the abundances of the parent molecular cloud where HR 8799 most likely originated. We find that the average abundances are consistent with solar photospheric abundances at the $<1\sigma$ level for carbon, oxygen, and sulfur, and $<1.5\sigma$ level for nitrogen. This finding is in line with abundance measurements for stars in other star-forming regions and young moving groups, such as Taurus, Scorpius–Centaurus, and the $\beta$ Pic moving group, whose members also have solar abundance values \citep[e.g.][]{Santos2008, DOrazi2011, Biazzo2017, Reggiani2024, Hejazi2025}. Based on these results, we choose to adopt the solar abundance as the reference for interpreting the measurements for the HR 8799 planets.
To be consistent with the retrievals, we adopt the \citet{asplund_Chemical_2009} solar photospheric abundance in this work. 
While this choice of stellar abundances differs from \citet{RuffioXuan2026}, it does not negate the general conclusions therein of an efficient solid accretion in HR~8799~cde based on an elevated sulfur abundance and near uniform metal enrichment (C, O, and S) within the larger uncertainties of the original analysis. 


\section{Results}\label{sec:results}

In this section, we present the atmospheric analysis results of the four HR 8799 planets. We first summarize the molecules detected in the NIRSpec data in Section~\ref{sec:ccf_molecule}, before discussing the bulk atmospheric properties in Section~\ref{sec:clouds_pt}. Next, we describe the constraints on the vertical diffusion coefficients of the planets' atmospheres from the observed disequilibrium  carbon chemistry (Section~\ref{sec:quench}). In Section~\ref{sec:nitrogen}, we describe the challenges with measuring N/H from NH$_3$ in retrievals, and present a self-consistent analysis with VULCAN to provide a better N/H estimate based on the retrieval results. Finally, Section~\ref{sec:abunds} summarizes the elemental abundances we measure for the planets. 

For HR 8799 b, the best-fit model and NIRSpec data are shown in Figure~\ref{fig:spec}, whereas the photometry data and model are shown in Figure~\ref{fig:phot}. The NIRSpec spectra for planets c, d, and e are shown in Appendix~\ref{app:pt_cde}.

\subsection{Molecular detections}\label{sec:ccf_molecule}
In the atmosphere of HR 8799 b, we confidently detect CO, $^{13}$CO, H$_2$O, CH$_4$, CO$_2$, H$_2$S, and NH$_3$ ($>4\sigma$), and obtain tentative detections of C$^{18}$O, HCN and HDO ($\approx2-3\sigma$). For HR 8799 c and d, we confidently detect CO, $^{13}$CO, C$^{18}$O, H$_2$O, CH$_4$, CO$_2$ and H$_2$S (see Figures~\ref{fig:hr8799c}, ~\ref{fig:hr8799d}). The strong detection of H$_2$S in HR 8799 d, a major improvement from the 1st epoch data in \citet{RuffioXuan2026}, now allows us to constrain the sulfur abundances across three different planets (d, c, b). Finally, HR 8799 e shows $>4\sigma$ of CO, $^{13}$CO, H$_2$O, CH$_4$, and CO$_2$ (Figures~\ref{fig:hr8799d}). H$_2$S is not detected in planet e even in the 2nd epoch observation, so longer integration times and a more optimal placement of the planet with respect to stellar diffraction features (see Section~\ref{sec:obs}) are needed to detect H$_2$S in this planet. 

To validate these detections, we run a set of free retrievals where we leave one molecular species out at a time. We perform these tests for CO$_2$, H$_2$S, $^{13}$CO and C$^{18}$O, NH$_3$, and HDO. We first compare the Bayesian evidence between these leave-one-out `reduced' models with the full model, which includes the full set of species. The Bayes factors for including each species are listed in Table~\ref{tab:bayes_free}. 

Next, we plot the residuals of the reduced models against a single molecular template, and calculate the cross-correlation function (CCF) between them (Figure~\ref{fig:spec}). For example, the reduced model for H$_2$S is a model where we did not include H$_2$S opacities, so the data residuals after subtracting this reduced model contains H$_2$S lines from the planet. In this example, the single molecular template for H$_2$S is constructed by taking the difference between the full model and an otherwise identical model in which the H$_2$S opacity is set to zero.

The CCFs are computed following the template matching S/N in \citet{Ruffio2017} (Eq.~18) while accounting for the inflated covariance matrices, which allows us to estimate the CCF S/N. The S/N of a molecular detection obtained this way represents an independent, and often more reliable, assessment of detection S/N \citep[e.g.][]{zhang_13COrich_2021, Xuan2024}. For HR 8799 b, we show the CCFs as insets in Figure~\ref{fig:spec} for H$_2$S, $^{13}$CO, CO$_2$, and NH$_3$, and also report the values in Table~\ref{tab:bayes_free}. The ambiguous case of HDO and the weaker detections of HCN and C$^{18}$O are discussed in Appendix~\ref{app:hdo}. 

We note that the isotopologue ratios of the planets will be reported and discussed in a follow-up paper by Kesseli \& Xuan et al. in prep.

\begin{deluxetable*}{lccccccc}
\tablecaption{Molecular detection significances in HR 8799 bcde \label{tab:bayes_free}}
\tablehead{
\colhead{Significance} &
\colhead{CO$_2$} &
\colhead{H$_2$S} &
\colhead{NH$_3$} &
\colhead{$^{13}$CO} &
\colhead{C$^{18}$O} &
\colhead{HCN} &
\colhead{HDO}
}
\startdata
\textbf{HR 8799 b} \\
\hline
$\Delta\ln B$ for full model &
234.6 & 144.1 & 8.8 & 32.6 & 2.4 & -0.5 & 13.1 \\
CCF S/N &
20.0 & 17.2 & 4.3 & 8.1 & 3.4 & 2.2 & 3.0 \\
\hline
\textbf{HR 8799 c} \\
\hline
$\Delta\ln B$ for full model &
372.2 & 61.3 & -- & 233.4 & 28.2 & -- & -- \\
CCF S/N &
28.5 & 11.2 & -- & 23.6 & 7.8 & -- & -- \\
\hline
\textbf{HR 8799 d} \\
\hline
$\Delta\ln B$ for full model &
213.2 & 15.6 & -- & 200.0 & 29.3 & -- & -- \\
CCF S/N &
21.0 & 5.8 & -- & 21.4 & 7.5 & -- & -- \\
\hline
\textbf{HR 8799 e} \\
\hline
$\Delta\ln B$ for full model &
32.8 & 0.7 & -- & 14.4 & 1.0 & -- & -- \\
CCF S/N &
7.4 & $<2$ & -- & 5.8 & $<2$ & -- & -- \\
\enddata
\tablecomments{
Log Bayes factor differences ($\Delta\ln B$) are computed by comparing the full model with the leave-one-out models. Positive values indicate that the full model is favored and therefore that a given molecule is supported by the data. CCF S/N values provide an independent estimate of detection significance. For HR 8799 b, HCN and HDO are only tentatively detected (see Appendix~\ref{app:hdo}). 
}
\end{deluxetable*}

\renewcommand{\arraystretch}{1.4} 
\begin{deluxetable*}{c|ccccccc}[t!]
\tablecaption{\textbf{Results of Spectral Retrievals for HR 8799 b, c, d, and e.}\label{table:spec_results}}
\tablewidth{\textwidth}
\tablehead{
\colhead{Planet} & \colhead{C/H} & \colhead{O/H} & \colhead{S/H} & \colhead{N/H} &  \colhead{Radius ($\Rj$)} & \colhead{$T_\textrm{eff}$ (K)} & \lbollsun }
\startdata
HR 8799 b & $4.2^{+0.9}_{-0.8}$ & $4.9^{+1.0}_{-0.8}$ & $5.2^{+0.8}_{-0.7}$ & $21.2^{+16.2}_{-8.8}\,^{a}$ & $1.03\pm0.03$ & $930\pm15$ & $-5.12\pm0.01$ \\
HR 8799 c & $4.4^{+0.8}_{-0.7}$ & $3.8^{+0.9}_{-0.8}$ & $2.6\pm0.4$ & ... & $1.14\pm0.03$ & $1120\pm15$ & $-4.70\pm0.01$  \\
HR 8799 d & $5.3^{+1.2}_{-1.1}$ & $4.1^{+1.1}_{-0.9}$ & $2.0^{+0.4}_{-0.3}$ & ... & $1.37\pm0.04$ & $1075\pm20$ & $-4.62\pm0.01$ \\
HR 8799 e & $3.3^{+1.2}_{-0.9}$ & $2.8^{+1.0}_{-0.7}$ & $1.4_{-1.1}^{+1.7}$ & ... & $0.98\pm0.05$ & $1230\pm35$ & $-4.68\pm0.02$ \\
\enddata
\tablecomments{Selected atmospheric parameters and their central 68\% credible interval with equal probability above and below the median. The C/H, O/H, S/H, and N/H values are relative to solar abundances from \citet{asplund_Chemical_2009}, and fold in uncertainties in the solar abundances. N/H is only reported for HR 8799 b, as we do not detect NH$_3$ to $>3\sigma$ confidence in the other planets. \\
$^{\rm (a)}$ The N/H value for HR 8799 b is derived from the VULCAN analysis in Section~\ref{sec:nitrogen}.}

\end{deluxetable*}

\begin{figure}[t!]
    \centering
    \includegraphics[width=\linewidth]{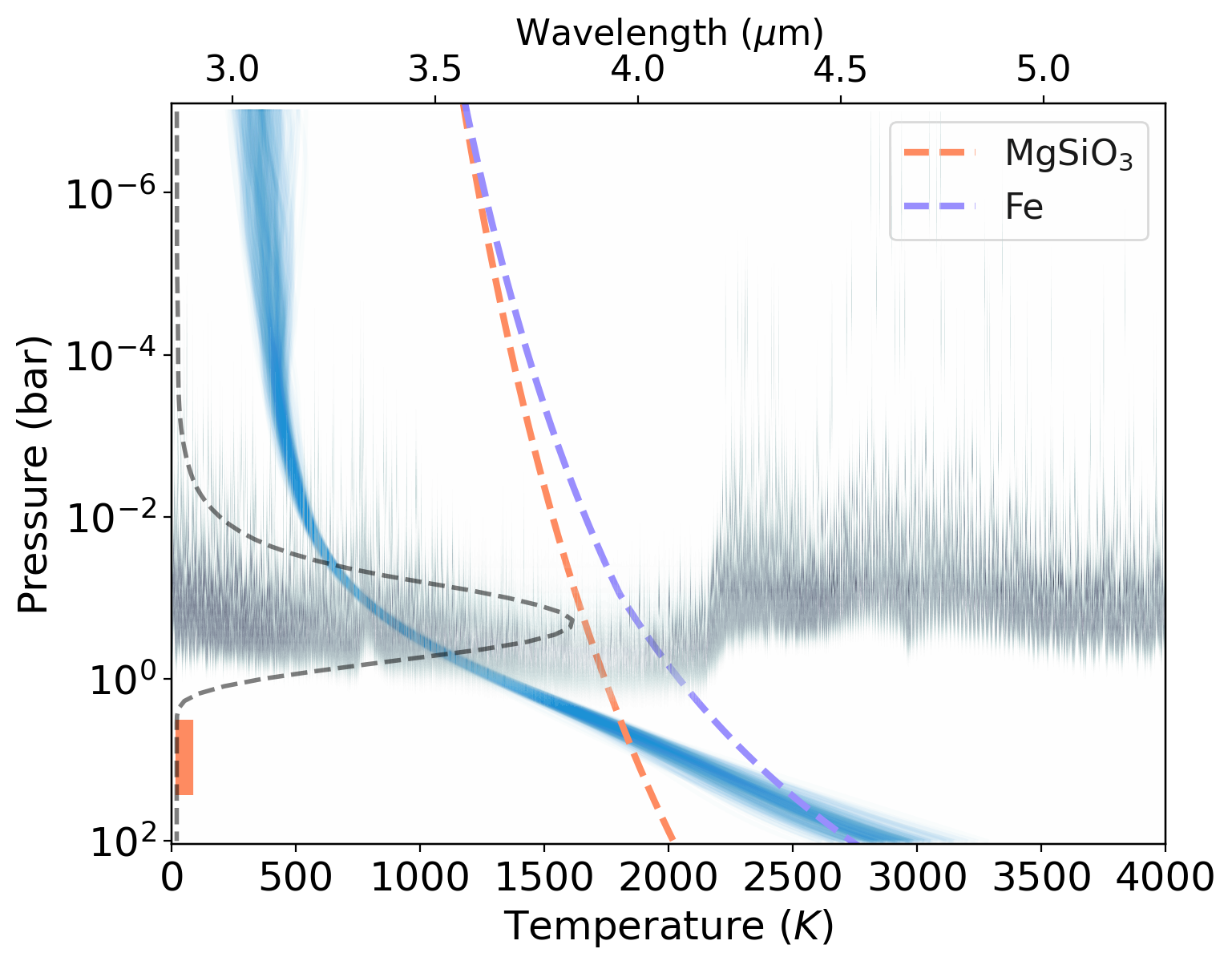}
    \caption{We show 200 random draws of the P-T profile for HR 8799 b in blue, and cloud condensation curves in dashed orange and purple lines. In the retrievals, the cloud base pressures are free parameters, and the retrieved MgSiO$_3$ cloud base ($2\sigma$ confidence interval) is marked by the thick orange line. The overlaid gray contour is the emission contribution function, which uses the top x-axis (wavelength in $\mu$m). The gray dashed histogram on the left is the wavelength-weighted emission contribution. }
    \label{fig:pt_emis}
\end{figure}

\subsection{Clouds, thermal structure, and bulk parameters}\label{sec:clouds_pt}
The retrieved $P$--$T$ profiles for HR 8799 b are shown in Figure~\ref{fig:pt_emis}, while those of planets c, d, and e are shown in Appendix~\ref{app:pt_cde}. Given the priors imposed by the \citet{Zhang2023} P-T parametrization that we use, the $P$--$T$ profile shapes mostly follow those of self-consistent models. 

For the planet radii, we obtain $1.03\pm0.03~\Rj$, $1.14\pm0.03~\Rj$, $1.37\pm0.04~\Rj$, and $0.98\pm0.05~\Rj$ for planets b, c, d and e, respectively. Assuming a stellar age of $30-40$ Myr \citep{Faramaz2021}, the ATMO 2020 and Saumon \& Marley evolutionary models \citep{Phillips2020, Saumon_2008} predict that $5-10~\Mj$ planets should have radii between $\approx1.1-1.4~\Rj$. The retrieved radii for the planets are consistent with these predictions at the $<10\%$ level. Planet d is found to have the largest radius, while planets b and e have radii somewhat lower than the expected range from evolutionary models. The orbit and stability analysis from \citet{Zurlo2022} find that planet d is the most massive of the four planets (see their Table 6). According to the \citet{Saumon_2008} models, at a young age of $30-40$ Myr, we expect the more massive planet to have the largest radius, which is consistent with that we find. 

We estimate the planets' $\Teff$ by generating $0.15-30~\mu$m low-resolution models from the posteriors using \texttt{petitRADTRANS}. Integrating these models yields the bolometric luminosity, which then gives $\Teff$ when using the retrieved radius posteriors. We infer luminosities in $\lbollsun$ units of $-5.12\pm0.01$, $-4.70\pm0.01$, $-4.62\pm0.01$, and $-4.68\pm0.02$ for planets b, c, d, and e, respectively. The $\Teff$ are $930\pm15~$K, $1120\pm15~$K, $1075\pm20~$K, and $1230\pm35~$K, for planets b, c, d, and e, respectively. The inferred luminosities for the planets are consistent at the $<1\sigma$ level with estimates from \citet{Nasedkin2024}, who performed retrievals on the $1-5\mu$m low-resolution spectra and MIRI photometry of these planets. 

For each planet, only one out of the two cloud species in the baseline model significantly impacts the spectra. For planets b, c, and e, the preferred cloud is MgSiO$_3$, while for planet d the retrievals prefer the inclusion of Fe clouds. To investigate what part of the data is sensitive to clouds, we ran alternative retrievals with one cloud species included at a time. We tested MgSiO$_3$, Fe, Mg$_2$SiO$_4$, Na$_2$S, ZnS, and KCl for planet b and MgSiO$_3$ and Fe for planet d. We found that different cloud species produce slightly different continuum shapes between $3-5\mu$m, due to the different wavelength dependencies of the cloud absorption and scattering cross sections (e.g. see Fig. 16 in \citealt{Nasedkin2024}). After high-pass filtering the spectral continuum, these differences are preserved as subtle wavelength-dependent line depths variations \citep{Xuan2024b}, which drive the modest preferences for different cloud species in our retrievals. The differences in line flux between clear and cloudy models can differ by $\sim20-40\%$ in certain narrow wavelength regions such as between $\approx3.8-3.9~\mu$m, which probes deeper in the atmosphere. From the single-cloud retrievals, HR 8799 b prefers MgSiO$_3$ with ln($B$)=7.8 over the next most favored clouds (ZnS or Mg$_2$SiO$_4$, which show nearly equal Bayesian evidence), and planet d prefers Fe clouds with ln($B$)=18.1 over MgSiO$_3$. For all planets, cloudy models are preferred over clear models. For HR 8799 d for example, ln($B$)=46.8 in favor of the Fe cloud model over the clear model. 

From our baseline retrievals, the MgSiO$_3$ cloud base pressures are determined to be $10^{+8}_{-4}$, $3^{+7}_{-1}$, and $8^{+9}_{-5}$ bars for planets b, c, and e, respectively. On the other hand, planet d prefers a Fe cloud base at $0.5\pm0.1$ bars. Interestingly, for all four planets, the retrieved cloud base pressures match to within $\approx1-2\sigma$ the expected MgSiO$_3$ cloud bases as determined by the intersection of the P-T profile and cloud condensation curves. For planet d, the retrieved location of the Fe cloud base is higher than the expected Fe cloud base based on equilibrium condensation (expected at $\approx3-10$ bars), and instead matches the expected MgSiO$_3$ cloud base ($\approx0.7-1.0$ bars). Thus, the planet d model is using the optical properties of Fe clouds but inconsistently finds a base pressure that equilibrium condensation would predict to be more appropriate for MgSiO$_3$. In a HR 8799 d retrieval with Fe clouds only, we obtained the same results for the retrieved cloud base pressure. We note the cloud base is at lower pressures for HR 8799 d compared to HR 8799 bce due to the hotter P-T profile of planet d (see Figure~\ref{fig:hr8799d}). 

While we chose to implement a more flexible model of the clouds than a classical EddySed model as a precaution (Section~\ref{sec:clouds}), for three out of four planets, the retrieved cloud bases matched the expected cloud base locations well. For those three planets (b, c, e), a classical EddySed model might have performed equally well. In addition, we note that the retrieved cloud particle radii from the models are large, on the order of $\sim10~\mu$m for all four planets. This suggests the clouds in these models are almost acting as a gray absorbers. 

Although the NIRSpec high-pass filtered spectra has some sensitivity to clouds, the differences between cloud species would be much more distinguishable in the spectral continuum. To confirm the cloud properties we measured for these planets, additional developments in improving the angular differential imaging or reference star differential imaging techniques \citep{Ruffio2024} with JWST/NIRSpec data are required. Ideally, these techniques should be applied to MIRI/MRS data (e.g. GO 4829) which cover silicate absorption features from $\sim8-11~\mu$m.

\subsection{Transport-induced disequilibrium chemistry}\label{sec:quench}
The atmosphere of HR 8799 b is far from chemical equilibrium. We measure quench pressures of $2.0\pm0.6$ bars for the CO-CH$_4$-H$_2$O system, and $17^{+40}_{-12}$ bars for the N$_2$-NH$_3$ system (though the latter is poorly constrained, see Section~\ref{sec:nitrogen}). Following the procedure in \citet{RuffioXuan2026}, we can estimate the vertical eddy diffusion coefficient, $K_{\rm zz}$, in the atmosphere. $K_{\rm zz}$ parameterizes the efficiency of vertical transport \citep{smith_Estimation_1998}. If mixing is strong (high $K_{\rm zz}$), the mixing ratios of various molecules can be determined by vertical transport from deeper in the atmosphere, instead of local chemical equilibrium. $K_{\rm zz}$ can be defined with respect to the mixing timescale ($\tau_{\rm mix}$) through 

\begin{equation}
    \tau_{\rm mix} = L^2 / K_{\rm zz} 
\end{equation}

where $L$ is the mixing length scale. In the retrievals, we account for transport-induced disequilibrium chemistry by fitting for carbon and nitrogen quench pressures (see Section~\ref{sec:chemistry}). From the P-T profile and carbon quench pressure posteriors, we estimated the $K_{\rm zz}$ by finding the point in the atmosphere where the mixing timescale is equal to the chemical reaction timescale relevant for the CO-CH$_4$ system \citep{Zahnle_methane_2014}. If assuming the mixing length scale $L$ is equal to the pressure scale height ($H$), we would obtain $\log_{\rm 10}(K_{\rm zz}~[{\rm cm^2~s^{-1}}])=7.8\pm0.7$. We note that in this analysis, $L$ is simply an effective length scale with which vertical transport takes place, and we follow the conventional choice of adopting $L=H$ in order to relate $K_{\rm zz}$ to a mixing timescale \citep{Zahnle_methane_2014}. $L$ is not necessarily identical to the mixing length parameter from mixing-length theory, which only applies in convective regions.

With the above assumptions, the $\log_{\rm 10}(K_{\rm zz}~[{\rm cm^2~s^{-1}}])$ of $7.8\pm0.7$ for HR 8799 b is consistent with the early estimate from \citet{barman2015} based on self-consistent PHOENIX models and Keck/OSIRIS spectra of HR 8799 b. In contrast to several previous works \citep{Xuan2022, deRegt2024}, which typically find very high $K_{\rm zz}$ values at or above the theoretical upper limit from mixing length theory \citep{Gierasch_convect1985, Zahnle_methane_2014}, our estimates for HR 8799 b (as well as the estimate from \citealt{barman2015}) are lower than the upper limit of $\log_{\rm 10}(K_{\rm zz}~[{\rm cm^2~s^{-1}}])\approx9.8$ for this planet. We note that in reality, $K_{\rm zz}$ should vary with pressure \citep{Mukherjee2022}, so our measurement mainly applies to pressures $\sim2$ bars in HR 8799 b, where the CH$_4$ and CO abundances quench. In Section~\ref{sec:nitrogen}, we compare the $P_{\rm quench}$-inferred $K_{\rm zz}$ to expectations from a self-consistent chemical calculation. 

\subsection{Measuring N/H from NH$_3$}\label{sec:nitrogen}
Among widely-separated planetary-mass companions ($>5\,$au), NH$_3$ was recently detected in GJ~504~b from MIRI photometry \citep{Malin2025} and in WD~0806-661 b from MIRI/LRS \citep{Voyer2025}. \citet{Meynardie2025} also report a constrained NH$_3$ abundance for Ross~458~c from NIRSpec $1-3\mu$m spectroscopy. \citet{Malin2025} estimated a NH$_3$ volume-mixing ratio of $\approx10^{-5.3}$ for GJ 504 b, but did not provide a N/H measurement, which is challenging because the bulk nitrogen reservoir is still mainly contained in spectrally inaccessible N$_2$ for this $\Teff\approx510~$K planet \citep{Ohno2023,Ohno2023b}. For Ross 458 c, which is hotter with $\Teff\approx770~$K \citep{Meynardie2025}, the same challenge applies. WD~0806-661 b has a lower $\Teff\approx350~$K, which makes NH$_3$ more abundant than N$_2$ at the $P<1$ bar region probed by MIRI/LRS for this companion. However, using \texttt{easyCHEM} we find that at $P>1$ bar, N$_2$ is a few times more abundant than NH$_3$ for a WD 0806-661 b-like companion. This means that N$_2$ cannot be neglected when estimating N/H, and this likely explains the low value of $\rm N/H < 0.1\times$ solar in \citet{Voyer2025} (as well as in \citealt{Meynardie2025}).

\begin{figure}
    \centering
    \includegraphics[width=0.8\linewidth]{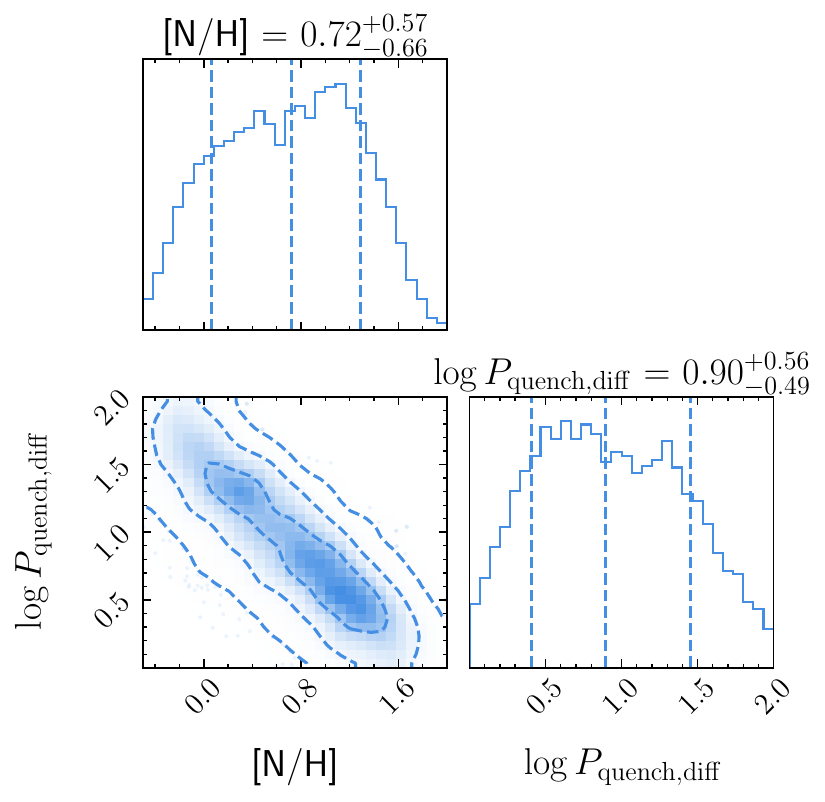}
    \caption{Posterior distributions for [N/H] and the $\log{P_{\rm quench,diff}}$ parameter that controls the quench pressure of NH$_3$. These values are directly taken from the \texttt{petitRADTRANS} retrieval of HR 8799 b, and do not incorporate the VULCAN analysis which provided an improved estimate of N/H (Section~\ref{sec:nitrogen}). The corner plot shows a strong correlation, demonstrating an inherent degeneracy between [N/H] and the quench pressure of NH$_3$ for this planet.}
    \label{fig:nh_nquench}
\end{figure}

In this paper, by constructing a chemical grid parametrized by C/H, O/H, and N/H (Section~\ref{sec:chemistry}), we account for N$_2$ in the nitrogen abundance for the \texttt{petitRADTRANS} retrievals of HR 8799 b. However, the retrievals give a large uncertainty with $\rm N/H=5.3^{+14.1}_{-4.1}\times$ solar. This is due to three reasons. First, we measure N/H from NH$_3$, which is detected at $4\sigma$ significance compared to $15\sigma$ for H$_2$S. Second, in the limit of a young planet with high intrinsic temperature and low surface gravity, the equilibrium NH$_3$ abundance scales sub-linearly as the square of the bulk nitrogen abundance \citep{Ohno2023b}. 
Third, and perhaps most crucially, the NH$_3$ quench pressure is a free parameter in the \texttt{petitRADTRANS} retrieval via the $\log P_{\rm quench, diff}$ parameter (see Section~\ref{sec:chemistry}).
Under the retrieved P-T profile for this planet, the NH$_3$ abundance in chemical equilibrium increases with increasing pressure at the deep atmosphere due to the enhanced net reaction of N$_2$+3H$_2\longrightarrow$2NH$_3$.
This causes a degeneracy between N/H and NH$_3$ quench pressure because the high NH$_3$ abundance inferred by the observation can be explained by either high N/H or high quench pressure. 
We do observe a strong correlation between these two parameters in Figure~\ref{fig:nh_nquench}, indicating the inherent degeneracy between them. This is the main reason for the large [N/H] error bars ($\approx0.6$ dex) from the petitRADTRANS retrievals in Section~\ref{sec:spec_analysis}. The fundamental issue here is that with only one observable (NH$_3$), we cannot simultaneously measure two parameters ([N/H] and the NH$_3$ quench pressure), unlike the situation for carbon disequilibrium chemistry involving CO and CH$_4$, where [C/H] can be inferred directly from the CO and CH$_4$ abundances.

\begin{figure*}
    \centering
    \includegraphics[width=\linewidth]{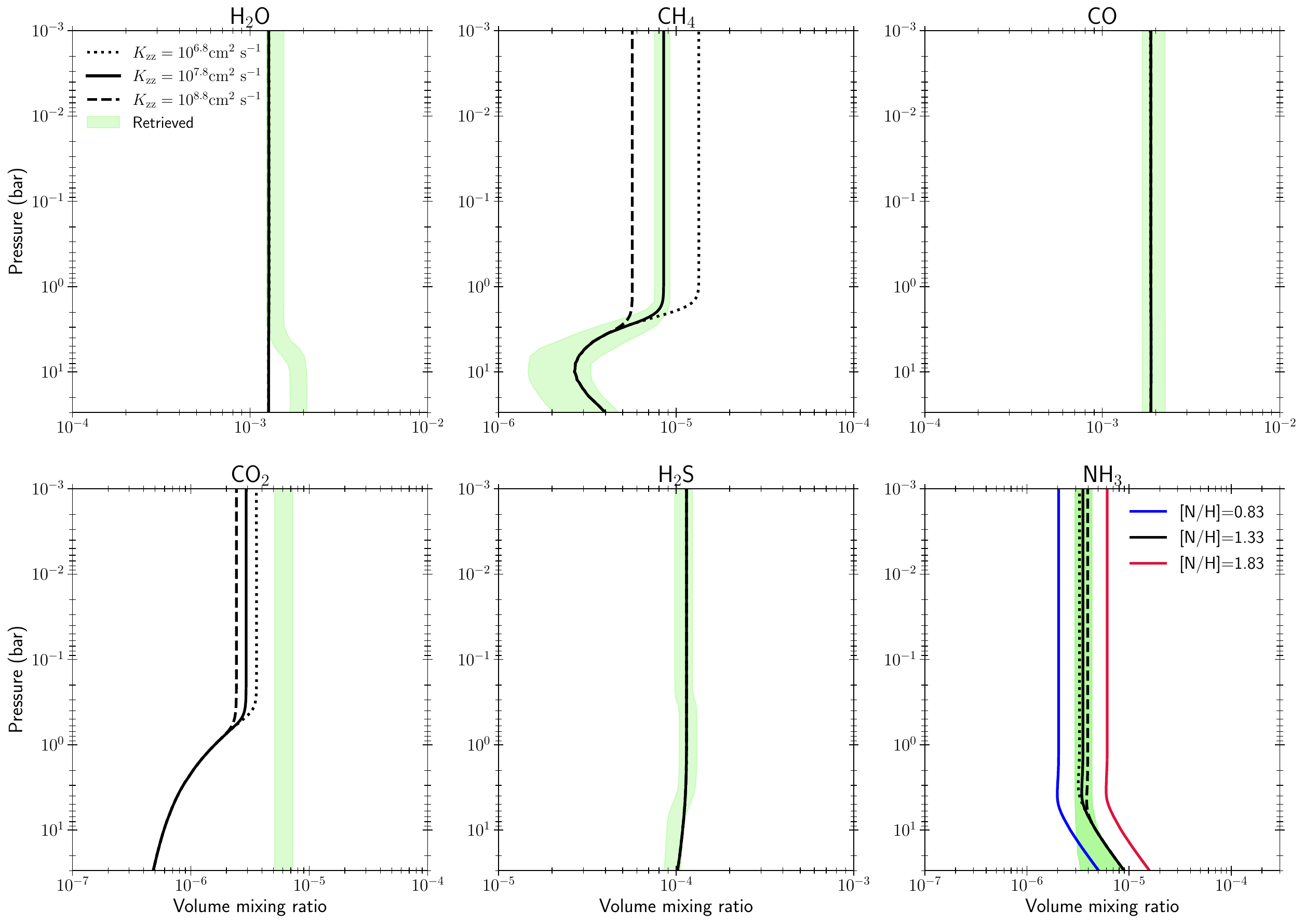}
    \caption{Volume-mixing ratio v.s. pressure profiles for the strongly detected molecules in HR 8799 b. The green shaded regions are the abundances derived from the fiducial \texttt{petitRADTRANS} retrieval. Black lines show model abundance profiles from the \texttt{VULCAN} analysis using the median retrieved elemental abundances. The solid black line uses the median retrieved $\log_{\rm 10}(K_{\rm zz}~[{\rm cm^2~s^{-1}}])=7.8$, while the dotted and dashed lines show model profiles where $K_{\rm zz}$ is perturbed from the median value by an order of magnitude. 
    The red and blue lines in the NH$_3$ panel (bottom right) show the NH$_3$ profiles if the bulk nitrogen abundance is perturbed from the median value of [N/H]=1.33 (or N/H$=21.2\times$ solar) by 0.5 dex.}
    \label{fig:VULCAN}
\end{figure*}

To break the degeneracy between [N/H] and the NH$_3$ quench pressure from the \texttt{petitRADTRANS} retrievals, and provide a better estimate of [N/H] that remains consistent with the retrieval results, we performed a self-consistent chemical kinetic analysis using the \texttt{VULCAN} code \citep{Tsai+17,Tsai+21}. Unlike the equilibrium chemistry grid used in the \texttt{petitRADTRANS} retrieval analysis (based on \texttt{easyCHEM}), \texttt{VULCAN} is able to self-consistently model the effect of disequilibrium chemistry for a given value of $K_{\rm zz}$.

\begin{figure*}[t!]
    \centering
    \includegraphics[width=0.85\linewidth]{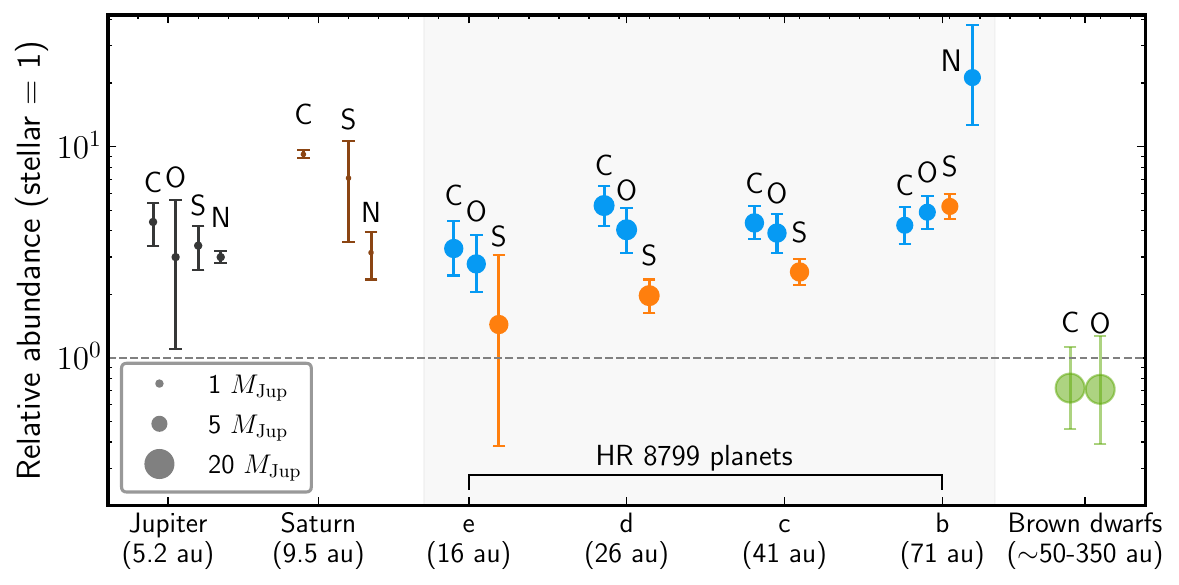}
    \caption{Elemental abundances relative to stellar values for the HR 8799 planets. We overplot measurements for Jupiter \citep{Wong2004, Li2020}, Saturn \citep{Briggs1989, Fletcher2009}, and the median C/H and O/H (with the scatter in values shown as uncertainties) for eight different low-mass brown dwarf companions ($m\approx10-30~\Mj$) from Keck/KPIC spectroscopy \citep{Xuan2024b}. The size of the points are proportional to mass; for the brown dwarfs, we adopt their median mass of $21.5~\Mj$. 
    We note that due to their cold temperatures, S/H for Saturn remains tentative due to H$_2$S condensation \citep{Atreya2018}, and O/H for Jupiter has large uncertainties due to water condensation. The S/H for HR 8799 e remains tentative due to insufficient data S/N to confidently detect H$_2$S.
    HR 8799 b shows a uniform enrichment pattern in C, O, and S, but its N is elevated compared to these three elements at the $2\sigma$ level. HR 8799 c, d, and e are similarly enriched in C and O as HR 8799 b ($3-5\times$ stellar). From planets d to b, there is a tentative trend of S/H increasing with increasing orbital distance.}
    \label{fig:abund_summary}
\end{figure*}

In the \texttt{VULCAN} analysis, we leverage the fact that NH$_3$ would be subject to approximately the same strength of vertical eddy diffusion as the carbon species (CO, CH$_4$) \footnote{\citet{zhang_Globalmean_2018} suggested that eddy diffusivity could vary with chemical species when the chemical timescale is shorter than the atmospheric dynamical timescale. However, we expect that each chemical species experiences comparable eddy diffusivity at and above the quench point because the dynamical timescale should be faster than the chemical timescale there by definition.}. Therefore, if we can constrain $K_{\rm zz}$ from the \texttt{petitRADTRANS} retrieved carbon quench pressure (based on CO and CH$_4$), we can then use this $K_{\rm zz}$ value as an input into \texttt{VULCAN} to find the [N/H] that best matches the \texttt{petitRADTRANS} retrieved NH$_3$ abundance profile. Knowledge of $K_{\rm zz}$ effectively allows us to better constrain the nitrogen quench pressure, and therefore better constrain the [N/H] compared to the \texttt{petitRADTRANS} retrieval, which varied the nitrogen quench pressure agnostically.

Specifically, we used the \texttt{petitRADTRANS} retrieved P-T profile, elemental abundances, NH$_3$ mixing ratio, and $K_{\rm zz}$ as inputs for the VULCAN analysis. We measured $\log (K_{\rm zz} / \rm{cm^2 s^{-1}})=7.8\pm0.7$ for HR 8799 b based on the carbon quench pressure (Section~\ref{sec:quench}), and the log10(NH$_3$) volume-mixing ratio at pressures below the quench pressure ($<2$ bars) is tightly constrained to be $-5.45\pm0.10$ from the \texttt{petitRADTRANS} retrieval (see Figure~\ref{fig:VULCAN}). The elemental abundances of [C/H], [O/H], and [S/H] are listed in Table~\ref{table:spec_results}.

Our procedure is as follows. First, we randomly draw 3000 pairs of P-T profiles and associated NH$_3$ abundance profiles from the \texttt{petitRADTRANS} posterior samples.
We then run \texttt{VULCAN} for each sampled P-T profile with bulk nitrogen abundances of [N/H]$=0.5$, $1.0$, $1.3$, $1.5$, and $2.0$, yielding 15000 chemical profiles computed by \texttt{VULCAN}.
In each simulation, we fix [O/H]=$0.69$, [C/H]=$0.63$, $K_{\rm zz}=10^{7.8}~{\rm cm^2~s^{-1}}$ from the median retrieved values.
For each sampled profile, we construct an interpolation function to convert the volume mixing ratio of NH$_3$ at $1~{\rm bar}$ to [N/H] using the outputs of \texttt{VULCAN}. A pressure of 1 bar is chosen since this is the pressure level mainly probed by the NIRSpec data (see Figure~\ref{fig:pt_emis}).
This interpolation function is used to convert the sampled NH$_3$ abundance at $1~{\rm bar}$ to [N/H] and reconstruct a posterior distribution of [N/H] from the \texttt{petitRADTRANS}-retrieved NH$_3$ and P-T profile posterior samples.
Our \texttt{VULCAN} grid with [N/H]$=0.5$--$2.0$ could find the [N/H] that reproduces the sampled NH$_3$ abundances for 2994 out of 3000 samples.
For the remaining $6$ samples, we extended the \texttt{VULCAN} grid to [N/H]=0.0 and 2.5 for deriving [N/H] without extrapolation. 
With this posterior reconstruction procedure, we obtain a nitrogen abundance of N/H=$21.2^{+16.2}_{-8.8}\times$ solar.

In Figure~\ref{fig:VULCAN}, we show the chemical profiles simulated by \texttt{VULCAN} for the median elemental abundances in Table~\ref{table:spec_results}.
The \texttt{VULCAN} chemical profiles are in excellent agreement with those derived by the \texttt{petitRADTRANS} retrieval, except for CO$_2$.
The discrepancy seen in CO$_2$ may be due to our simplified assumption of a vertically constant CO$_2$ abundance in the \texttt{petitRADTRANS} retrieval, or the constant $K_{\rm zz}$ assumed in the \texttt{VULCAN} analysis. Because CO$_2$ quenches at a lower pressure than CO, CH$_4$, and NH$_3$, having a lower $K_{\rm zz}$ at the CO$_2$ quench pressure could explain the observed discrepancy. However, this requires further investigation in future studies.
Note that the possible reduced $K_{\rm zz}$ in pressures where CO$_2$ quenches should not affect the N/H estimation presented above, as NH$_3$ quenching takes place at much deeper pressures in the atmosphere than CO$_2$ quenching. In the \texttt{VULCAN} analysis, we fixed $\log_{\rm 10}(K_{\rm zz}~[{\rm cm^2~s^{-1}}])=7.8$, since it is tightly constrained by the CH$_4$ abundance profile. However, we also show that the quenched NH$_3$ abundance barely changes even if we perturb $K_{\rm zz}$ by an order of magnitude (see Figure~\ref{fig:VULCAN}), in agreement with previous work suggesting that the quenched abundance of NH$_3$ is insensitive to $K_{\rm zz}$ \citep[][]{Zahnle_methane_2014,Ohno2023}.
This analysis further strengthens our inference for the N/H value.

The N/H value estimated from the VULCAN analysis (N/H=$21.2^{+16.2}_{-8.8}\times$ solar) is consistent but at the upper end of the \texttt{petitRADTRANS} retrieval value ($\rm N/H=5.3^{+14.1}_{-4.1}\times$ solar). At the $2\sigma$ level, it indicates that HR 8799 b has a higher N abundance than C, O, and S, with $\rm N/S=4.1^{+3.2}_{-1.7}$ and $\rm N/C = 5.1^{+4.0}_{-2.2}$ ($\times$ stellar). We adopt the VULCAN value of $\rm N/H=21.2^{+16.2}_{-8.8}$ for the rest of this paper, since the retrieval value is artificially broadened by the inherent degeneracy between NH$_3$ quench pressure and [N/H] for this planet, discussed above.

\subsection{Elemental abundances}\label{sec:abunds}
HR 8799 b shows remarkably uniform enrichment across C, O, and S, with $\rm C/H=4.2\pm0.8$, $\rm O/H=4.9\pm0.9$, and $\rm S/H=5.2\pm0.7$ relative to solar. As justified in Section~\ref{sec:stellar_ab}, we approximate the abundances of HR 8799 A as solar. 

We also update the abundances for HR 8799 c, d, and e using the higher S/N, second epoch NIRSpec observation presented in this paper, and plot their abundances along with planet b in Figure~\ref{fig:abund_summary}. The elemental abundances are also listed in Table~\ref{table:spec_results}. For c, d, and e, the new C/H, S/H, and O/H values are consistent at the $1-2\sigma$ level with the first epoch results \citep{RuffioXuan2026}. One improvement is that HR 8799 d now has a much better constrained S/H compared to what was reported in \citet{RuffioXuan2026}. In addition, while HR~8799~e appeared consistent with a stellar composition in \citet{RuffioXuan2026}, it is now confidently enriched in C and O with reduced C and O uncertainties similarly to the other planets in the system. We caution that the S/H measurement in HR 8799 e remains tentative however, as this planet lacks a confident detection of H$_2$S (Section~\ref{sec:ccf_molecule}).

\begin{figure*}[t!]
    \centering
    \includegraphics[width=\linewidth]{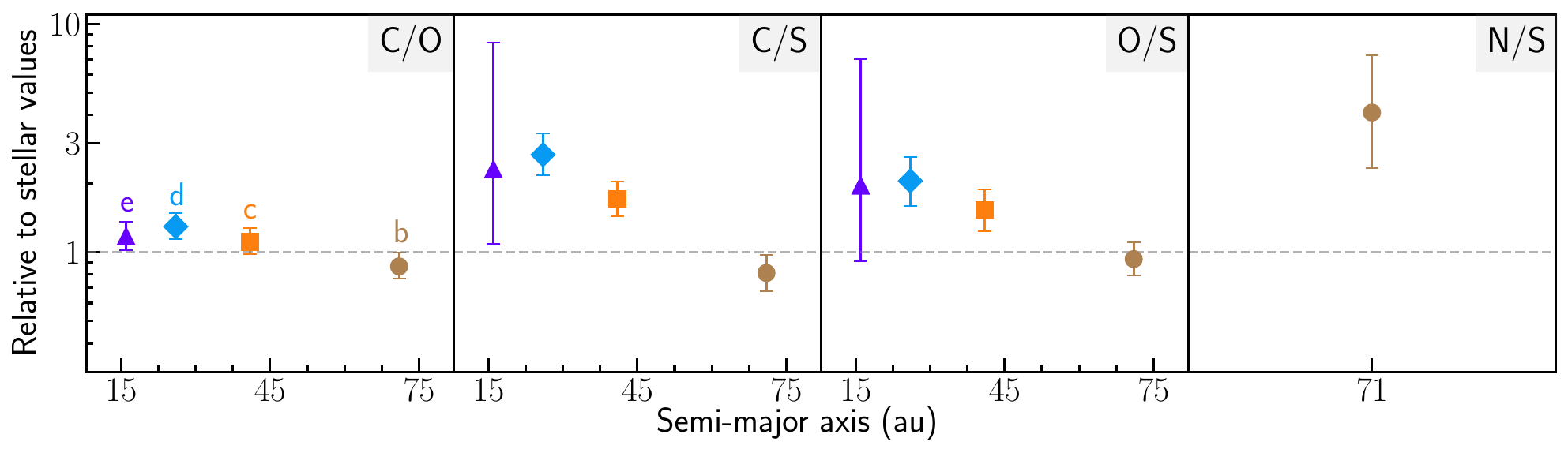}
    \caption{Relative abundances between volatile (C, O, N) and refractory (S) elements for the HR 8799 planets, as a function of planet semi-major axis. The gray dashed line at 1 indicates a stellar abundance ratio. The N/H, and hence N/S, is only measured for planet b.}
    \label{fig:c_s_o_ratios}
\end{figure*}

Across the four planets, the enrichment in C and O is uniform and ranges between $3-5\times$ stellar. While the C/S ratio only varies from $0.8-2.7$ across the planets, there is a tentative trend of decreasing C/S with larger semi-major axis based on planets d to b (the C/S of HR 8799 e has large error bars). Planets d, c, and b all show enriched sulfur, which is a tracer of solid accretion. We discuss the implications of these results in Section~\ref{sec:discuss_abunds}.

\section{Discussion}\label{sec:discuss}

\subsection{The metal-enriched atmospheres of the HR 8799 planets}\label{sec:discuss_abunds}
In a recent retrieval study with ground-based, low-resolution spectroscopy, \citet{Nasedkin2024} reported high atmospheric metallicities for the HR 8799 planets, as inferred for carbon and oxygen bearing species (mainly CO and H$_2$O). In other recent literature, \citet{molliere_Retrieving_2020} and \citet{Wang2023} also found elevated metallicities for HR 8799 e and c, respectively. Finally, \citet{Balmer2025b} reported the detection of CO$_2$ in the four HR 8799 planets from JWST/NIRCam photometry, which pointed to elevated atmospheric metallicities compared to several brown dwarfs, which are assumed to have solar metallicities.

In this work, and in \citet{RuffioXuan2026}, we leverage the moderate spectral resolution ($R\sim2700$) and high S/N of the JWST/NIRSpec spectra to provide unambiguous molecular detections of CO, CH$_4$, H$_2$O, H$_2$S, CO$_2$, and now in planet b, NH$_3$ (Figure~\ref{fig:spec}). In addition, by jointly fitting $1-5~\mu$m photometry (Figure~\ref{fig:phot}) with the NIRSpec spectra, we are able to break some of the degeneracies between cloud properties, bulk atmospheric parameters, and molecular abundances \citep{Wang2023}. 

We confirm that the atmospheres of the HR 8799 planets are enhanced in metallicity compared to their star (which has solar metallicity, see Section~\ref{sec:stellar_ab}). The carbon and oxygen abundances for all four planets are found to be about $3-5\times$ stellar, while the sulfur abundances are also enriched by $2-5\times$ stellar across planets d, c, and b (Figure~\ref{fig:abund_summary}). Planet e's sulfur abundance is marginally elevated, but consistent with stellar at the $1\sigma$ level. These results represent an update to the abundances reported in \citet{RuffioXuan2026}. The enrichment in sulfur for at least the outer three planets suggests they accreted a significant amount of solids from the circumstellar disk (see Figure~\ref{fig:Metal_vs_semimajor}, Table~\ref{table:metal_mass}).

We observe potential trends in C/S and O/S with orbital distance. The C/S ratios (relative to stellar) increase with decreasing orbital distance, from $0.8\pm0.2$, $1.7\pm0.3$, to $2.7\pm0.6$ for planets b, c, and d, respectively (Figure~\ref{fig:c_s_o_ratios}). A similar trend is observed in O/S from b to d given the stellar-like C/O ratios for all four planets. This suggests that these planets accreted metals from material that had different volatile-to-refractory ratios. Furthermore, we measure N/H for the first time in HR~8799~b from NH$_3$, and find an elevated value of $21.2^{+16.2}_{-8.8}\times$ stellar. At the $2\sigma$ level, nitrogen is the most enriched metal in HR 8799 b among the four elements measured, with N/S=$4.1^{+3.2}_{-1.7}$ for this outermost planet. 

\subsection{Implications for planet formation}\label{sec:formation}

\subsubsection{Building blocks of the HR 8799 planets}
\label{sec:building_blocks}

\begin{figure*}[t!]
    \centering
    \includegraphics[width=\linewidth]{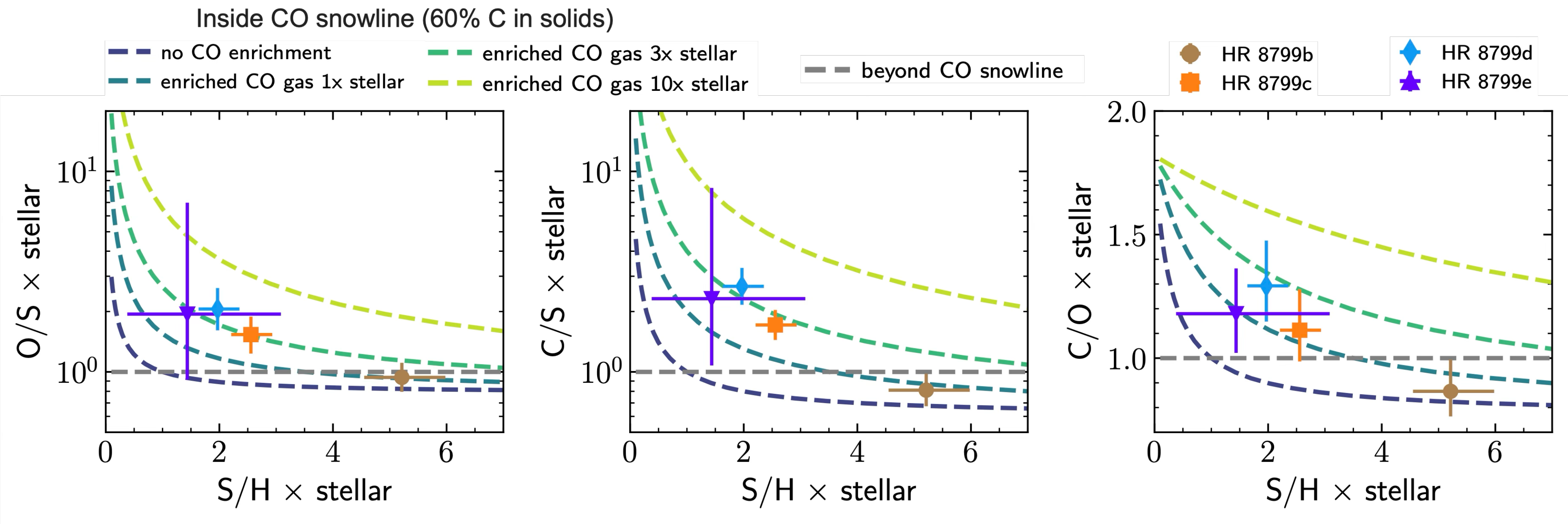}
    \caption{The measured elemental ratios for the four HR 8799 planets compared to composition expected for planetary building blocks in different regions of the disk. The x-axis corresponds to the solid-to-gas accretion ratio, which is constrained by the S/H of each planet. The y-axis shows the elemental ratios expected for accretion beyond the CO snowline (gray dashed line) and for different levels of CO enrichment inside the CO snowline (colored dashed lines). Planet b is most uniform and compatible with formation beyond the CO snowline. The three inner planets have higher volatile-to-refractory ratios and their composition likely necessitates the accretion of CO-enriched disk gas.}
    \label{fig:comp_curves}
\end{figure*}

Our measurements of multiple volatile-to-refractory ratios (C/S, O/S, N/S) across four different planets provide unique constraints on the formation history of this system. The abundance patterns we observe can be interpreted under a simple model, which accounts for the different C, O, S, and N compositions in disk solids and gas (planet-building blocks) relative to the snowline locations. 
In the outermost regions of disks beyond the snowlines, we expect all the metals to be condensed out and present in stellar proportions in the solids. As we move closer in to the host star and cross the snowline of a volatile, the elements comprising the volatile species leave the solid phase and enter the gas phase \citep[e.g.,][]{oberg_effects_2011}. 
In addition, dust grows into pebble-sized solids and drifts in radially due to the slight difference in the orbital velocity of the pebbles and the pressure-supported gas \citep[e.g.,][]{Birnstiel2024}. When these pebbles cross a volatile's snowline, they release that volatile into the gas phase, which then evolves with the disk gas \citep[e.g.,][]{Oberg2016, Booth2017}. This differential movement of dust and gas leads to a redistribution of metals and local deviations from stellar composition. 

For the HR 8799 system, sulfur is a refractory species and traces the accretion of solids by the planets. Solids accreted by the planets, whether they originally formed locally or drifted in from more distant regions of the disk, contributed to the planets' overall complements of C, O, and N. The accretion of solids should yield C/S, O/S, and N/S $\leq$ stellar for the planets because part of the volatiles (C, N, O) may be in the gas phase while all the refractories (S) are in solids. However, for all four planets, we have evidence of super-stellar volatile-to-refractory ratios. Since the planets' volatile reservoirs must be enriched by more than what can be achieved through solid accretion, the gas accreted by the planets must have been itself enriched in the volatiles C, O, and N.  A straightforward way to achieve this enhancement is to appeal to pebble drift.  In this scenario, pebbles from outside the N$_2$ and CO snowlines drifted inward, sublimated, and enriched the disk gas in the volatile species.  Because gas and solids are accreted by the growing planets via different physical processes, this volatile-enhanced gas can separately contribute to the metal composition of the planetary atmospheres, and the metal contributions of solids and gas need not occur in proportions commensurate with their proportions in the protoplanetary disk.

\begin{figure*}[t!]
    \centering
    \includegraphics[width=\linewidth]{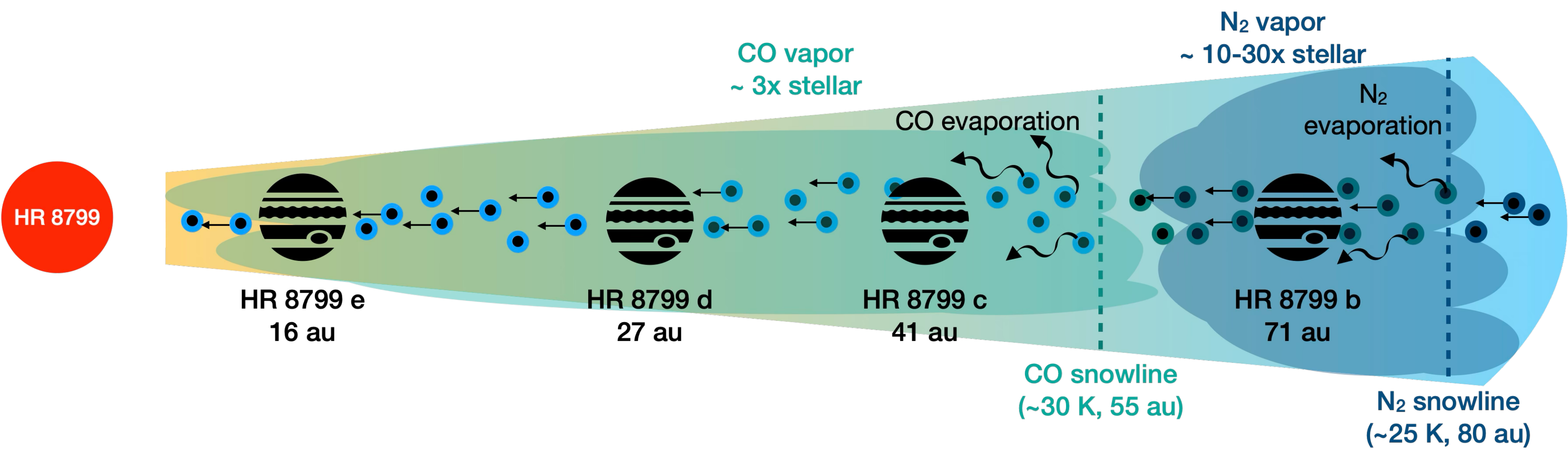}
    \caption{A cartoon illustrating pebble drift and evaporation in the context of the HR 8799 planets and their protoplanetary disk. The composition (similar C/H and O/H but with lower and varying amounts of S/H) of the three inner planets is compatible with the accretion of disk gas enriched in CO. Planet b's composition implies accretion of solids enriched uniformly in C, O, and S but gas enriched in N$_2$. Remarkably, the planets' locations relative to the CO and N$_2$ snowlines are compatible with their measured composition (Figure~\ref{fig:abund_summary})}
    \label{fig:cartoon}
\end{figure*}

In Figure~\ref{fig:comp_curves}, we compare the elemental ratios of the HR 8799 planets with models from \citet{Chachan2023}, which outline the expected planetary composition based on formation inside or outside the CO snowline. Outside the CO snowline, all of C, O, and S are assumed to be in the solids, so that the planet is enriched only by solids with stellar C/S and O/S ratios.  Hence, the planetary C/S, O/S, and C/O are constant at stellar values. Inside the CO snowline, CO is removed from solids. We break the total volatile abundance of each planet into (1) that accreted by solids, determined by the total solid accretion as measured from S/H and the fraction of C and O remaining in the solids after CO removal, plus (2) that accreted by gas, determined by the level of CO enhancement in the gas. We let the CO enrichment for the disk gas vary to capture the variation expected from the interplay between the pebble flux and viscous diffusion of the gas. We plot the expected and measured elemental ratios (O/S, C/S, and C/O) as a function of the solid-to-gas accretion ratio (S/H) for the planets. For the fiducial case, we assume that 40\% of C (and a corresponding amount of O) is in the form of CO ($f_{\rm [CO]/[C]}$ = 40\%) based on measurements from the interstellar medium, molecular clouds, protostellar cores, and solar system comets \citep{oberg_bergin2021}. However, we note that the split in the volatile C budget is uncertain and CO could constitute only 20\% of the C budget with the other 20\% contained in CO$_2$ \citep{McClure2023}. We vary this value later to quantify the effect of this uncertainty on the implied pebble mass that passed through the planets' location.

The outermost planet HR 8799 b's uniform C, O, and S enrichment is compatible with formation beyond the CO snowline, where C and O are both condensed into solids along with S. On the other hand, the compositions of the three inner planets (c, d, e) are consistent with the accretion of solids depleted in CO + disk gas that is $\sim3\times$ enriched in CO relative to stellar. The current data do not require different levels of CO enrichment in disk gas at the locations of planets c, d, and e. More precise measurements in the future may reveal such differences, which could point to spatially non-uniform enrichment of disk gas, as expected at early times before uniformity in gas enrichment is achieved by viscous diffusion \citep{Booth2017, Booth2019, Schneider2021a, Ohno2025}.

Although the C, O, and S enrichment for HR 8799 b is uniform and indicates the accretion of these metals via solids beyond the CO snowline, this planet's N enrichment is significantly larger and requires the accretion of disk gas that is enriched in N at the $\sim 10-30 \times$ stellar level. A priori, it is not surprising that the measured enrichment of N$_2$ and CO are different. First, we expect different fractions of N and C to be in N$_2$ and CO, respectively. Recent observations of comet 67P suggest that ammonium salts contain a larger fraction of N than previously known \citep{Poch2020, Altwegg2022}\footnote{These ammonium salts could also be the carrier of refractory sulfur given the abundance of NH$^{+}_{4}$SH$^{-}$ in comet 67P. In solar composition material, just 20\% of N can tie up all available S in this form.} and this reduces the fraction of N in N$_2$. We assume that 60\% of N is in N$_2$ ($f_{\rm [N_2]/[N]}$ = 60\%) for our fiducial calculations and vary it in the range of $50-80$\% (upper limit based on \citealt{oberg_bergin2021}) later. 
Second, the enrichment of disk gas inside a snowline is a function of location and time, with the regions just interior to the snowline having the largest enrichment at early times (e.g. $< 1$ Myr). 
HR 8799 b's N enrichment could be higher if N$_2$ vapor was more localized than CO vapor at the time of the planets' accretion. The enrichment in N$_2$ and CO required to explain the planets' composition could be compatible with a single value of the mass flux of pebbles through the N$_2$ and CO snowlines: for a $3 \times $ solar CO enrichment of the disk gas, we would expect the N enrichment to be $\sim (f_{\rm [N_2]/[N]} / f_{\rm [CO]/[C]}) \, 3 \sim (0.6/0.4) \, 3 \sim 4.5 \times$ stellar, which agrees with our measurements for HR 8799 b at the $\sim 1.9\sigma$ level. Notably, if only $f_{\rm [CO]/[C]} = 20$\% of C is in CO and $f_{\rm [N_2]/[N]} = 80$\% of N is in N$_2$, the calculation above would yield an N enrichment of $12 \times$ stellar, which is compatible with our measurement at $< 1 \sigma$. 

\begin{figure*}
    \centering
    \includegraphics[width=0.49\linewidth]{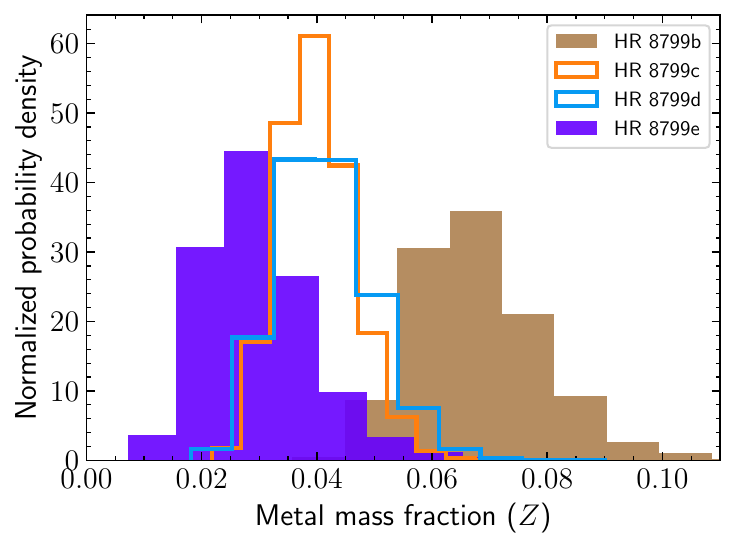}
    \includegraphics[width=0.49\linewidth]{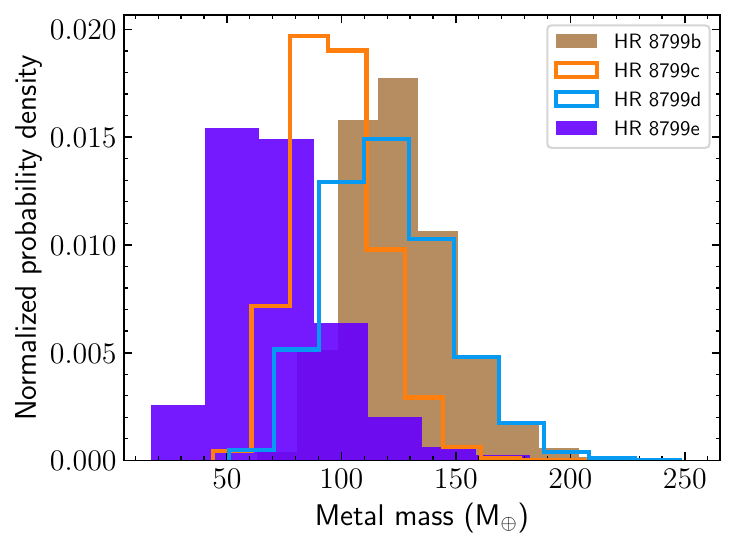}
    \caption{The metal mass fraction and the total metal mass of the four HR 8799 planets calculated from the measured atmospheric abundances. The total metal mass is estimated assuming planet masses from \cite{Zurlo2022}.}
    \label{fig:planets_metallicity}
\end{figure*}

\begin{deluxetable*}{c|ccccc}[t!]
\tablecaption{Estimated metal mass in the HR~8799 planets\label{table:metal_mass}}
\tablewidth{\textwidth}
\tablehead{
\colhead{Planet} & \colhead{Metal mass fraction} & \colhead{Total metal mass (\Me)} & \colhead{Metals from solids (\Me)} & \colhead{Metals from gas (\Me)}}
\startdata
HR 8799 b & $0.067^{+0.012}_{-0.010}$ & $123_{-19}^{+24}$ & $98_{-14}^{+16}$ & $22_{-10}^{+18}$ \\
HR 8799 c & $0.040^{+0.007}_{-0.006}$ & $96_{-16}^{+19}$ & $55_{-9}^{+10}$ & $41_{-13}^{+15}$ \\
HR 8799 d & $0.041^{+0.009}_{-0.008}$ & $119_{-24}^{+28}$ & $51_{-9}^{+11}$ & $67_{-19}^{+23}$ \\
HR 8799 e & $0.028^{+0.010}_{-0.008}$ & $68_{-19}^{+26}$ & $31_{-22}^{+34}$ & $35_{-22}^{+22}$ \\
Total  & $-$ & $412^{+48}_{-43}$ & $239^{+40}_{-31}$ & $170^{+39}_{-36}$ \\ 
\enddata
\end{deluxetable*}

How well do these inferences sit with the expected thermal structure of the HR 8799 protoplanetary disk? Since the host star is more massive and luminous than the Sun, the temperature of the irradiated part of its disk would have been higher than in the protosolar disk. Using a simple estimate for the disk's temperature $T_{\rm irr} = 150 \, {\rm K} \, (L_{\star}/L_{\odot})^{2/7} \, (M_{\star}/M_{\odot})^{-1/7} \, (r / {\rm 1 au})^{-3/7}$ and assuming that $L_{\star} = L_{\odot} (M_{\star}/M_\odot)^{1.5}$ for a pre-main sequence star \citep{Ida2016} and $M_{\star} = 1.47 \, M_{\odot}$ \citep{Sepulveda2022}, we find that
\begin{equation}
    T \simeq 34 \, \bigg(\frac{r}{{\rm 40 \, au}}\bigg)^{-3/7} \, {\rm K},
\end{equation}
where we normalize to $r=40$ au, which is close to planet c's location and roughly in the middle of the planetary system. Such a temperature structure is appropriate for $1 - 5$ Myr. For such a disk, the CO ($\sim 30$ K) and N$_2$ ($\sim 25$ K) snowline locations would have been located at $\sim 55$ au and $\sim 80$ au, respectively. Our inferences about the planetary building blocks are in remarkable confluence with expected locations of the CO and N$_2$ snowlines (Figure~\ref{fig:cartoon}). Planet b at 71 au likely accreted its metals between the CO and N$_2$ snowlines and as a result it is uniformly enriched in C, O, and S from solid accretion and enhanced in N due to accretion of N$_2$-enriched gas. Planets c, d, and e all lie interior to the CO snowline and their compositions are compatible with accretion of a combination of local solids and CO-enriched disk gas. Therefore, the compositions we measured do not require the planets to have undergone significant migration. However, there is inherent uncertainty in the sublimation temperature and snowline locations (e.g., depending on whether sublimation is of pure ice or mixed ices, \citealt{Fayolle2016, Piso+16}) and some migration may be required considering the plausible range of these quantities (e.g., if the CO snowline were at $\sim 25$ K and $\sim 80$ au, planet b's composition would suggest it acquired most of its mass beyond this distance). Indeed, a resonant configuration for the system would likely require at least some migration \citep[e.g.,][]{Zurlo2022, Poblete2025}. Finally, we note that all planets most likely formed outside the CO$_2$ snowline ($\sim 10$ au for the assumed disk properties) although planet e's semimajor axis is close to this snowline's location.

\subsubsection{Estimated metal content of the planets}
The measured atmospheric metallicity of the planet likely provides a lower limit on its total bulk metallicity \citep{Thorngren2019}. We use the measured atmospheric metal abundances of the HR 8799 planets to estimate their metal mass fraction assuming that the atmospheric metallicity is equal to their bulk metallicity. For this calculation, we include the following metals: C, N, O, Mg, Si, S, and Fe. These metals contribute $\sim 90$\% ($Z \simeq 0.012$) of the total metal content in solar composition material ($Z = 0.014$, \citealt{asplund_Chemical_2009}, Neon constitutes the majority of the remaining metal mass fraction). The enrichment of refractory elements Si, Mg, and Fe is assumed to be the same as that of sulfur. We only use the measured abundance of N for planet b and assume it is absent in three inner planets for the purpose of these calculations. Figure~\ref{fig:planets_metallicity} and Table~\ref{table:metal_mass} show the  metal mass fraction and metal mass for the HR 8799 planets estimated with these assumptions. The total metal mass in all four planets is $408^{+45}_{-40} \, {\rm M}_\oplus$.

Given that sulfur traces solid accretion, we can go beyond the total metal mass estimates and calculate the metal mass that came in via solids and the remaining that came in through metal-enriched gas. We can then use these estimated masses to understand how the planets accreted their metals and the amount of metals required in the disk to satisfy these estimates. For this calculation, we assume that planet b's C, O, and S (and additional refractory) inventory came from solid accretion while its N inventory came from the accretion of disk gas. For planets c, d, and e, the solids bring in all of the measured sulfur as well as part of the C and O that remains in solids after the removal of CO, which is assumed to constitute 40\% of the total C repository (in line with the earlier assumption in Section~\ref{sec:building_blocks}).

Figure~\ref{fig:Metal_vs_semimajor} shows the metal mass accreted from solids (solid markers) as well as the total metal mass (open markers) for each planet as a function of their semi-major axes. These values are listed in Table~\ref{table:metal_mass}. For planet b, most of the metal mass likely comes in via solids with N making up the difference. Since planets c and d are significantly more enriched in C and O relative to S, the fraction of their metal mass accreted via solids is correspondingly lower. For planet e, the uncertainty in S abundance places much looser constraints on how the metals acquired by the planet were partitioned between solids and gas. Nonetheless, its C and O abundances are compatible with the accretion of $3 \times$ stellar CO-enriched gas, in agreement with inferences for planets c and d (Figure~\ref{fig:comp_curves}). A more precise measurement of planet e's S abundance would be useful for pinning down the origin of its C and O enrichment.

\subsubsection{How did these planets accrete these metals?}

\begin{figure}
    \centering
    \includegraphics[width=\linewidth]{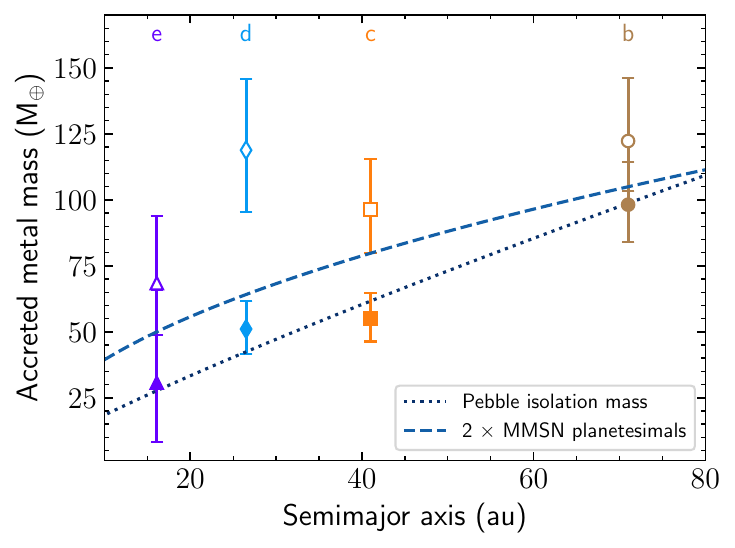}
    \caption{The total metal mass (empty markers) as well as the metal mass accreted via solids (solid markers) for the HR 8799 planets as a function of their semimajor axes. The dashed and dotted lines show the expected amount of solids accreted by the planet in the planetesimal and pebble accretion paradigms.}
    \label{fig:Metal_vs_semimajor}
\end{figure}

The estimated metal mass accreted via solids increases with planet semi-major axis, mirroring the trend in S/H. We compare the solid mass accreted with simple estimates of the expected amounts from the planetesimal and pebble accretion paradigms. Assuming that planets accrete all the planetesimals in their feeding zone $\Delta$ ($ = \pm 3.5 \times R_{\rm Hill}$ the Hill radius, \citealt{Lissauer1993}) and that the surface density of planetesimals $\Sigma_{\rm pls} \propto r^{-3/2}$, we find the total planetesimal mass $M_{\rm pls, \, fz}$ accreted to be:
\begin{equation}
    M_{\rm pls, \, fz} \simeq 79 \, \bigg(\frac{\Sigma_{{\rm pls,} \, r = 40~{\rm au}}}{0.26~{\rm g~cm^{-2}}} \bigg) \, \bigg(\frac{\Delta}{3.5~ R_{\rm Hill}} \bigg)\bigg(\frac{r}{40~{\rm au}} \bigg)^{0.5} \, M_\oplus. 
\end{equation}
This estimate is shown in Figure~\ref{fig:Metal_vs_semimajor} with a dashed line and matches the estimated solid mass in the planets well. The required $\Sigma_{\rm pls}$ is just $2 \times$ Minimum Mass Solar Nebula (MMSN) value of $0.13~{\rm g~cm^{-2}}$ at 40 au \citep{Chiang2010} and smaller than the Minimum Mass Extrasolar Nebula (MMEN $\sim 5 \times$ MMSN, \citealt{Chiang2013}). The corresponding Toomre $Q = 6.2 \, (r/{\rm 40 \, au})^{-3/14}$ is $> 3$ even as far out as 1000 au, so such a disk is gravitationally stable. We note that given the planets' proximity in terms of mutual Hill radii (3.2 $R_{\rm Hill, ed}$, 2.8 $R_{\rm Hill, dc}$, 3.8 $R_{\rm Hill, cb}$, where the subscript indicates planet pairs), their feeding zones partially overlap. Completely non-overlapping feeding zones would require $\Delta \approx 1.8$ and $\Sigma_{\rm pls}$ to be $4 \times$ MMSN, which is gravitationally stable and reasonable for a HR 8799-like system. Even for a $2\times$ MMSN disk, the total planetesimal mass in the relevant region ($\approx 9.5 - 98$ au, including the $3.5~R_{\rm Hill}$ region within 16 au and beyond 71 au) is 211 \Me, which is commensurate with the measured metal mass accreted by solids ($238^{+40}_{-31}$ \Me), implying that a $2\times$ MMSN disk with some redistribution of solids amongst the planets would match the observations. 
In the solar system, planetesimals were present out to at least 40~au. Given a more massive host star, it is quite plausible that the planetesimal disk of HR 8799 extended much further out \citep[e.g.][]{Andrews2018}, potentially beyond planet b's orbit at $\sim70\,$au. 

The main challenge with planetesimal accretion is that the accretion timescale $t_{\rm acc}$ becomes very long at large distances, especially if the collision cross-section is just the geometric cross-section:
\begin{align}
    t_{\rm acc} &\simeq \frac{M_{\rm pls, \, fz}}{\pi R_{\rm p}^2\Sigma \Omega F} \nonumber \\
    &\sim \bigg(\frac{1}{F}\bigg) \, \bigg(\frac{2~R_{\rm Jup}}{R_{\rm p}}\bigg)^2 \, \bigg(\frac{r}{40~{\rm au}}\bigg)^{3.5} \, 10^{11}~{\rm years},
\end{align}
where $\Omega$ is the Keplerian frequency, $R_{\rm p}$ is the planet radius, and $F = 1 + (v_{\rm esc} / v_{\rm rel})^2$ is the gravitational focusing factor that can boost the accretion cross-section if the relative velocity $v_{\rm rel}$ of the planet-planetesimal encounters is much less than the escape velocity of the accreting planet $v_{\rm esc}$ . For $t_{\rm acc}$ to be less than the disk lifetime $\lesssim 10 ~{\rm Myr}$, $F \times (R_{\rm p} / 2 R_{\rm Jup})^2 \gtrsim 10^4$. This requires significant damping of planetesimal random velocities -- by disk gas or smaller solids -- so that $v_{\rm rel}\ll v_{\rm esc}$, and/or a significantly larger accretion cross-section than the geometric cross-section of a $5-10$ Jupiter mass object at 1 Myr, which is expected to be $\approx2.0-2.4~R_{\rm Jup}$ from evolutionary models \citep{Phillips2020, Morley2024, Davis2025}.

The metals accreted via solids may have been accreted in the form of small mm-cm sized `pebbles' instead. The advantage pebbles have over planetesimals is that they are typically accreted much more rapidly because their accretion cross-section is much larger than the geometric cross-section of the growing planet \citep{Ormel2017}. The classical expectation for the amount of pebbles accreted before planets reach pebble isolation mass for our adopted disk and stellar parameters, and an assumed dimensionless Shakura-Sunyaev parameter for turbulence $\alpha = 3 \times 10^{-4}$ (which is intermediate in log-space between the accepted range of $10^{-4} - 10^{-3}$ for this parameter)  is \citep{Bitsch2018}:
\begin{equation}
    M_{\rm pebble, iso} \simeq 60 \, \bigg(\frac{r}{40~{\rm au}} \bigg)^{6/7} \, M_\oplus.
\end{equation}
This is shown as the dotted line in Figure~\ref{fig:Metal_vs_semimajor} and matches the solid mass accreted by the planets reasonably well too. However, numerous observational \citep{Gasman2025, Krijt2025} and theoretical \citep{Zhu2012, Drazkowska2019, Lee2022, Stammler2023, Huang2025} studies have shown that the disk sub-structures created by planets when they reach pebble isolation mass are leaky and let pebbles through to the inner disk. Some of these leaking pebbles may be accreted by planets \citep{VanClepper2025}, meaning the pebble isolation mass could be considered as a minimum mass of pebbles accreted by a planet. The measured trend is consistent with expectations from pebble isolation alone with chosen parameters, although to explain the observed atmospheric abundances, it would require the interior to be well-mixed.

\begin{figure*}
    \centering
    \includegraphics[width=\linewidth]{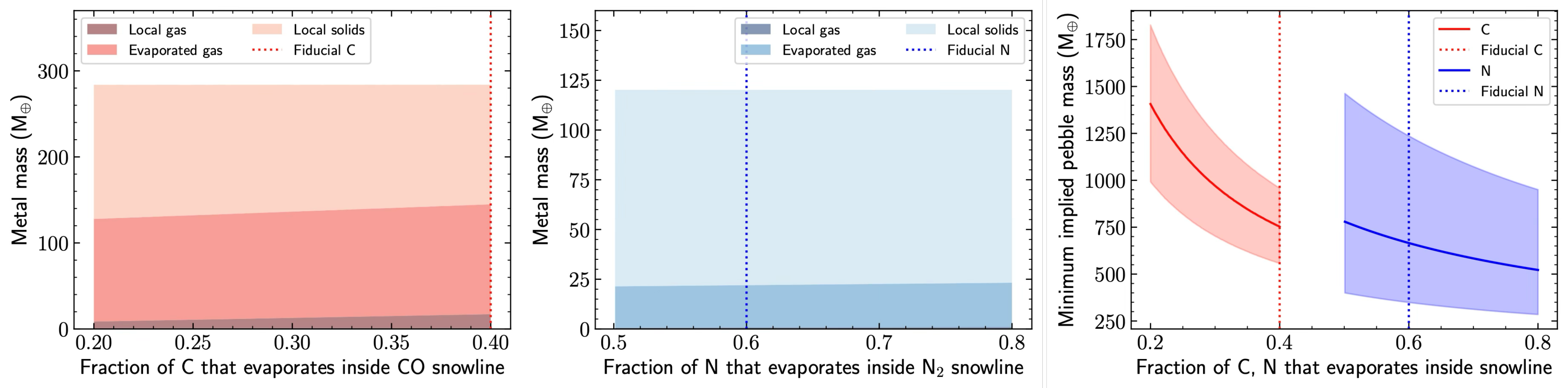}
    \caption{Left panel: Metal mass of the three inner planets (c, d, e) from different sources (solids and gas with local composition, and excess metals from gas due to pebble drift and evaporation) as a function of the assumed fraction of C that evaporates inside CO snowline. We only plot the median values in this panel. Middle panel: Similar to left panel but for metals from different sources for planet b with varying amounts of N that evaporates inside N$_2$ snowline. Right panel: Pebble mass implied by the excess (evaporated) metals accreted from disk gas as a function of how much C and N evaporate inside the CO and N$_2$ snowlines, respectively.}
    \label{fig:implied_pebbles}
\end{figure*}

The excess metals implied by the super-stellar volatile-to-refractory ratios (C/S, O/S, and N/S) require the accretion of metal-enriched gas. Figure~\ref{fig:comp_curves} shows that the accreted gas was likely enriched in CO at the $3\times$ solar level. The elevated N/H for planet b also suggests that it accreted disk gas that was enriched in N at the $10-30 \times$ solar level. The total mass of metals acquired by planets c, d, and e via accretion of CO-enriched gas that evaporated off drifting pebbles is $128^{+35}_{-33}$ \Me. For planet b, the amount of N accreted via N$_2$-enriched gas from drifting pebbles is $21^{+18}_{-10}$ \Me. These estimates only include the excess CO or N$_2$ due to pebble drift and evaporation, and exclude the CO and N$_2$ gas that would have evaporated from local solids ($0.4 \times$ solar and $0.6 \times$ solar, respectively, for our fiducial choice), which is why they are lower than the values for metals accreted from the gas in Table~\ref{table:metal_mass}. The excess metal mass accreted from gas enables us to calculate two independent estimates (from CO and N$_2$) of the minimum pebble mass that must have passed through the snowlines to produce the required amount of CO and N$_2$ vapor. Excitingly, this provides a novel constraint on the progenitor disk mass that goes beyond the minimum mass nebula estimate that is typically derived solely from the mass of metals present in a planet and assumes 100\% accretion efficiency.

The minimum pebble mass required depends on the fraction of C ($f_{\rm [CO]/[C]}$) in CO and N ($f_{\rm [N_2]/[N]}$) in N$_2$. Changing these fractions has a relatively minor effect on the amount of metals coming from different sources (solids and gas with local composition, metal-enriched gas, left two panels in Figure~\ref{fig:implied_pebbles}) but it has a linear effect on the fraction of the pebble mass that evaporates at a given snowline. Figure~\ref{fig:implied_pebbles} right panel shows the implied minimum pebble mass as a function of the fraction of C in CO and N in N$_2$. For our fiducial value of $f_{\rm [CO]/[C]}$ = 40\%, the evaporation of CO removes 17\% of the mass of the pebble flux. Dividing the excess CO mass in the three inner planets ($128^{+35}_{-33}$ \Me) by mass fraction of pebbles that evaporates at the CO snowline, we find that a minimum $750 \pm 200$ \Me of pebbles need to pass through the CO snowline.

The fiducial fraction of N in N$_2$ $f_{\rm [N_2]/[N]} =$ 60\% and its evaporation removes 3.2\% of the pebble flux mass. N$_2$ has only been measured in planet b and its abundance is less well constrained, so we get a looser constraint of $670^{+570}_{-320}$ \Me on the pebble mass. The minimum pebble mass constrained by both CO and N$_2$ vapors agree quite well. Tighter compositional constraints, especially for N in all the planets, and a more stringent comparison of the pebble mass implied by these two different volatiles would provide a powerful test of the hypothesis that pebble drift and evaporation enriched the disk gas that was accreted by these planets. 

The overall dust mass required to fit the estimated metal abundances of the HR 8799 planets is $\gtrsim 1000$ \Me. This need not be the instantaneous disk mass (which can be lower) at any point during the star's formation, especially if the planets started forming early enough when the disk was fed with more material from the protostellar environment. Typical disk mass estimates that assume optically thin emission would put such a disk in the top few \% of Class 0/I disks (\citealt{Tobin20, Tychoniec2020}; though see \citealt{Xu2022} who revised these disk masses upwards). However, emerging evidence suggests that disks are not optically thin at ALMA wavelengths \citep{Zhu2019, Xin2023, Garufi2025}. Using longer wavelength data (e.g. from the VLA, at which disks are more likely to be optically thin), \cite{Painter2025} find that eight bright sources in the nearby Taurus star forming region all likely contain $\gtrsim 1000$ \Me of solids, which is $\sim 10 \times$ greater than the previously estimated mass of these disks (see their Figure 12). The disks in their sample are mostly Class II disks so we would expect younger Class 0/I disks to be even more massive. A disk with Toomre $Q = 2$ that is marginally gravitationally stable would need to be $\sim 500$ au in size to contain $\sim 1000$ \Me of solids beyond the CO snowline (55 au). A more stable disk (larger $Q$) would have to be even larger. The disk that formed the HR 8799 planets was likely similar to the brightest and largest disks that we find in star-forming regions.

We have demonstrated that the abundance ratios of the HR 8799 planets can be self-consistently explained by a simple model of solid accretion in quantities constrained by sulfur abundances plus accretion of gas enhanced in volatile metals (C, O, N) due to pebble drift. The metal enhancement in the gas required for all four planets can be explained with a self-consistent disk model. While more complicated physical processes likely impact the detailed abundances of the planets, the fact that such a simple model is able to explain the fundamental trends observed in this system is evidence in favor of the overall paradigm of metal enhancement due to a combination of solid accretion plus enhanced gas metallicity due to pebble drift.

It is our wish that the detailed volatile-to-refractory abundances of the HR 8799 planets will motivate follow-up work in several areas. For example, while the planet formation model we present above represents a starting framework, it does not consider all possible complexities in the planet formation process. For example, the disk's chemical composition will evolve not only due to pebble drift and evaporation, but also from chemical processing of the disk gas and the volatile ice on grain surfaces. The latter effect has been explored in the literature and is important to consider for more detailed planet formation inferences \citep{Eistrup2016, Eistrup2018, molliere_Interpreting_2022a}. In addition, the model above does not consider the impact of planetary migration on the final composition, and importantly, assumes that the planets do not migrate across snowlines. We defer detailed considerations of these effects to more complex formation models in future work. We also note that the model presented above is agnostic to whether the refractory sulfur observed in the planetary atmospheres comes from erosion of solids from the core, or solids that are accreted in the gas accretion phase. While this does not impact the conclusions we presented, it is worth studying as the implied metal mass from solids is higher ($50-100~\Me$ for planets d, c, b) than the required core mass for runaway gas accretion (see also \citealt{Chachan2025b} who reach similar conclusions for transiting giant planets). There are also alternative models to explain enhanced volatile abundances in giant planet atmospheres \citep{VanClepper2025}, which could be tested against these observations. Finally, uncertainty is also present in the disk chemical composition, such as the amount of C in CO, and N in N$_2$ which we have highlighted. Future studies that better constrain these values would be valuable for planet formation inferences.

\subsection{Comparison to hot Jupiters}\label{sec:hjs}
While the measurements of H$_2$S in the HR 8799 planets from \citet{RuffioXuan2026} and this analysis represent the first such measurement in widely-separated, self-luminous planets, there have been a heightened interest in measuring refractory abundances in hot Jupiters over the past few years. H$_2$S was detected in the canonical hot Jupiter HD 189733 b with both JWST/NIRCam and MIRI data \citep{Fu2024, Inglis2024b}. For this planet, \citet{Fu2024} report $\rm S/H=10.4_{-2.4}^{+3.3}$, $\rm O/H=3.4_{-0.4}^{+0.5}$, and $\rm C/H=1.0_{-0.14}^{+0.17}$ (relative to stellar). The star HD 189733 A has solar C and O abundances \citep{brewer_SPECTRAL_2016}. Assuming the star also has solar S abundance,\footnote{We note that for HD 189733A, \citet{Kolecki2022} report $\rm [S/H]=-0.30\pm0.08$, but \citet{Luck2017} report $\rm [S/H]=+0.60\pm0.39$, so there is inconsistency in the sulfur abundance of this star.} this implies a very low $\rm C/S$ of $\approx0.1$, but also a low C/O ratio. This led \citet{Fu2024} to propose the planet may have accreted water-rich planetesimals just outside the water snowline ($\sim2$ AU), which would enrich the planet in O and S. The very different volatile-to-refractory composition of this hot Jupiter compared to the HR 8799 planets is consistent with these planets having very different formation histories.

Ultra-hot Jupiters with ($T_{\rm eq}\gtrsim2000~$K) are also ideal targets for measuring refractory abundances \citep{Lothringer2021, Chachan2023}, since elements like Fe, Mg, and Si can be in the gas phase on the daysides of these planets. Indeed, high-resolution spectrographs operating at visible wavelengths (e.g., HARPS, ESPRESSO, Maroon-X) have detected and constrained abundances of these key refractory species in many ultra-hot Jupiters \citep{Gandhi2023_uhj}. However, the majority of the results constraining the abundances of refractories in ultra-hot Jupiter atmospheres lacked access to volatiles (which primarily absorb in the infrared), and so refractory-to-volatile ratios ($\mathcal{V}/\mathcal{R}$) have typically not been reported, except in a few cases. WASP-121~b has been the target of multiple studies which have jointly constrained the abundances of volatiles and refractories \citep{Lothringer2021, Smith2024, Pelletier2025_highres, Pelletier2025_jwst, Evans-Soma2025}. While all studies found enhanced refractory abundances compared to stellar and super-stellar C/O, \citet{Lothringer2021} and \citet{Smith2024} found low volatile-enrichment compared to refractories (both finding $\mathcal{V}/\mathcal{R}\sim0.4 \pm0.4 \times$stellar), while \citet{Pelletier2025_highres} found slightly enhanced $\mathcal{V}/\mathcal{R}$ on the order of $\sim 2$, and the most recent $JWST$ data preferred a $\mathcal{V}/\mathcal{R}$ consistent with stellar or slightly super-stellar values \citep{Pelletier2025_jwst, Evans-Soma2025}. The two other ultra-hot Jupiters with constraints on both their volatile and refractory content are KELT-20 b and WASP-178 b. Both planets have stellar or sub-stellar $\mathcal{R}$/H, likely requiring super-stellar O/H to explain their water absorption feature while accounting for its dissociation at the high temperatures of the atmosphere, and sub-stellar C/O, thereby implying super-stellar $\mathcal{V}/\mathcal{R}$  \citep{Chachan2025, Finnerty2025, Lothringer2025}. 

Together, these results do not paint a clear picture of a shared formation and migration pathway for all hot Jupiters as the results suggest varying C/O, O/H, C/H, and $\mathcal{V}/\mathcal{R}$ ratios. These differences may be due to true compositional differences or may also be driven by the variety of assumptions in modeling frameworks \citep[e.g., water dissociation, cold trapping, etc.;][]{Pelletier2025_jwst}. Larger samples and more sophisticated or physically-motivated modeling frameworks may lead to a more coherent picture of abundance patterns in hot Jupiters. 

\section{Conclusion}
\label{sec:conclusion}

In this work, we aimed to measure detailed atmospheric abundances of the four gas giant planets orbiting the star HR~8799 ($5$--$10~\Mj$; 16--71~au) and inform their formation pathways. 
We used JWST/NIRSpec IFU observations from the GTO program 1188 with the moderate-resolution spectroscopy mode ($R\sim2,700$) to acquire $3$--$5\,\mu$m spectra of the four planets in the system. 
Compared to \citet{RuffioXuan2026}, this new analysis includes HR~8799~b, which was outside of the field-of-view in the previous set of NIRSpec observations, as well as improved spectra of HR~8799~cde with an S/N per spectral bin increased by more than a factor two. 
The NIRSpec data post-processing was performed with the Python package \texttt{BREADS} \citep{breads} and the atmospheric inferences used the radiative transfer code \texttt{petitRADTRANS} \citep{molliere_petitRADTRANS_2019}.
In addition to carbon and oxygen-bearing molecules (CO, CH$_4$, H$_2$O, CO$_2$, $^{13}$CO, C$^{18}$O), we constrain the sulfur (H$_2$S) abundances  of HR~8799~bcd to within 0.1~dex, and detect nitrogen (NH$_3$) in HR~8799~b. Due to the degeneracy between [N/H] and the NH$_3$ quench pressure in the \texttt{petitRADTRANS} retrievals, we use the chemical model \texttt{VULCAN} to infer a nitrogen enrichment of N/H$=21.2^{+16.2}_{-8.8}\times$ stellar for HR 8799 b; see all retrieved abundances in Table~\ref{table:spec_results}. 

The super-stellar O/S, C/S in HR~8799~cd, and tentative super-stellar N/S in HR~8799~b suggest accretion of metal-enriched gas during the formation of the planets, which could be explained by pebble drift and evaporation. In this context, the enrichment pattern of C, O, S, and N across the system would be consistent with the outer planet forming between the N$_2$ and CO snowlines, and the three inner planets forming within the CO snowline. In this work, we devised a relatively simple disk model with pebble drift and evaporation that can explain the compositions of the four planets. However, the planet formation process is highly complex, and our model does not consider a variety of potential complications from time-dependent disk chemistry and planetary migration. It would be valuable for future studies to incorporate such effects and assess whether they also reproduce the formation history inferred here.

Overall, this work demonstrates JWST's capability to precisely constrain refractory and volatile species in distant gas giant planets, which are essential to determine the origin of metal enrichment by disentangling solid accretion from gas accretion \citep{Chachan2023}.
In the future, additional modeling studies, and observations of exoplanet atmospheric compositions from transiting planets to distant gas giants that go beyond carbon and oxygen-bearing molecules are needed to better constrain the accretion processes underlying giant planet formation.

\begin{acknowledgments}
J.W.X. thanks Sergey Yurchenko, Maria Pettyjohn, Jonathan Tennyson, Caroline Morley, Brianna Lacy, Paul Mollière, Tamara Molyarova, Evert Nasedkin, Jonathan Fortney, Sagnick Mukherjee, Rixin Li, Eugene Chiang, Peter Gao, and Jason Wang for helpful discussions. 

Part of this work was supported by the National Aeronautics and Space Administration under Grants/Contracts/Agreements No. 80NSSC25K7300 (J.-B.R.) issued through the Astrophysics Division of the Science Mission Directorate. Any opinions, findings, and conclusions or recommendations expressed in this work are those of the author(s) and do not necessarily reflect the views of the National Aeronautics and Space Administration.
J.W.X. acknowledges support from NASA through grants (program \#5342, \#4982) from the Space Telescope Science Institute, which is operated by the Association of Universities for Research in Astronomy, Inc., under NASA contract NAS 5-03127. J.W.X is also thankful for support from the Heising-Simons Foundation 51 Pegasi b Fellowship (grant \#2025-5887). J.~M. acknowledges support from the NASA Exoplanet Research Program grant 80NSSC23K0281. 
D.J.\ is supported by NRC Canada and by an NSERC Discovery Grant. This work is based on observations made with the NASA/ESA/CSA James Webb Space Telescope. The data were obtained from the Mikulski Archive for Space Telescopes at the Space Telescope Science Institute, which is operated by the Association of Universities for Research in Astronomy, Inc., under NASA contract NAS 5-03127 for JWST. These observations are associated with program 1188.
Part of this work was carried out at the Jet Propulsion Laboratory, California Institute of Technology, under a contract with the National Aeronautics and Space Administration (\# 80NM0018D0004)

\end{acknowledgments}

\facilities{JWST(NIRSpec)}

\software{\texttt{petitRADTRANS} \citep{molliere_Retrieving_2020}; \texttt{pymultinest} \citep{Buchner2014, Feroz2019}; VULCAN \citep{Tsai+17} }

\appendix

\newpage

\section{Fitted parameters and priors for HR 8799 c, d, e}\label{app:cde_table}

Here we list the fitted parameters and priors for the retrievals of HR 8799 c, d, and, e, which are nearly identical to those of planet b, except we do not include a parameter for [N/H] since NH$_3$ is not detected in these inner planets from the current data. 

\begin{deluxetable*}{ll|ll}[b!]
\tablecaption{Fitted Parameters and Priors for HR 8799 c, d, e Retrievals\label{tab:param_prior_cde}}
\tabletypesize{\small}
\tablehead{
\colhead{Parameter} & \colhead{Prior} & \colhead{Parameter} & \colhead{Prior}
}
\startdata
Mass ($\Mj$)                 & $\mathcal{N}(\mu_{\rm M, dyn}, \sigma_{\rm M, dyn})^{\rm (a)}$ 
  & $T_{\rm ref}$ [$P=10^{2}$] (K) & $\mathcal{U}(2000, 4000)$ \\
Radius ($\Rj$)               & $\mathcal{U}(0.6, 2.0)$ 
  & $\left(d\ln{T}/d\ln{P}\right)_{1}$ [$10^2$] & $\mathcal{N}(0.15, 0.01)$ \\
$[{\rm C/H}]$                & $\mathcal{U}(0.0,1.2)$ 
  & $\left(d\ln{T}/d\ln{P}\right)_{2}$ [$10^1$] & $\mathcal{N}(0.18, 0.04)$ \\
${\rm C/O}$                  & $\mathcal{U}(0.0,1.2)$ 
  & $\left(d\ln{T}/d\ln{P}\right)_{3}$ [$10^0$] & $\mathcal{N}(0.21, 0.05)$ \\
\logco                       & $\mathcal{U}(0, 8)$ 
  & $\left(d\ln{T}/d\ln{P}\right)_{4}$ [$10^{-1}$] & $\mathcal{N}(0.16, 0.06)$ \\
\logcoo                      & $\mathcal{U}(0, 8)$ 
  & $\left(d\ln{T}/d\ln{P}\right)_{5}$ [$10^{-2}$] & $\mathcal{N}(0.08, 0.025)$ \\
log(CO$_2$) mass–mixing ratio & $\mathcal{U}(-10, -2)$ 
  & $\left(d\ln{T}/d\ln{P}\right)_{6}$ [$10^{-3}$] & $\mathcal{N}(0.06, 0.02)$ \\
log(HCN) mass–mixing ratio   & $\mathcal{U}(-10, -2)$ 
  & $\left(d\ln{T}/d\ln{P}\right)_{7}$ [$10^{-4}$] & $\mathcal{U}(-0.05, 0.10)$ \\
log(H$_2$S scale factor)     & $\mathcal{U}(-3, 2)$ 
  & $\left(d\ln{T}/d\ln{P}\right)_{8}$ [$10^{-5}$] & $\mathcal{U}(-0.05, 0.10)$ \\
log($P_{\rm quench, C}$/bar) & $\mathcal{U}(-5, 2)$ 
  & $\left(d\ln{T}/d\ln{P}\right)_{9}$ [$10^{-6}$] & $\mathcal{U}(-0.05, 0.10)$ \\
$\log(r_{\rm cloud} / \rm {cm})$  & $\mathcal{U}(-7, 1)$ 
  & $\left(d\ln{T}/d\ln{P}\right)_{10}$ [$10^{-7}$] & $\mathcal{U}(-0.05, 0.10)$ \\
$\log(P_{\rm cloud}/{\rm bar})$ & $\mathcal{U}(-6, 1.5)$ 
  & RV (\kms) & $\mathcal{U}(-50 , 50)$ \\
$\sigma_{\rm g}$             & $\mathcal{U}(1.05, 3)$ 
  & Error multiple & $\mathcal{U}(1, 5)$ \\
${\rm log}(X_{\rm cloud})$   & $\mathcal{U}(-8, 0)$ 
  & $r$ & $\mathcal{U}(300, 1400)$ \\
 & & $r_0$ & $\mathcal{U}(-800, 1200)$ \\
\enddata
\tablecomments{
Here we list parameters and priors for the HR 8799 cde retrievals (Section~\ref{sec:cde_setup}). See Table~\ref{tab:param_prior} for comments. \\
$^{\rm (a)}$ Mass priors from \citet{Zurlo2022}. The adopted values are $7.7\pm0.7~\Mj$, $9.2\pm0.7~\Mj$, and $7.6\pm0.9~\Mj$ for planets c, d, and e respectively. \\
}
\end{deluxetable*}

\section{NIRSpec spectra and P-T profiles for HR 8799 c, d, e}\label{app:pt_cde}
Here we show the NIRSpec spectra and model fits for HR 8799 c, d, and e, as well as their retrieved P-T profiles and emission contribution functions. 

\begin{figure*}
    \centering
    \includegraphics[width=\linewidth]{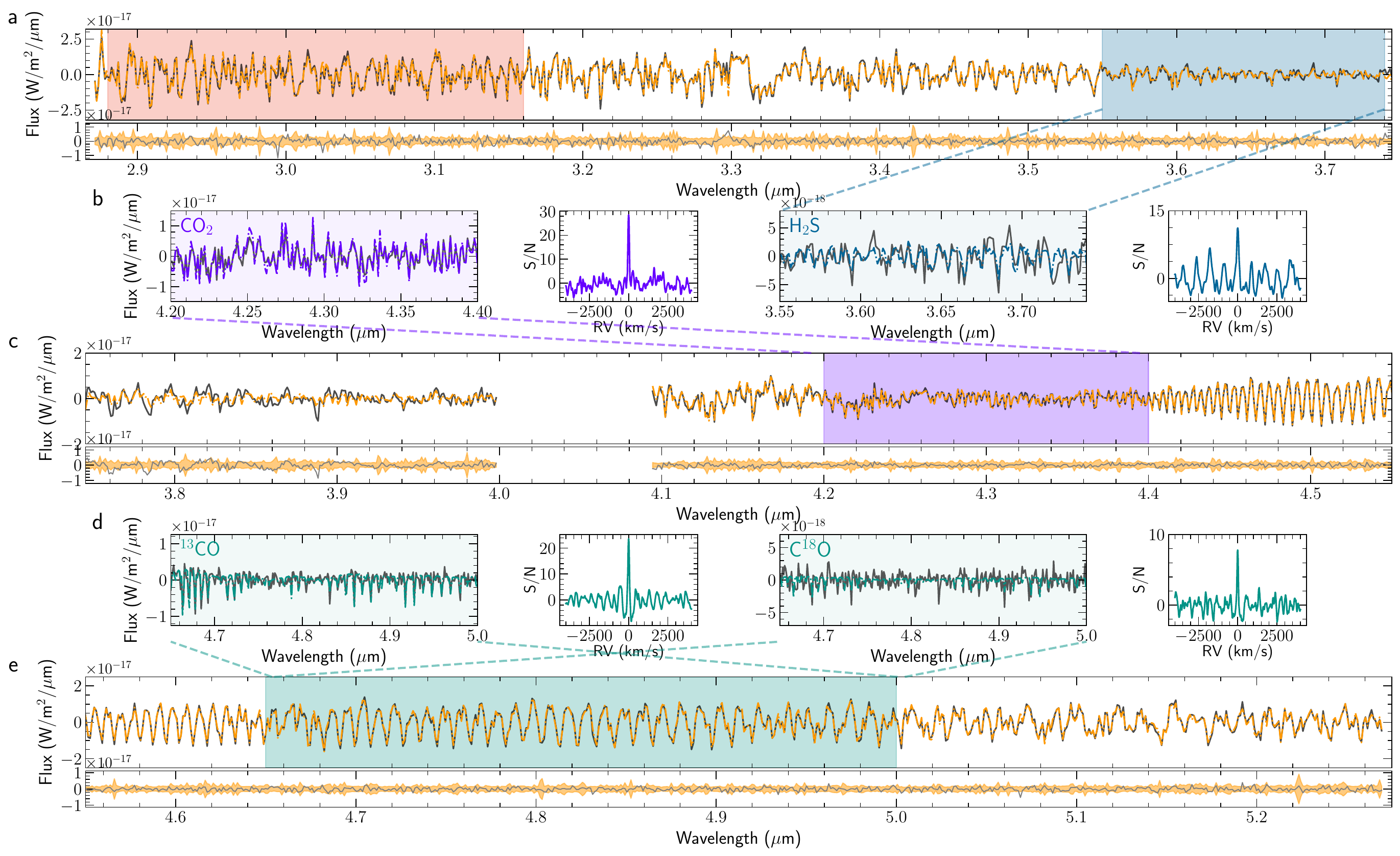}
    
    \includegraphics[width=0.45\linewidth]{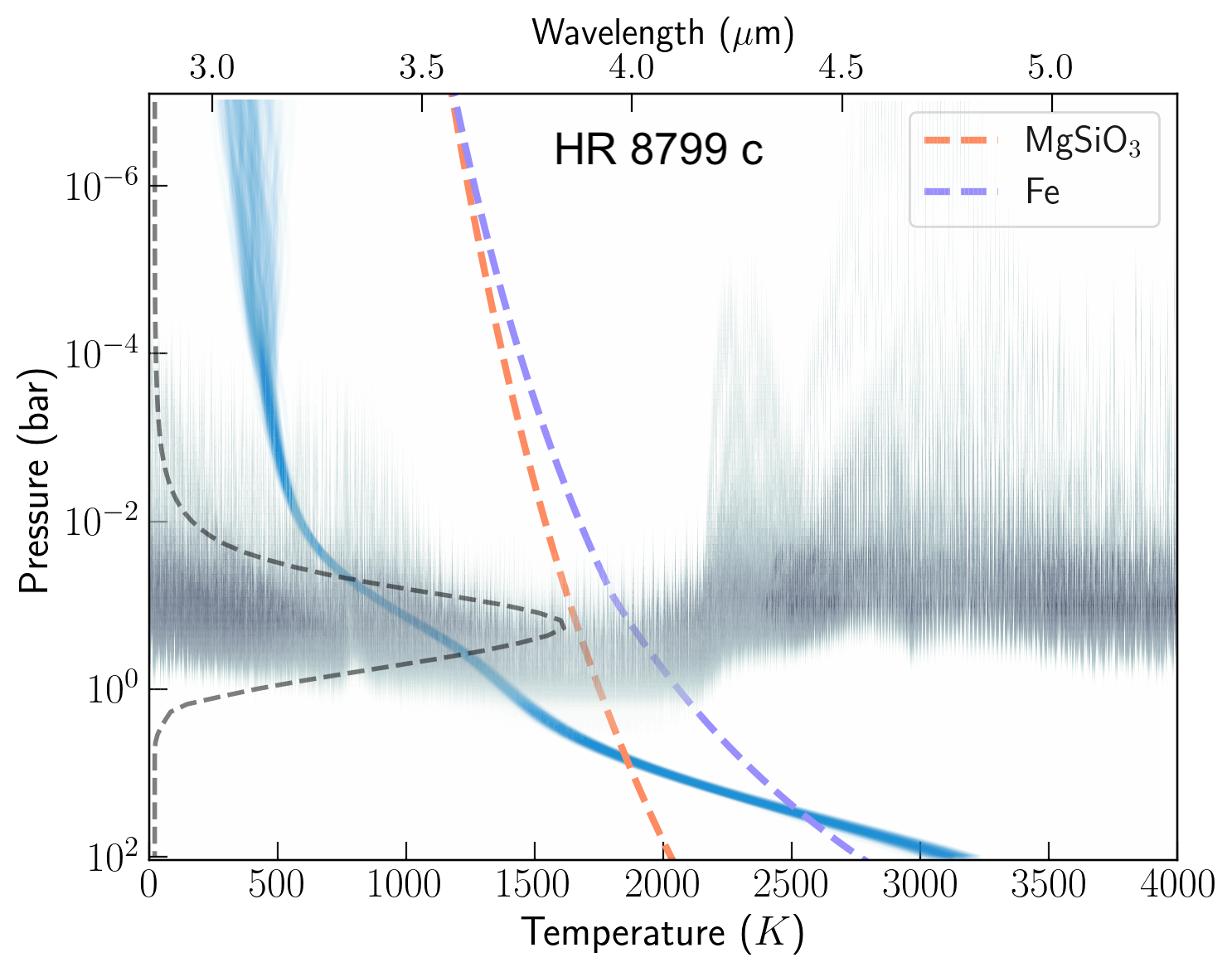}
    \caption{\textbf{Top}: Same as Figure~\ref{fig:spec}, but for HR 8799 c. Because NH$_3$ is not detected in planet c, we highlight detections of CO$_2$, H$_2$S, $^{13}$CO, and C$^{18}$O. \textbf{Bottom}: Retrieved P-T profiles for HR 8799 c, see legend of Figure~\ref{fig:pt_emis}.}
    \label{fig:hr8799c}
\end{figure*}

\begin{figure*}
    \centering
    \includegraphics[width=\linewidth]{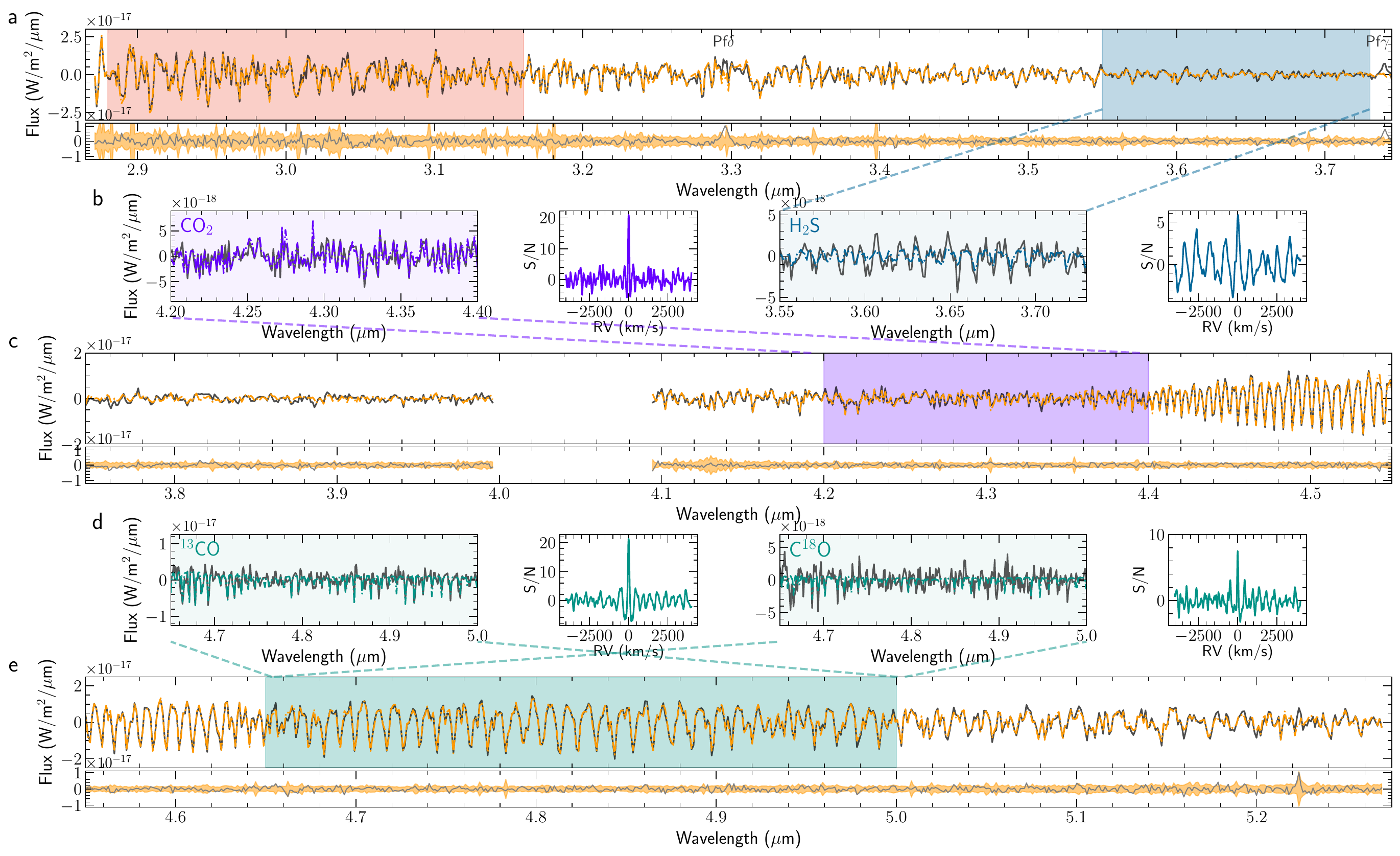}
    \includegraphics[width=0.45\linewidth]{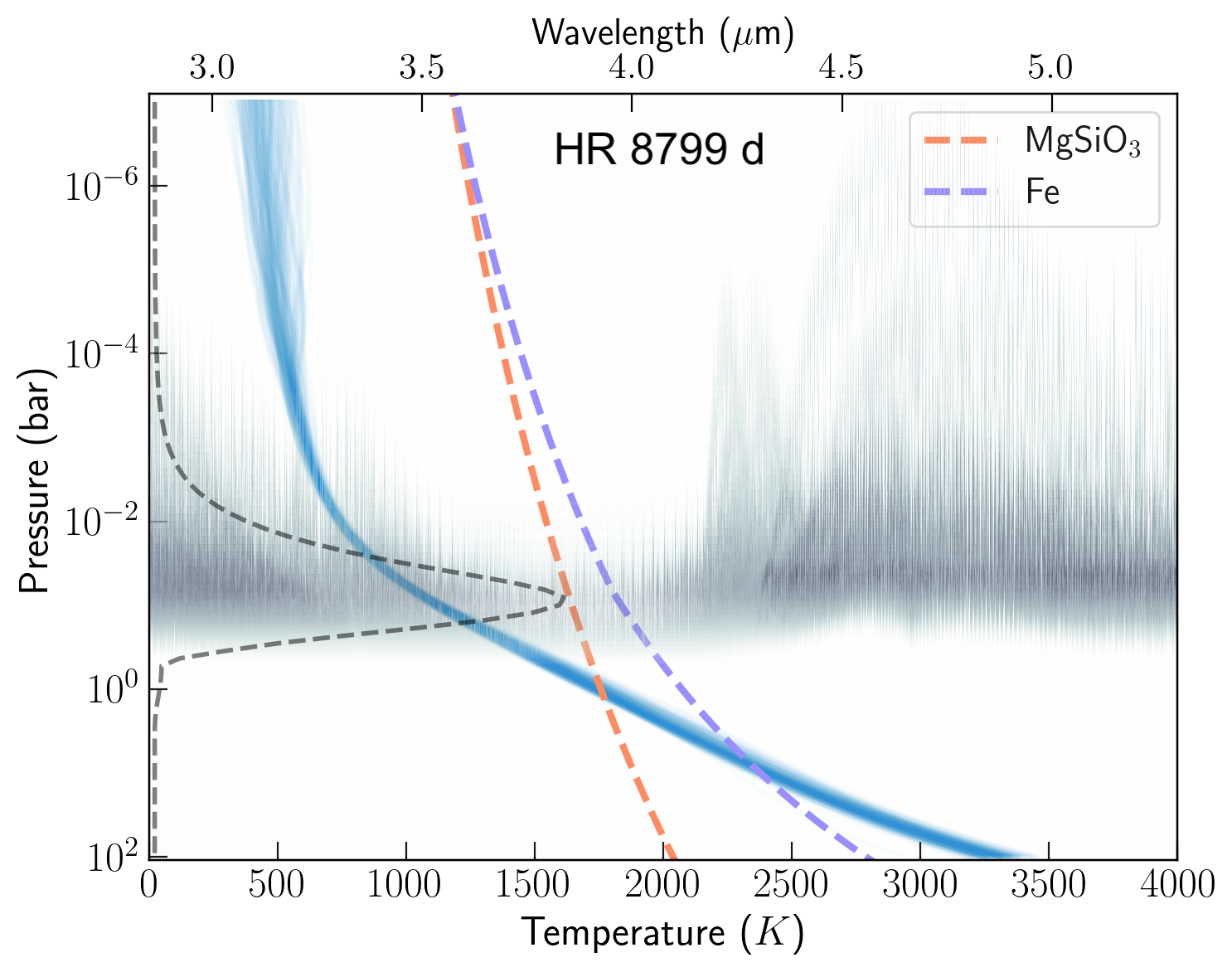}
    \caption{\textbf{Top}: Same as Figure~\ref{fig:spec}, but for HR 8799 d. Because NH$_3$ is not detected in planet d, we highlight detections of CO$_2$, H$_2$S, $^{13}$CO, and C$^{18}$O. We also indicate locations of positive residuals in panel a, which result from over-subtraction of Pf$\delta$ and Pf$\gamma$ lines in HR 8799 A. Since these stellar emission lines are sparse, they do not affect the planet retrievals \citep{RuffioXuan2026}. \textbf{Bottom}: Retrieved P-T profiles for HR 8799 d, see legend of Figure~\ref{fig:pt_emis}.}
    \label{fig:hr8799d}
\end{figure*}

\begin{figure*}
    \centering
    \includegraphics[width=\linewidth]{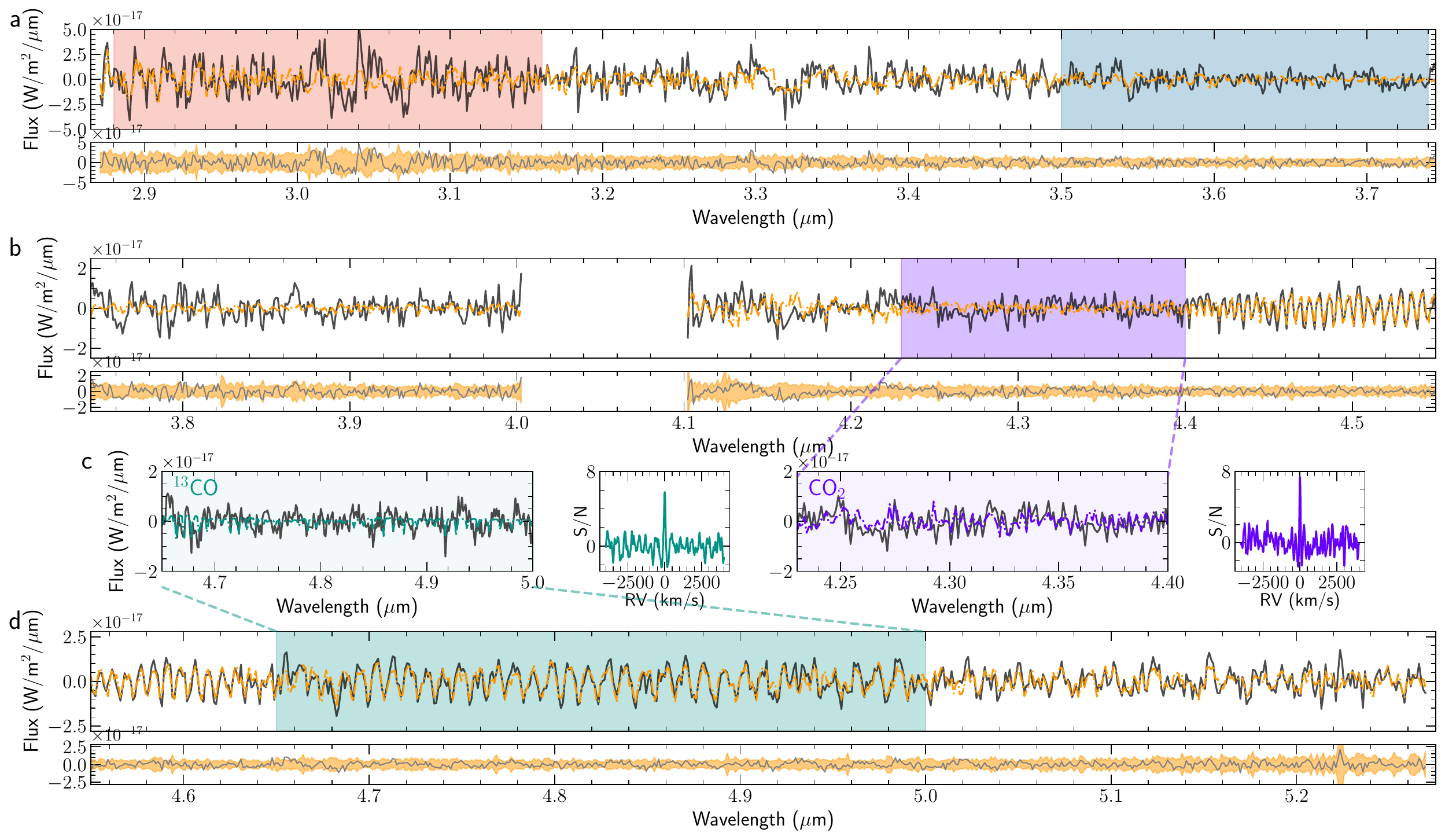}
    \includegraphics[width=0.45\linewidth]{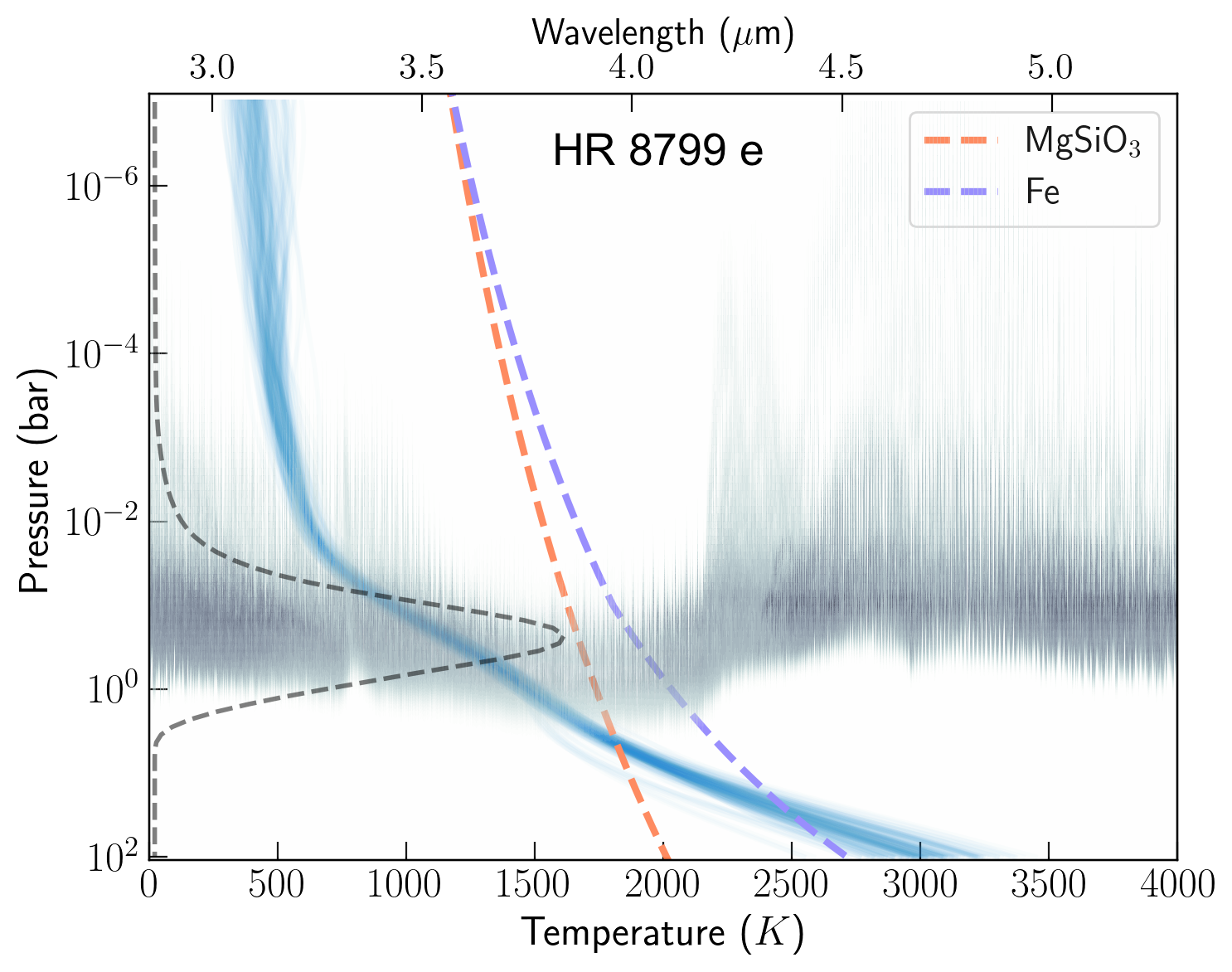}
    \caption{\textbf{Top}: Same as Figure~\ref{fig:spec}, but for HR 8799 e. From our CCF analysis, H$_2$S and C$^{18}$O are not independently detected in planet e, so we highlight detections of CO$_2$ and $^{13}$CO. \textbf{Bottom}: Retrieved P-T profiles for HR 8799 e, see legend of Figure~\ref{fig:pt_emis}.}
    \label{fig:hr8799e}
\end{figure*}

\clearpage

\section{Unidentified spectral features}\label{app:resid_features}

\begin{figure*}
    \centering
    \includegraphics[width=0.9\linewidth]{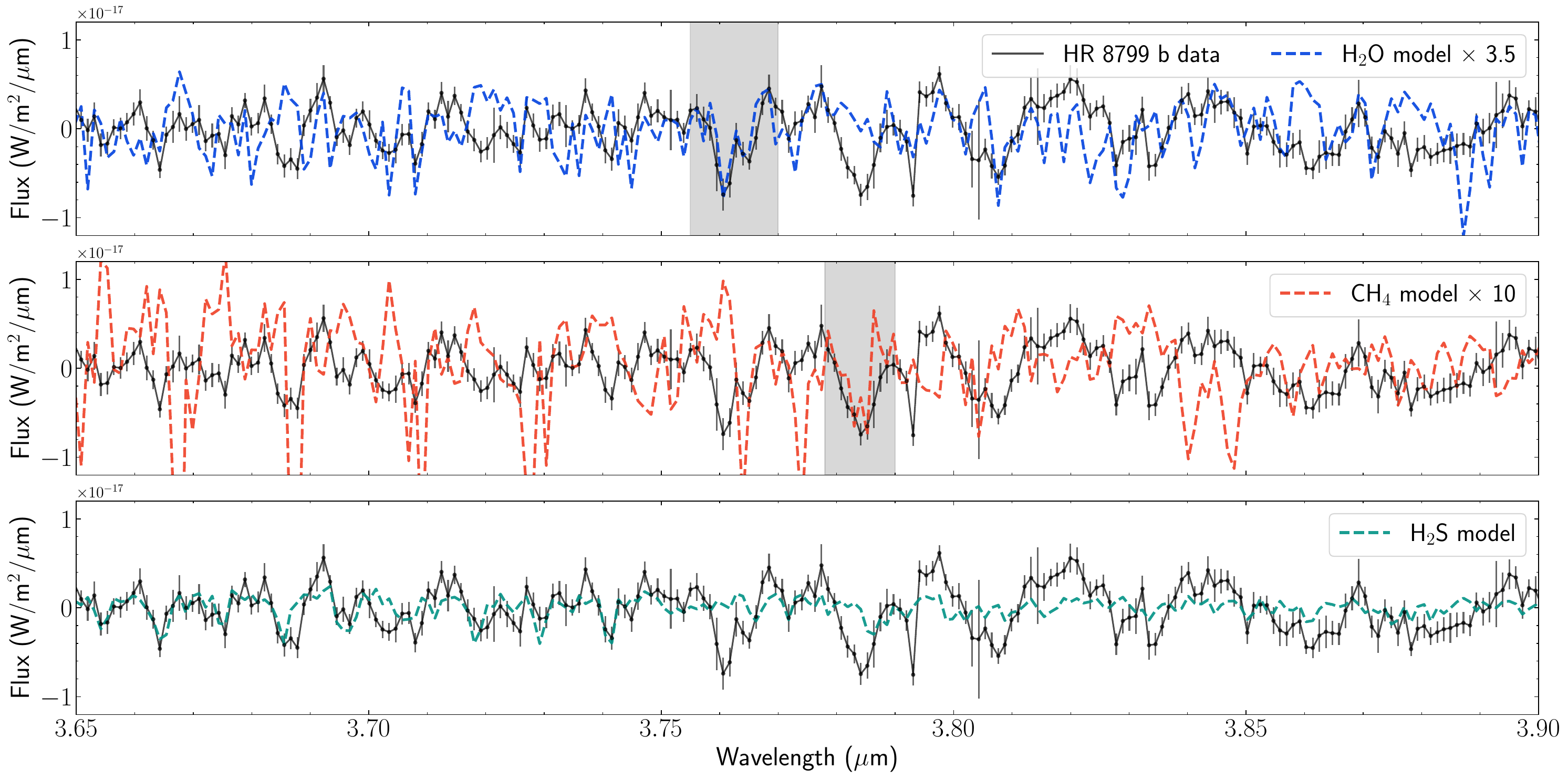}
    \caption{NIRSpec spectra of HR 8799 b compared to single molecule templates. The templates for H$_2$O and CH$_4$ are manually scaled by factors of 3.5 and 10 respectively. While certain lines in the gray shaded regions appear similar to H$_2$O and CH$_4$ absorption lines, these molecules do not match the data well at other wavelengths.}
    \label{fig:b_vs_molecules}
\end{figure*}

For HR 8799 b, a portion of the NIRSpec data between $3.65-3.9~\mu$m is not well fit by the \texttt{petitRADTRANS} models, and show elevated residuals compared to the other wavelengths (see Figure~\ref{fig:spec}). For HR 8799 b, the standard deviation of the residuals is $\approx2.4$ times higher between $3.65-3.9~\mu$m than elsewhere. The main absorption lines between $3.65-3.9~\mu$m come from H$_2$O, CH$_4$, and H$_2$S. Some of the residual features appear to match certain weak H$_2$O, or perhaps CH$_4$, lines. However, to match the depths of these residuals, we need to drastically increase the molecular opacity of H$_2$O or CH$_4$ (see Figure~\ref{fig:b_vs_molecules}). Doing so would grossly overpredict the H$_2$O and CH$_4$ line strengths at all other wavelengths, where these molecules have stronger opacity. The fact that at least some residuals line up with H$_2$O and CH$_4$ lines suggest that potential inaccuracies in the line lists may contribute to the residuals. The opacities we use are computed from the following line lists: \citet{Polyansky2018} for H$_2$O, \citet{Azzam2016} for H$_2$S, and \citet{hargreaves_Accurate_2020} for CH$_4$. 

\begin{figure*}
    \centering
    \includegraphics[width=0.9\linewidth]{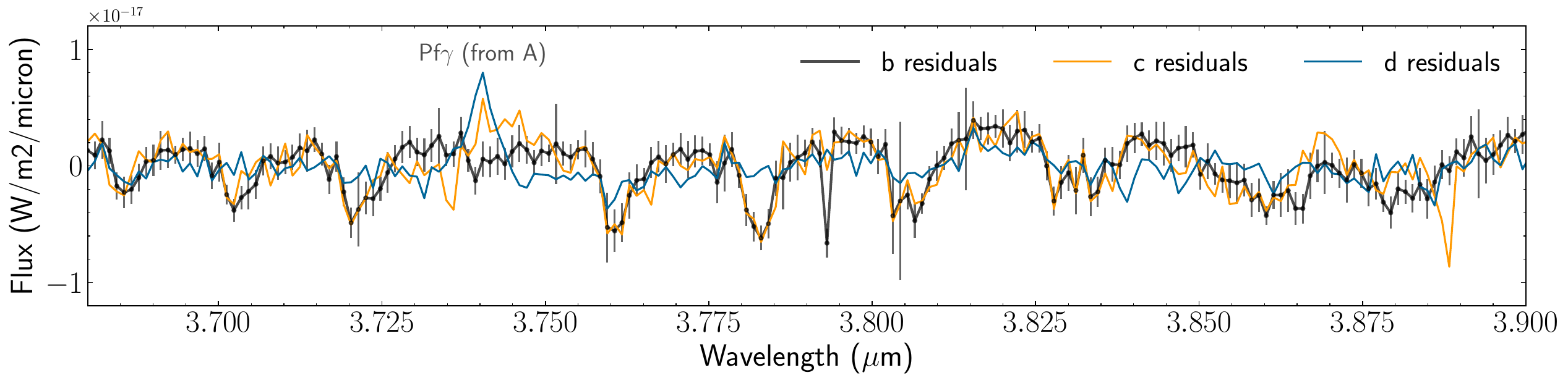}
    \caption{The data residuals (data-model) for the HR 8799 b, c, and d spectra presented in this work. Planets b and c show remarkably similar residuals features, whereas planet d does not show the same features. The data for planet e, not shown, is too noisy in this region to provide a useful comparison.}
    \label{fig:b_vs_c_resids}
\end{figure*}

Interestingly, similar residual features appear in the published NIRSpec data of HR 8799 c \citep{RuffioXuan2026}, as well as the 2nd epoch NIRSpec data of c we analyzed in this work. We illustrate the similarity of the residual features between planets b and c in Figure~\ref{fig:b_vs_c_resids}. On the other hand, planet d or e do not show the same residual features. If these features are real absorption lines from the planets, we have not found a convincing match based on the molecules we tried. The molecules we tested include CH$_3$, C$_2$H$_2$, C$_2$H$_4$, HDO, SO$_2$, HF, SiH$_4$, H$_3$+, SiO, NH, CH, PH, SH, HCl, and NaH. None of them were able to explain the residual features we see. We also checked that the NIRSpec spectrum of the star, HR 8799 A, does not contain these features. Furthermore, when analyzing spectra across all spatial dimensions of the NIRSpec IFU, we found that these residuals features were co-located with planets b and c and were not found in the speckle field of the star. Therefore, what we observe is unlikely to arise from stellar contamination. 

\begin{figure*}
    \centering
    \includegraphics[width=0.9\linewidth]{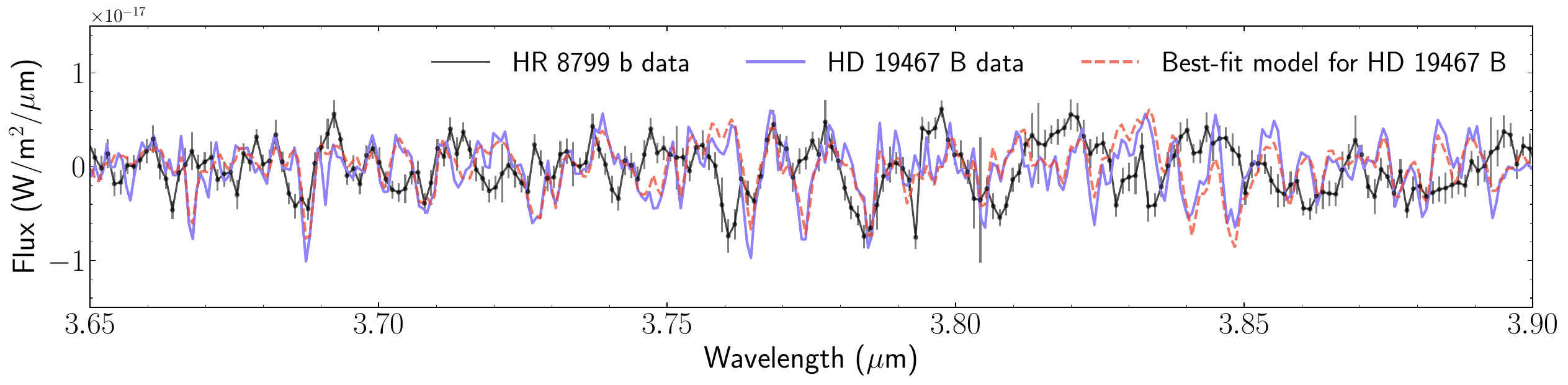}
    \caption{NIRSpec spectra of HR 8799 b and HD 19467 B, a brown dwarf companion with similar $\Teff$. We show a best-fit model for HD 19467 B from the same retrieval framework used for the HR 8799 planets. The brown dwarf's spectra is relatively well matched by the model, and does not show similar residual features as HR 8799 b does.}
    \label{fig:b_vs_hd19467B}
\end{figure*}

Finally, we show in Figure~\ref{fig:b_vs_hd19467B} that the brown dwarf companion, HD 19467 B, which has similar \Teff as HR 8799 b but higher surface gravity ($\logg$ of 5.0 vs 4.0) does not show the same residual features. The data for HD 19467 B \citep{Ruffio2024} were taken using the same instrument mode as the HR 8799 data in this paper. We ran a free retrieval with vertically-constant abundances on the brown dwarf's NIRSpec spectrum using \texttt{petitRADTRANS} and the same analysis framework. The brown dwarf's spectrum in the $3.65-3.9~\mu$m region are relatively well-fit by the model. Most of the lines in the HD 19467 B model are from H$_2$O and CH$_4$; H$_2$S was not detected in the brown dwarf. 

As shown by the emission contribution function (Figure~\ref{fig:pt_emis}), the region of elevated residuals coincides with a region of weaker molecular opacity, and probes deeper atmospheric pressure levels (higher temperatures) compared to the rest of the data. Besides line list issues and unidentified absorbers, an alternative explanation for the elevated residuals between $3.65-3.9~\mu$m could be inaccuracy in the cloud models or the flux continuum level. Depending on the cloud base pressure, cloud opacity impacts various pressure levels in the atmosphere differently. Therefore, it may be possible that the data and model mis-matches we see are caused by insufficient treatment of the clouds. However, in the various cloud models we tried, including two-component cloud models following \citet{Zhang2025}, we did not observe noticeable improvement in the fit quality between $3.65-3.9~\mu$m, which may argue against this explanation. Ultimately, the lack of continuum shape information in the NIRSpec data makes it challenging to fully assess whether clouds might play a role in explaining these unidentified features. While we use archival photometry to anchor the spectral continuum in this work, future work should look into recovering the planet's spectral continuum in NIRSpec using angular differential imaging. 

\clearpage

\section{Potential for Photochemistry}\label{app:photchem_plot}

We investigated whether photochemical products may be responsible for the unidentified spectral features discussed in the previous section. To do so, we used the finite-difference techniques with the \texttt{KINETICS} model \citep{allen81,yung84,moses11} to solve the coupled continuity equations that describe the chemical production, loss, and vertical transport of H-C-N-O-S-Cl species in the planet's atmosphere. The chemical reaction list is presented in \citet{tsai23jwst}. Vertical transport is assumed to occur through diffusive mixing, considering both molecular and ``eddy'' diffusion. The eddy diffusion coefficient $K_{\mathrm{zz}}$ in the deep convective region of the atmosphere is freely adjusted to better fit the observations. At pressures less than 0.3 bar, we assume that $K_{\mathrm{zz}}$ = $1\times10^{5} (0.3~{\rm bar}/P)^{0.5}$ cm$^2$ s$^{-1}$. The initial conditions are set to thermochemical equilibrium assuming the retrieved abundances of the various elements from Table~\ref{table:spec_results} (with Cl assumed at 5x solar). Further details of the modeling procedure can be found in \citet{moses13gj436} and \citep{Moses2016}.

\begin{figure}[t!]
    \centering
    \includegraphics[width=\linewidth]{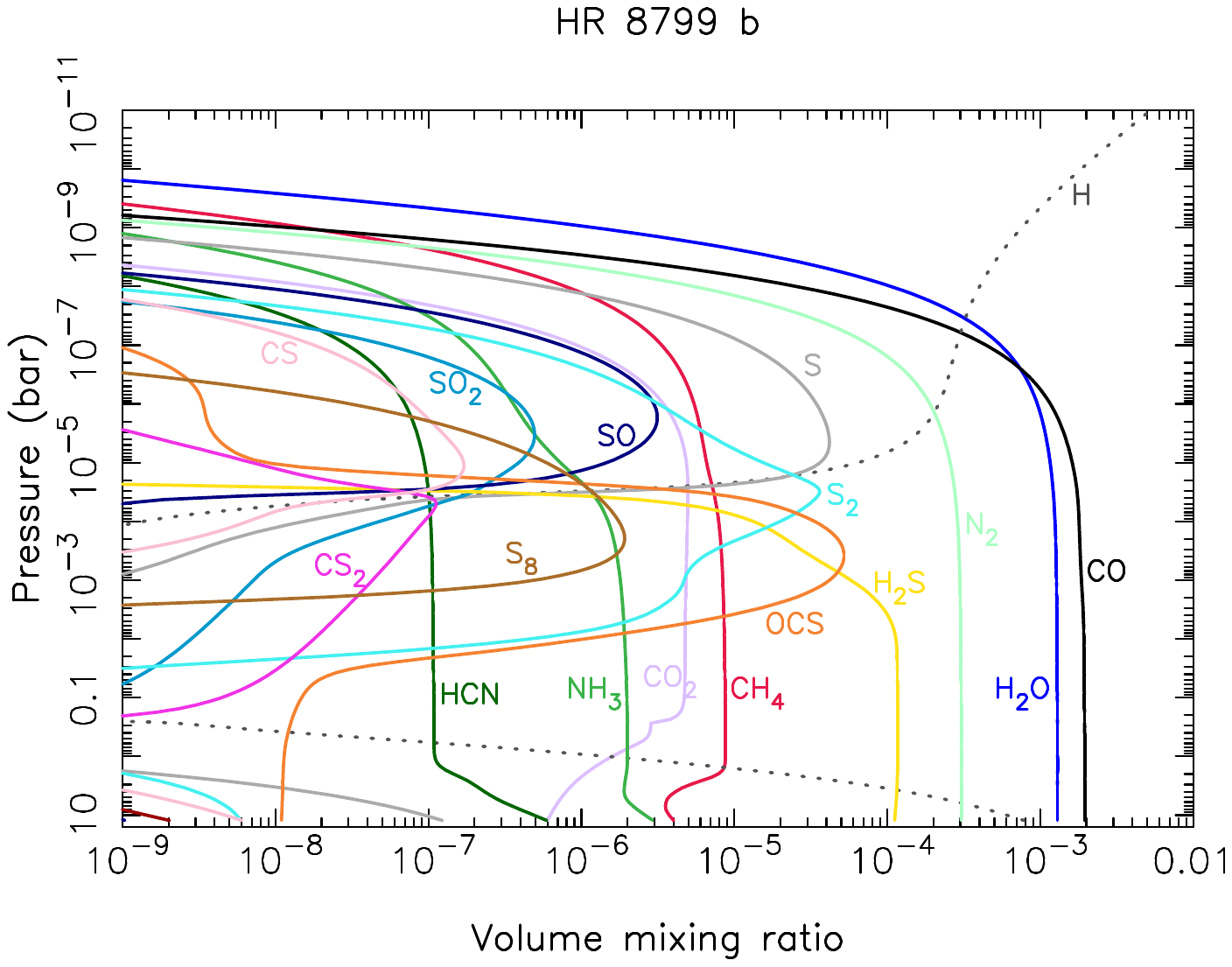}
    \caption{Volume-mixing ratios of various species from the KINETICS photochemical model. Sulfur photochemistry is more important than carbon photochemistry, due to the much higher H$_2$S abundance than CH$_4$ abundance for this planet in its photosphere.}
    \label{fig:photchem}
\end{figure}

The model is an update to that presented in \citet{Moses2016}, which explored photochemistry in HR 8799 b, as we now include sulfur and chlorine chemistry. Because the H-SH bond is relatively weak and the H$_2$S mixing ratio exceeds that of CH$_4$ on HR 8799~b, we find that sulfur photochemistry is more active and important in the radiative region of the atmosphere, despite the planet's large orbital distance ($71$ au). OCS is potentially an important observable photochemical product worth considering in future spectroscopic analyses, although this hinges on uncertainties in OCS kinetics in reducing environments. Other spectroscopically active species worth noting include CS$_2$, CS, SO, and SO$_2$ (see Figure~\ref{fig:photchem}), although these molecules are produced in lesser quantities than OCS. While our modeling suggests that sulfur-based photochemical products may be important in the $\sim$1 mbar region \citep[see also][]{zahnle16}, additional retrievals are needed to investigate whether they can re-produce the features from $3.75-3.95~\mu$m. 

The \texttt{KINETICS} modeling indicates that the assumed $K_{\mathrm{zz}}$ profile with a deep $K_{\mathrm{zz}}$ value of 10$^{8}$ cm$^{2}$ s$^{-1}$ reproduces the retrieved mixing ratios of CH$_4$, CO, H$_2$O, and CO$_2$, again consistent with the derived  $K_{\mathrm{zz}}$ from the retrieval (Section~\ref{sec:quench}). However, as with the \texttt{VULCAN} modeling described in Section~\ref{sec:nitrogen}, both NH$_3$ and HCN are underpredicted by a factor of $\sim$2, suggesting that an N/H ratio of at least 10$\times$ solar would provide a better fit to the data. We note that this conclusion does depends on certain key reaction rate coefficients involved with NH$_3$-N$_2$ quenching \citep[]{moses14}.

\section{Tentative molecular detections and assessing the presence of HDO}\label{app:hdo}

\begin{figure*}[t!]
    \centering
    \includegraphics[width=0.8\linewidth]{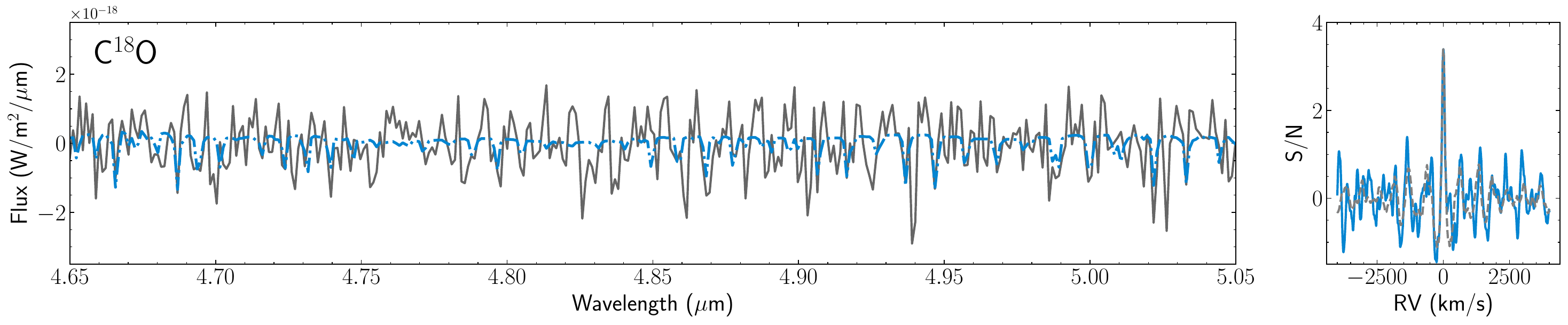}
    \includegraphics[width=0.8\linewidth]{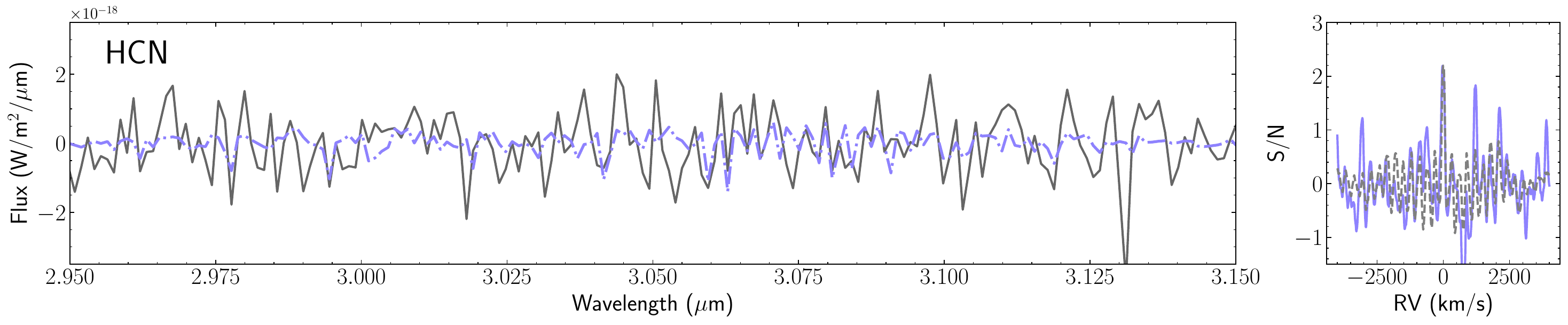}
    \includegraphics[width=0.8\linewidth]{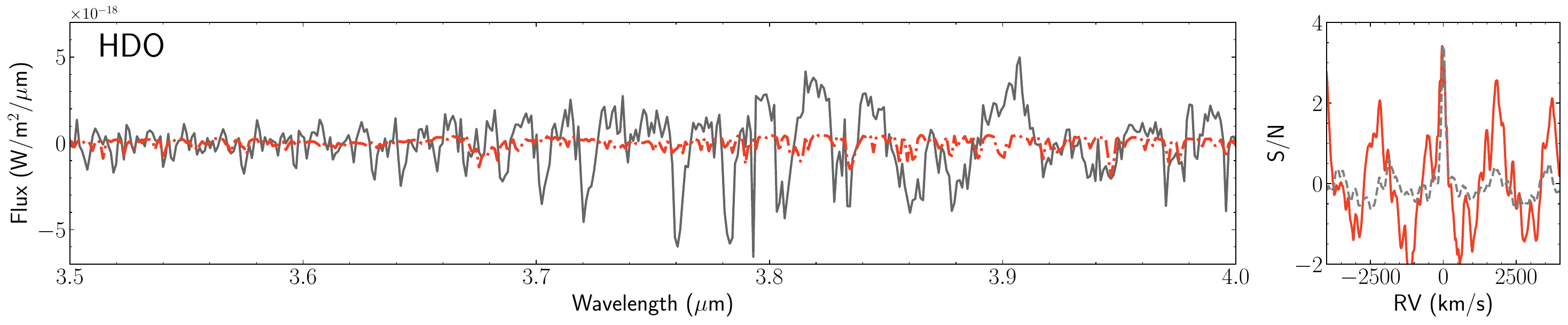}
    \caption{Left: The data residuals from leave-one-out retrievals for C$^{18}$O, HCN, and HDO (in black), and molecular templates for the corresponding species (in color). Right: CCFs between the data residuals and molecular templates, which indicate $2-3\sigma$ detections of these species.}
    \label{fig:hdo_vs_resids}
\end{figure*}

A few species are weakly or tentatively detected in the JWST/NIRSpec data of HR 8799 b, including C$^{18}$O, HCN, and HDO. From our retrievals, the HDO abundance (or H$_2$O/HDO ratio) is consistently well-bounded. However, we do not consider this a detection of HDO, because HDO has its strongest features in the same wavelength region where we have elevated residuals (i.e. $3.5-4.0~\mu$m). As shown in Figure~\ref{fig:hdo_vs_resids}, the strong residual features here are much larger in amplitude than the HDO lines. This causes significant structure in the wings of the HDO CCF (Figure~\ref{fig:hdo_vs_resids}). 

From the baseline retrieval, the mass fraction ratio of H$_2$O/HDO = $4400^{+760}_{-630}$, which implies a D/H ratio of $\approx10^{-4}$. This is about five times higher than the solar D/H value, as well as the value recently measured for an isolated brown dwarf \citep{Rowland2024}. If future analyses are able to fit the large residual features in the vicinity of the HDO lines, it would be useful to re-visit the HDO detection and D/H constraints for HR 8799 b. We note that the CH$_3$D abundance is not constrained by our retrievals, but this is expected since the CH$_4$ mass-mixing ratio is nearly two orders of magnitude lower than the H$_2$O mass-mixing ratio in HR 8799 b. This suggests that for warmer planets like HR 8799 b, HDO may be a useful tracer for deuterium as well \citep{Morley2019}. 

\clearpage

\bibliography{main}{}

@ARTICLE{Piso+16,
       author = {{Piso}, Ana-Maria A. and {Pegues}, Jamila and {{\"O}berg}, Karin I.},
        title = "{The Role of Ice Compositions for Snowlines and the C/N/O Ratios in Active Disks}",
      journal = {\apj},
         year = 2016,
        month = dec,
       volume = {833},
       number = {2},
          eid = {203},
        pages = {203},
          doi = {10.3847/1538-4357/833/2/203},
archivePrefix = {arXiv},
       eprint = {1611.00741},
 primaryClass = {astro-ph.EP},
       adsurl = {https://ui.adsabs.harvard.edu/abs/2016ApJ...833..203P}
}

@ARTICLE{Lissauer1993,
       author = {{Lissauer}, Jack J.},
        title = "{Planet formation.}",
      journal = {\araa},
         year = 1993,
        month = jan,
       volume = {31},
        pages = {129-174},
          doi = {10.1146/annurev.aa.31.090193.001021},
       adsurl = {https://ui.adsabs.harvard.edu/abs/1993ARA&A..31..129L}
}

@ARTICLE{Davis2025,
       author = {{Davis}, C. Evan and {Fortney}, Jonathan J. and {Iyer}, Aishwarya and {Mukherjee}, Sagnick and {Morley}, Caroline V. and {Marley}, Mark S. and {Line}, Michael and {Muirhead}, Philip S.},
        title = "{The Sonora Substellar Atmosphere Models. VI. Red Diamondback: Extending Diamondback with SPHINX for Brown Dwarf Early Evolution}",
      journal = {\apj},
         year = 2025,
        month = dec,
       volume = {994},
       number = {2},
          eid = {198},
        pages = {198},
          doi = {10.3847/1538-4357/ae1015},
       adsurl = {https://ui.adsabs.harvard.edu/abs/2025ApJ...994..198D}
}

@ARTICLE{Morley2024,
       author = {{Morley}, Caroline V. and {Mukherjee}, Sagnick and {Marley}, Mark S. and {Fortney}, Jonathan J. and {Visscher}, Channon and {Lupu}, Roxana and {Gharib-Nezhad}, Ehsan and {Thorngren}, Daniel and {Freedman}, Richard and {Batalha}, Natasha},
        title = "{The Sonora Substellar Atmosphere Models. III. Diamondback: Atmospheric Properties, Spectra, and Evolution for Warm Cloudy Substellar Objects}",
      journal = {\apj},
         year = 2024,
        month = nov,
       volume = {975},
       number = {1},
          eid = {59},
        pages = {59},
          doi = {10.3847/1538-4357/ad71d5},
archivePrefix = {arXiv},
       eprint = {2402.00758},
 primaryClass = {astro-ph.SR},
       adsurl = {https://ui.adsabs.harvard.edu/abs/2024ApJ...975...59M}
}

@article{Fayolle2016,
  title = {N2 and {{CO Desorption Energies}} from {{Water Ice}}},
  author = {Fayolle, Edith C. and Balfe, Jodi and Loomis, Ryan and Bergner, Jennifer and Graninger, Dawn and Rajappan, Mahesh and Öberg, Karin I.},
  year = 2016,
  month = jan,
  date = {2016-01-01},
  journaltitle = {The Astrophysical Journal},
  volume = {816},
  pages = {L28},
  publisher = {IOP},
  issn = {0004-637X},
  doi = {10.3847/2041-8205/816/2/L28},
  url = {https://ui.adsabs.harvard.edu/abs/2016ApJ...816L..28F},
  urldate = {2025-02-11}
}

@ARTICLE{Li+20,
       author = {{Li}, Cheng and {Ingersoll}, Andrew and {Bolton}, Scott and {Levin}, Steven and {Janssen}, Michael and {Atreya}, Sushil and {Lunine}, Jonathan and {Steffes}, Paul and {Brown}, Shannon and {Guillot}, Tristan and {Allison}, Michael and {Arballo}, John and {Bellotti}, Amadeo and {Adumitroaie}, Virgil and {Gulkis}, Samuel and {Hodges}, Amoree and {Li}, Liming and {Misra}, Sidharth and {Orton}, Glenn and {Oyafuso}, Fabiano and {Santos-Costa}, Daniel and {Waite}, Hunter and {Zhang}, Zhimeng},
        title = "{The water abundance in Jupiter's equatorial zone}",
      journal = {Nature Astronomy},
         year = 2020,
        month = feb,
       volume = {4},
        pages = {609-616},
          doi = {10.1038/s41550-020-1009-3},
archivePrefix = {arXiv},
       eprint = {2012.10305},
 primaryClass = {astro-ph.EP},
       adsurl = {https://ui.adsabs.harvard.edu/abs/2020NatAs...4..609L}
}

@ARTICLE{Li+17,
       author = {{Li}, Cheng and {Ingersoll}, Andrew and {Janssen}, Michael and {Levin}, Steven and {Bolton}, Scott and {Adumitroaie}, Virgil and {Allison}, Michael and {Arballo}, John and {Bellotti}, Amadeo and {Brown}, Shannon and {Ewald}, Shawn and {Jewell}, Laura and {Misra}, Sidharth and {Orton}, Glenn and {Oyafuso}, Fabiano and {Steffes}, Paul and {Williamson}, Ross},
        title = "{The distribution of ammonia on Jupiter from a preliminary inversion of Juno microwave radiometer data}",
      journal = {\grl},
         year = 2017,
        month = jun,
       volume = {44},
       number = {11},
        pages = {5317-5325},
          doi = {10.1002/2017GL073159},
       adsurl = {https://ui.adsabs.harvard.edu/abs/2017GeoRL..44.5317L}
}

@ARTICLE{Mousis+19,
       author = {{Mousis}, Olivier and {Ronnet}, Thomas and {Lunine}, Jonathan I.},
        title = "{Jupiter{\textquoteright}s Formation in the Vicinity of the Amorphous Ice Snowline}",
      journal = {\apj},
         year = 2019,
        month = apr,
       volume = {875},
       number = {1},
          eid = {9},
        pages = {9},
          doi = {10.3847/1538-4357/ab0a72},
archivePrefix = {arXiv},
       eprint = {1902.08924},
 primaryClass = {astro-ph.EP},
       adsurl = {https://ui.adsabs.harvard.edu/abs/2019ApJ...875....9M}
}

@ARTICLE{Owen+99,
       author = {{Owen}, Tobias and {Mahaffy}, Paul and {Niemann}, H.~B. and {Atreya}, Sushil and {Donahue}, Thomas and {Bar-Nun}, Akiva and {de Pater}, Imke},
        title = "{A low-temperature origin for the planetesimals that formed Jupiter}",
      journal = {\nat},
         year = 1999,
        month = nov,
       volume = {402},
       number = {6759},
        pages = {269-270},
          doi = {10.1038/46232},
       adsurl = {https://ui.adsabs.harvard.edu/abs/1999Natur.402..269O}
}

@ARTICLE{Aguichine+22,
       author = {{Aguichine}, Artyom and {Mousis}, Olivier and {Lunine}, Jonathan I.},
        title = "{The Possible Formation of Jupiter from Supersolar Gas}",
      journal = {\psj},
         year = 2022,
        month = jun,
       volume = {3},
       number = {6},
          eid = {141},
        pages = {141},
          doi = {10.3847/PSJ/ac6bf1},
archivePrefix = {arXiv},
       eprint = {2204.14102},
 primaryClass = {astro-ph.EP},
       adsurl = {https://ui.adsabs.harvard.edu/abs/2022PSJ.....3..141A}
}

@ARTICLE{Nakazawa&Okuzumi25,
       author = {{Nakazawa}, Kanon and {Okuzumi}, Satoshi},
        title = "{Nitrogen transport in protoplanetary disks by ammonium salts: A possible origin of Jupiter's nitrogen enrichment}",
      journal = {\pasj},
         year = 2025,
        month = jun,
       volume = {77},
       number = {3},
        pages = {539-555},
          doi = {10.1093/pasj/psaf021},
archivePrefix = {arXiv},
       eprint = {2410.05743},
 primaryClass = {astro-ph.EP},
       adsurl = {https://ui.adsabs.harvard.edu/abs/2025PASJ...77..539N}
}

@ARTICLE{OhnoUeda21,
       author = {{Ohno}, Kazumasa and {Ueda}, Takahiro},
        title = "{Jupiter's ``cold'' formation in the protosolar disk shadow. An explanation for the planet's uniformly enriched atmosphere}",
      journal = {\aap},
         year = 2021,
        month = jul,
       volume = {651},
          eid = {L2},
        pages = {L2},
          doi = {10.1051/0004-6361/202141169},
archivePrefix = {arXiv},
       eprint = {2106.09084},
 primaryClass = {astro-ph.EP},
       adsurl = {https://ui.adsabs.harvard.edu/abs/2021A&A...651L...2O}
}

@ARTICLE{Bosman+19,
       author = {{Bosman}, A.~D. and {Cridland}, A.~J. and {Miguel}, Y.},
        title = "{Jupiter formed as a pebble pile around the N$_{2}$ ice line}",
      journal = {\aap},
         year = 2019,
        month = dec,
       volume = {632},
          eid = {L11},
        pages = {L11},
          doi = {10.1051/0004-6361/201936827},
archivePrefix = {arXiv},
       eprint = {1911.11154},
 primaryClass = {astro-ph.EP},
       adsurl = {https://ui.adsabs.harvard.edu/abs/2019A&A...632L..11B}
}

@ARTICLE{Gao2018,
       author = {{Gao}, Peter and {Marley}, Mark S. and {Ackerman}, Andrew S.},
        title = "{Sedimentation Efficiency of Condensation Clouds in Substellar Atmospheres}",
      journal = {\apj},
         year = 2018,
        month = mar,
       volume = {855},
       number = {2},
          eid = {86},
        pages = {86},
          doi = {10.3847/1538-4357/aab0a1},
archivePrefix = {arXiv},
       eprint = {1802.06241},
 primaryClass = {astro-ph.EP},
       adsurl = {https://ui.adsabs.harvard.edu/abs/2018ApJ...855...86G}
}

@ARTICLE{Poch2020,
       author = {{Poch}, Olivier and {Istiqomah}, Istiqomah and {Quirico}, Eric and {Beck}, Pierre and {Schmitt}, Bernard and {Theul{\'e}}, Patrice and {Faure}, Alexandre and {Hily-Blant}, Pierre and {Bonal}, Lydie and {Raponi}, Andrea and {Ciarniello}, Mauro and {Rousseau}, Batiste and {Potin}, Sandra and {Brissaud}, Olivier and {Flandinet}, Laur{\`e}ne and {Filacchione}, Gianrico and {Pommerol}, Antoine and {Thomas}, Nicolas and {Kappel}, David and {Mennella}, Vito and {Moroz}, Lyuba and {Vinogradoff}, Vassilissa and {Arnold}, Gabriele and {Erard}, St{\'e}phane and {Bockel{\'e}e-Morvan}, Dominique and {Leyrat}, C{\'e}dric and {Capaccioni}, Fabrizio and {De Sanctis}, Maria Cristina and {Longobardo}, Andrea and {Mancarella}, Francesca and {Palomba}, Ernesto and {Tosi}, Federico},
        title = "{Ammonium salts are a reservoir of nitrogen on a cometary nucleus and possibly on some asteroids}",
      journal = {Science},
         year = 2020,
        month = mar,
       volume = {367},
       number = {6483},
          eid = {aaw7462},
        pages = {aaw7462},
          doi = {10.1126/science.aaw7462},
archivePrefix = {arXiv},
       eprint = {2003.06034},
 primaryClass = {astro-ph.EP},
       adsurl = {https://ui.adsabs.harvard.edu/abs/2020Sci...367.7462P}
}

@ARTICLE{Altwegg2022,
       author = {{Altwegg}, K. and {Combi}, M. and {Fuselier}, S.~A. and {H{\"a}nni}, N. and {De Keyser}, J. and {Mahjoub}, A. and {M{\"u}ller}, D.~R. and {Pestoni}, B. and {Rubin}, M. and {Wampfler}, S.~F.},
        title = "{Abundant ammonium hydrosulphide embedded in cometary dust grains}",
      journal = {\mnras},
         year = 2022,
        month = nov,
       volume = {516},
       number = {3},
        pages = {3900-3910},
          doi = {10.1093/mnras/stac2440},
archivePrefix = {arXiv},
       eprint = {2208.11396},
 primaryClass = {astro-ph.EP},
       adsurl = {https://ui.adsabs.harvard.edu/abs/2022MNRAS.516.3900A}
}

@ARTICLE{Tsai+21,
       author = {{Tsai}, Shang-Min and {Malik}, Matej and {Kitzmann}, Daniel and {Lyons}, James R. and {Fateev}, Alexander and {Lee}, Elspeth and {Heng}, Kevin},
        title = "{A Comparative Study of Atmospheric Chemistry with VULCAN}",
      journal = {\apj},
         year = 2021,
        month = dec,
       volume = {923},
       number = {2},
          eid = {264},
        pages = {264},
          doi = {10.3847/1538-4357/ac29bc},
archivePrefix = {arXiv},
       eprint = {2108.01790},
 primaryClass = {astro-ph.EP},
       adsurl = {https://ui.adsabs.harvard.edu/abs/2021ApJ...923..264T}
}

@ARTICLE{Painter2025,
       author = {{Painter}, Caleb and {Andrews}, Sean M. and {Chandler}, Claire J. and {Ueda}, Takahiro and {Wilner}, David J. and {Long}, Feng and {Macias}, Enrique and {Carrasco-Gonzalez}, Carlos and {Chung}, Chia-Ying and {Liu}, Hauyu Baobab and {Birnstiel}, Tilman and {Hughes}, A. Meredith},
        title = "{Detailed Microwave Continuum Spectra from Bright Protoplanetary Disks in Taurus}",
      journal = {The Open Journal of Astrophysics},
         year = 2025,
        month = sep,
       volume = {8},
          eid = {134},
        pages = {134},
          doi = {10.33232/001c.144268},
archivePrefix = {arXiv},
       eprint = {2507.21268},
 primaryClass = {astro-ph.SR},
       adsurl = {https://ui.adsabs.harvard.edu/abs/2025OJAp....8E.134P}
}

@ARTICLE{Garufi2025,
       author = {{Garufi}, A. and {Carrasco-Gonz{\'a}lez}, C. and {Mac{\'\i}as}, E. and {Testi}, L. and {Curone}, P. and {Ricci}, L. and {Facchini}, S. and {Long}, F. and {Manara}, C.~F. and {Pascucci}, I. and {Rosotti}, G. and {Zagaria}, F. and {Clarke}, C. and {Herczeg}, G.~J. and {Isella}, A. and {Rota}, A. and {Mauc{\'o}}, K. and {van der Marel}, N. and {Tazzari}, M.},
        title = "{The centimeter emission from planet-forming disks in Taurus}",
      journal = {\aap},
         year = 2025,
        month = feb,
       volume = {694},
          eid = {A290},
        pages = {A290},
          doi = {10.1051/0004-6361/202452496},
archivePrefix = {arXiv},
       eprint = {2501.11686},
 primaryClass = {astro-ph.EP},
       adsurl = {https://ui.adsabs.harvard.edu/abs/2025A&A...694A.290G}
}

@ARTICLE{Chiang2010,
       author = {{Chiang}, E. and {Youdin}, A.~N.},
        title = "{Forming Planetesimals in Solar and Extrasolar Nebulae}",
      journal = {Annual Review of Earth and Planetary Sciences},
         year = 2010,
        month = may,
       volume = {38},
        pages = {493-522},
          doi = {10.1146/annurev-earth-040809-152513},
archivePrefix = {arXiv},
       eprint = {0909.2652},
 primaryClass = {astro-ph.EP},
       adsurl = {https://ui.adsabs.harvard.edu/abs/2010AREPS..38..493C}
}

@ARTICLE{Chiang2013,
       author = {{Chiang}, Eugene and {Laughlin}, Gregory},
        title = "{The minimum-mass extrasolar nebula: in situ formation of close-in super-Earths}",
      journal = {\mnras},
         year = 2013,
        month = jun,
       volume = {431},
       number = {4},
        pages = {3444-3455},
          doi = {10.1093/mnras/stt424},
archivePrefix = {arXiv},
       eprint = {1211.1673},
 primaryClass = {astro-ph.EP},
       adsurl = {https://ui.adsabs.harvard.edu/abs/2013MNRAS.431.3444C}
}

@ARTICLE{Birnstiel2024,
       author = {{Birnstiel}, Tilman},
        title = "{Dust Growth and Evolution in Protoplanetary Disks}",
      journal = {\araa},
         year = 2024,
        month = sep,
       volume = {62},
       number = {1},
        pages = {157-202},
          doi = {10.1146/annurev-astro-071221-052705},
archivePrefix = {arXiv},
       eprint = {2312.13287},
 primaryClass = {astro-ph.EP},
       adsurl = {https://ui.adsabs.harvard.edu/abs/2024ARA&A..62..157B}
}

@ARTICLE{Fabrycky2010,
       author = {{Fabrycky}, Daniel C. and {Murray-Clay}, Ruth A.},
        title = "{Stability of the Directly Imaged Multiplanet System HR 8799: Resonance and Masses}",
      journal = {\apj},
         year = 2010,
        month = feb,
       volume = {710},
       number = {2},
        pages = {1408-1421},
          doi = {10.1088/0004-637X/710/2/1408},
archivePrefix = {arXiv},
       eprint = {0812.0011},
 primaryClass = {astro-ph},
       adsurl = {https://ui.adsabs.harvard.edu/abs/2010ApJ...710.1408F}
}

@INPROCEEDINGS{Ormel2017,
       author = {{Ormel}, Chris W.},
        title = "{The Emerging Paradigm of Pebble Accretion}",
    booktitle = {Formation, Evolution, and Dynamics of Young Solar Systems},
         year = 2017,
       editor = {{Pessah}, Martin and {Gressel}, Oliver},
       series = {Astrophysics and Space Science Library},
       volume = {445},
        month = jan,
        pages = {197},
          doi = {10.1007/978-3-319-60609-5_7},
       adsurl = {https://ui.adsabs.harvard.edu/abs/2017ASSL..445..197O}
}

@ARTICLE{Bitsch2018,
       author = {{Bitsch}, Bertram and {Morbidelli}, Alessandro and {Johansen}, Anders and {Lega}, Elena and {Lambrechts}, Michiel and {Crida}, Aur{\'e}lien},
        title = "{Pebble-isolation mass: Scaling law and implications for the formation of super-Earths and gas giants}",
      journal = {\aap},
         year = 2018,
        month = apr,
       volume = {612},
          eid = {A30},
        pages = {A30},
          doi = {10.1051/0004-6361/201731931},
archivePrefix = {arXiv},
       eprint = {1801.02341},
 primaryClass = {astro-ph.EP},
       adsurl = {https://ui.adsabs.harvard.edu/abs/2018A&A...612A..30B}
}

@ARTICLE{Xin2023,
       author = {{Xin}, Z. and {Espaillat}, C.~C. and {Rilinger}, A.~M. and {Ribas}, {\'A}. and {Mac{\'\i}as}, E.},
        title = "{Measuring the Dust Masses of Protoplanetary Disks in Lupus with ALMA: Evidence That Disks Can Be Optically Thick at 3 mm}",
      journal = {\apj},
         year = 2023,
        month = jan,
       volume = {942},
       number = {1},
          eid = {4},
        pages = {4},
          doi = {10.3847/1538-4357/aca52b},
archivePrefix = {arXiv},
       eprint = {2212.00599},
 primaryClass = {astro-ph.SR},
       adsurl = {https://ui.adsabs.harvard.edu/abs/2023ApJ...942....4X}
}

@ARTICLE{Zhu2019,
       author = {{Zhu}, Zhaohuan and {Zhang}, Shangjia and {Jiang}, Yan-Fei and {Kataoka}, Akimasa and {Birnstiel}, Tilman and {Dullemond}, Cornelis P. and {Andrews}, Sean M. and {Huang}, Jane and {P{\'e}rez}, Laura M. and {Carpenter}, John M. and {Bai}, Xue-Ning and {Wilner}, David J. and {Ricci}, Luca},
        title = "{One Solution to the Mass Budget Problem for Planet Formation: Optically Thick Disks with Dust Scattering}",
      journal = {\apjl},
         year = 2019,
        month = jun,
       volume = {877},
       number = {2},
          eid = {L18},
        pages = {L18},
          doi = {10.3847/2041-8213/ab1f8c},
archivePrefix = {arXiv},
       eprint = {1904.02127},
 primaryClass = {astro-ph.EP},
       adsurl = {https://ui.adsabs.harvard.edu/abs/2019ApJ...877L..18Z}
}

@article{Drazkowska2019,
  title = {Including {{Dust Coagulation}} in {{Hydrodynamic Models}} of {{Protoplanetary Disks}}: {{Dust Evolution}} in the {{Vicinity}} of a {{Jupiter-mass Planet}}},
  shorttitle = {Including {{Dust Coagulation}} in {{Hydrodynamic Models}} of {{Protoplanetary Disks}}},
  author = {Dr{\k a}{\.z}kowska, Joanna and Li, Shengtai and Birnstiel, Til and Stammler, Sebastian M. and Li, Hui},
  year = 2019,
  month = nov,
  journal = {The Astrophysical Journal},
  volume = {885},
  pages = {91},
  publisher = {IOP},
  issn = {0004-637X},
  doi = {10.3847/1538-4357/ab46b7},
  urldate = {2025-03-01}
}

@ARTICLE{Huang2025,
       author = {{Huang}, Pinghui and {Yu}, Fangyuan and {Lee}, Eve J. and {Dong}, Ruobing and {Bai}, Xue-Ning},
        title = "{Leaky Dust Traps in Planet-embedded Protoplanetary Disks}",
      journal = {\apj},
         year = 2025,
        month = jul,
       volume = {988},
       number = {1},
          eid = {94},
        pages = {94},
          doi = {10.3847/1538-4357/addd1f},
archivePrefix = {arXiv},
       eprint = {2503.19026},
 primaryClass = {astro-ph.EP},
       adsurl = {https://ui.adsabs.harvard.edu/abs/2025ApJ...988...94H}
}

@ARTICLE{Stammler2023,
       author = {{Stammler}, Sebastian Markus and {Lichtenberg}, Tim and {Dr{\k{a}}{\.z}kowska}, Joanna and {Birnstiel}, Tilman},
        title = "{Leaky dust traps: How fragmentation impacts dust filtering by planets}",
      journal = {\aap},
         year = 2023,
        month = feb,
       volume = {670},
          eid = {L5},
        pages = {L5},
          doi = {10.1051/0004-6361/202245512},
archivePrefix = {arXiv},
       eprint = {2301.05505},
 primaryClass = {astro-ph.EP},
       adsurl = {https://ui.adsabs.harvard.edu/abs/2023A&A...670L...5S}
}

@ARTICLE{Lee2022,
       author = {{Lee}, Eve J. and {Fuentes}, J.~R. and {Hopkins}, Philip F.},
        title = "{Establishing Dust Rings and Forming Planets within Them}",
      journal = {\apj},
         year = 2022,
        month = oct,
       volume = {937},
       number = {2},
          eid = {95},
        pages = {95},
          doi = {10.3847/1538-4357/ac8cfe},
archivePrefix = {arXiv},
       eprint = {2206.01219},
 primaryClass = {astro-ph.EP},
       adsurl = {https://ui.adsabs.harvard.edu/abs/2022ApJ...937...95L}
}

@ARTICLE{VanClepper2025,
       author = {{Van Clepper}, Eric R. and {Alarc{\'o}n}, Felipe and {Bergin}, Edwin and {Ciesla}, Fred J.},
        title = "{Dust Recycling and Icy Volatile Enhancement (DRIVE): A Novel Method of Volatile Enrichment in Cold Giant Planets}",
      journal = {\apjl},
         year = 2025,
        month = dec,
       volume = {994},
       number = {2},
          eid = {L44},
        pages = {L44},
          doi = {10.3847/2041-8213/ae1d81},
archivePrefix = {arXiv},
       eprint = {2511.07590},
 primaryClass = {astro-ph.EP},
       adsurl = {https://ui.adsabs.harvard.edu/abs/2025ApJ...994L..44V}
}

@ARTICLE{Chachan2025b,
       author = {{Chachan}, Yayaati and {Fortney}, Jonathan J. and {Ohno}, Kazumasa and {Thorngren}, Daniel and {Murray-Clay}, Ruth},
        title = "{Revising the Giant Planet Mass─Metallicity Relation: Deciphering the Formation Sequence of Giant Planets}",
      journal = {\apj},
         year = 2025,
        month = nov,
       volume = {994},
       number = {1},
          eid = {43},
        pages = {43},
          doi = {10.3847/1538-4357/ae0cbf},
archivePrefix = {arXiv},
       eprint = {2509.20428},
 primaryClass = {astro-ph.EP},
       adsurl = {https://ui.adsabs.harvard.edu/abs/2025ApJ...994...43C}
}

@ARTICLE{Krijt2025,
       author = {{Krijt}, Sebastiaan and {Banzatti}, Andrea and {Zhang}, Ke and {Pinilla}, Paola and {Kaeufer}, Till and {Bergin}, Edwin A. and {Salyk}, Colette and {Pontoppidan}, Klaus and {Blake}, Geoffrey A. and {Long}, Feng and {Huang}, Jane and {Colmenares}, Mar{\'\i}a Jos{\'e} and {Williams}, Joe and {Houge}, Adrien and {Narang}, Mayank and {Vioque}, Miguel and {Lambrechts}, Michiel and {Cleeves}, L. Ilsedore and {{\"O}berg}, Karin and {The Jdiscs Collaboration}},
        title = "{Cosmic Cascades: How Disk Substructure Regulates the Flow of Water to Inner Planetary Systems}",
      journal = {\apjl},
         year = 2025,
        month = sep,
       volume = {990},
       number = {2},
          eid = {L72},
        pages = {L72},
          doi = {10.3847/2041-8213/adfbe3},
archivePrefix = {arXiv},
       eprint = {2508.10402},
 primaryClass = {astro-ph.EP},
       adsurl = {https://ui.adsabs.harvard.edu/abs/2025ApJ...990L..72K}
}

@ARTICLE{Ida2016,
       author = {{Ida}, S. and {Guillot}, T. and {Morbidelli}, A.},
        title = "{The radial dependence of pebble accretion rates: A source of diversity in planetary systems. I. Analytical formulation}",
      journal = {\aap},
         year = 2016,
        month = jun,
       volume = {591},
          eid = {A72},
        pages = {A72},
          doi = {10.1051/0004-6361/201628099},
archivePrefix = {arXiv},
       eprint = {1604.01291},
 primaryClass = {astro-ph.EP},
       adsurl = {https://ui.adsabs.harvard.edu/abs/2016A&A...591A..72I}
}

@ARTICLE{Gasman2025,
       author = {{Gasman}, Danny and {Temmink}, Milou and {van Dishoeck}, Ewine F. and {Kurtovic}, Nicolas T. and {Grant}, Sierra L. and {Sellek}, Andrew and {Tabone}, Beno{\^\i}t and {Henning}, Thomas and {Kamp}, Inga and {G{\"u}del}, Manuel and {Barrado}, David and {Caratti o Garatti}, Alessio and {Glauser}, Adrian M. and {Waters}, Laurens B.~F.~M. and {Arabhavi}, Aditya M. and {Jang}, Hyerin and {Kanwar}, Jayatee and {Lienert}, Julia L. and {Perotti}, Giulia and {Schwarz}, Kamber and {Vlasblom}, Marissa},
        title = "{MINDS: The influence of outer dust disc structure on the volatile delivery to the inner disc}",
      journal = {\aap},
         year = 2025,
        month = feb,
       volume = {694},
          eid = {A147},
        pages = {A147},
          doi = {10.1051/0004-6361/202452152},
archivePrefix = {arXiv},
       eprint = {2501.04587},
 primaryClass = {astro-ph.EP},
       adsurl = {https://ui.adsabs.harvard.edu/abs/2025A&A...694A.147G}
}

@ARTICLE{Gwang2025,
       author = {{Wang}, Gavin and {Xuan}, Jerry W. and {Gonz{\'a}lez Picos}, Dar{\'\i}o and {Zhang}, Zhoujian and {Zhang}, Yapeng and {Mawet}, Dimitri and {Hsu}, Chih-Chun and {Wang}, Jason J. and {Blake}, Geoffrey A. and {Ruffio}, Jean-Baptiste and {Horstman}, Katelyn and {Sappey}, Ben and {Xin}, Yinzi and {Finnerty}, Luke and {Echeverri}, Daniel and {Jovanovic}, Nemanja and {Baker}, Ashley and {Bartos}, Randall and {Calvin}, Benjamin and {Cetre}, Sylvain and {Delorme}, Jacques-Robert and {Doppmann}, Gregory W. and {Fitzgerald}, Michael P. and {Liberman}, Joshua and {L{\'o}pez}, Ronald A. and {Morris}, Evan and {Pezzato-Rovner}, Jacklyn and {Phillips}, Caprice L. and {Schofield}, Tobias and {Skemer}, Andrew and {Wallace}, J. Kent and {Wang}, Ji},
        title = "{Chemical and Isotopic Homogeneity between the L Dwarf CD-35 2722 B and Its Early M Host Star}",
      journal = {\apj},
         year = 2026,
        month = feb,
       volume = {997},
       number = {2},
          eid = {195},
        pages = {195},
          doi = {10.3847/1538-4357/ae232f},
archivePrefix = {arXiv},
       eprint = {2511.19588},
 primaryClass = {astro-ph.EP},
       adsurl = {https://ui.adsabs.harvard.edu/abs/2026ApJ...997..195W}
}

@ARTICLE{Lothringer2025,
       author = {{Lothringer}, Joshua D. and {Bennett}, Katherine A. and {Sing}, David K. and {Kehoe-Seamons}, Brian and {Rustamkulov}, Zafar and {Reggiani}, Henrique and {Schlaufman}, Kevin C. and {McCreery}, Patrick and {Norris}, Seti and {Hauschildt}, Peter and {Cacho-Negrete}, Ceiligh and {Gressier}, Am{\'e}lie and {Espinoza}, N{\'e}stor and {Gapp}, Cyril and {Evans-Soma}, Thomas M. and {Stevenson}, Kevin B. and {Wakeford}, Hannah and {Gibson}, Neale and {Wilson}, Jamie and {Nikolov}, Nikolay},
        title = "{Refractory and Volatile Species in the UV-to-IR Transmission Spectrum of Ultra-hot Jupiter WASP-178b with HST and JWST}",
      journal = {\aj},
         year = 2025,
        month = may,
       volume = {169},
       number = {5},
          eid = {274},
        pages = {274},
          doi = {10.3847/1538-3881/adc117},
archivePrefix = {arXiv},
       eprint = {2503.15472},
 primaryClass = {astro-ph.EP},
       adsurl = {https://ui.adsabs.harvard.edu/abs/2025AJ....169..274L}
}

@ARTICLE{Smith2024,
       author = {{Smith}, Peter C.~B. and {Sanchez}, Jorge A. and {Line}, Michael R. and {Rauscher}, Emily and {Mansfield}, Megan Weiner and {Kempton}, Eliza M.-R. and {Savel}, Arjun and {Wardenier}, Joost P. and {Pino}, Lorenzo and {Bean}, Jacob L. and {Beltz}, Hayley and {Panwar}, Vatsal and {Brogi}, Matteo and {Malsky}, Isaac and {Fortney}, Jonathan and {D{\'e}sert}, Jean-Michel and {Pelletier}, Stefan and {Parmentier}, Vivien and {Kanumalla}, Sai Krishna Teja and {Welbanks}, Luis and {Meyer}, Michael and {Monnier}, John},
        title = "{The Roasting Marshmallows Program with IGRINS on Gemini South. II. WASP-121 b has Superstellar C/O and Refractory-to-volatile Ratios}",
      journal = {\aj},
         year = 2024,
        month = dec,
       volume = {168},
       number = {6},
          eid = {293},
        pages = {293},
          doi = {10.3847/1538-3881/ad8574},
archivePrefix = {arXiv},
       eprint = {2410.19017},
 primaryClass = {astro-ph.EP},
       adsurl = {https://ui.adsabs.harvard.edu/abs/2024AJ....168..293S}
}

@ARTICLE{Evans-Soma2025,
       author = {{Evans-Soma}, Thomas M. and {Sing}, David K. and {Barstow}, Joanna K. and {Piette}, Anjali A.~A. and {Taylor}, Jake and {Lothringer}, Joshua D. and {Reggiani}, Henrique and {Goyal}, Jayesh M. and {Ahrer}, Eva-Maria and {Mayne}, Nathan J. and {Rustamkulov}, Zafar and {Kataria}, Tiffany and {Christie}, Duncan A. and {Gapp}, Cyril and {Dong}, Jiayin and {Foreman-Mackey}, Daniel and {Hattori}, Soichiro and {Marley}, Mark S.},
        title = "{SiO and a super-stellar C/O ratio in the atmosphere of the giant exoplanet WASP-121 b}",
      journal = {Nature Astronomy},
         year = 2025,
        month = jun,
       volume = {9},
        pages = {845-861},
          doi = {10.1038/s41550-025-02513-x},
archivePrefix = {arXiv},
       eprint = {2506.01771},
 primaryClass = {astro-ph.EP},
       adsurl = {https://ui.adsabs.harvard.edu/abs/2025NatAs...9..845E}
}

@ARTICLE{Ohno2025,
       author = {{Ohno}, Kazumasa and {Ikoma}, Masahiro and {Okuzumi}, Satoshi and {Kimura}, Tadahiro},
        title = "{A dichotomy of the mass-metallicity relation of exoplanetary atmospheres demarcated by their birthplace}",
      journal = {\pasj},
         year = 2026,
        month = feb,
          doi = {10.1093/pasj/psaf157},
archivePrefix = {arXiv},
       eprint = {2506.16060},
 primaryClass = {astro-ph.EP},
       adsurl = {https://ui.adsabs.harvard.edu/abs/2026PASJ..tmp...16O}
}

@ARTICLE{Booth2019,
       author = {{Booth}, R.~A. and {Ilee}, J.~D.},
        title = "{Planet-forming material in a protoplanetary disc: the interplay between chemical evolution and pebble drift}",
      journal = {\mnras},
         year = 2019,
        month = aug,
       volume = {487},
       number = {3},
        pages = {3998-4011},
          doi = {10.1093/mnras/stz1488},
archivePrefix = {arXiv},
       eprint = {1905.12639},
 primaryClass = {astro-ph.EP},
       adsurl = {https://ui.adsabs.harvard.edu/abs/2019MNRAS.487.3998B}
}

@ARTICLE{Zhu2012,
       author = {{Zhu}, Zhaohuan and {Nelson}, Richard P. and {Dong}, Ruobing and {Espaillat}, Catherine and {Hartmann}, Lee},
        title = "{Dust Filtration by Planet-induced Gap Edges: Implications for Transitional Disks}",
      journal = {\apj},
         year = 2012,
        month = aug,
       volume = {755},
       number = {1},
          eid = {6},
        pages = {6},
          doi = {10.1088/0004-637X/755/1/6},
archivePrefix = {arXiv},
       eprint = {1205.5042},
 primaryClass = {astro-ph.SR},
       adsurl = {https://ui.adsabs.harvard.edu/abs/2012ApJ...755....6Z}
}

@ARTICLE{Xu2022,
       author = {{Xu}, Wenrui},
        title = "{Testing a New Model of Embedded Protostellar Disks against Observations: The Majority of Orion Class 0/I Disks Are Likely Warm, Massive, and Gravitationally Unstable}",
      journal = {\apj},
         year = 2022,
        month = aug,
       volume = {934},
       number = {2},
          eid = {156},
        pages = {156},
          doi = {10.3847/1538-4357/ac7b94},
archivePrefix = {arXiv},
       eprint = {2203.00941},
 primaryClass = {astro-ph.SR},
       adsurl = {https://ui.adsabs.harvard.edu/abs/2022ApJ...934..156X}
}

@ARTICLE{Chachan2025,
       author = {{Chachan}, Yayaati and {Lothringer}, Joshua and {Inglis}, Julie and {Beltz}, Hayley and {Knutson}, Heather A. and {Spake}, Jessica and {Benneke}, Bjorn and {Wong}, Ian and {Rustamkulov}, Zafar and {Sing}, David and {Bennett}, Katherine A.},
        title = "{Strong NUV Refractory Absorption and Dissociated Water in the Hubble Transmission Spectrum of the Ultra Hot Jupiter KELT-20 b}",
      journal = {\aj},
         year = 2025,
        month = oct,
       volume = {170},
       number = {4},
          eid = {234},
        pages = {234},
          doi = {10.3847/1538-3881/adfbeb},
archivePrefix = {arXiv},
       eprint = {2508.10092},
 primaryClass = {astro-ph.EP},
       adsurl = {https://ui.adsabs.harvard.edu/abs/2025AJ....170..234C}
}

@ARTICLE{Finnerty2025,
       author = {{Finnerty}, Luke and {Xin}, Yinzi and {Xuan}, Jerry W. and {Inglis}, Julie and {Fitzgerald}, Michael P. and {Agrawal}, Shubh and {Baker}, Ashley and {Bartos}, Randall and {Blake}, Geoffrey A. and {Calvin}, Benjamin and {Cetre}, Sylvain and {Delorme}, Jacques-Robert and {Doppmann}, Greg and {Echeverri}, Daniel and {Horstman}, Katelyn and {Hsu}, Chih-Chun and {Jovanovic}, Nemanja and {Liberman}, Joshua and {L{\'o}pez}, Ronald A. and {Martin}, Emily C. and {Mawet}, Dimitri and {Morris}, Evan and {Pezzato}, Jacklyn and {Ruffio}, Jean-Baptiste and {Sappey}, Ben and {Schofield}, Tobias and {Skemer}, Andrew and {Venenciano}, Taylor and {Wallace}, J. Kent and {Wallack}, Nicole L. and {Wang}, Jason J. and {Wang}, Ji},
        title = "{Water Dissociation and Rotational Broadening in the Atmosphere of KELT-20 b from High-resolution Spectroscopy}",
      journal = {\aj},
         year = 2025,
        month = jun,
       volume = {169},
       number = {6},
          eid = {333},
        pages = {333},
          doi = {10.3847/1538-3881/adce02},
archivePrefix = {arXiv},
       eprint = {2503.01946},
 primaryClass = {astro-ph.EP},
       adsurl = {https://ui.adsabs.harvard.edu/abs/2025AJ....169..333F}
}

@ARTICLE{Pelletier2025_jwst,
       author = {{Pelletier}, S. and {Coulombe}, L.-P. and {Splinter}, J. and {Benneke}, B. and {MacDonald}, R.~J. and {Lafreni{\`e}re}, D. and {Cowan}, N.~B. and {Allart}, R. and {Rauscher}, E. and {Frazier}, R.~C. and {Meyer}, M.~R. and {Albert}, L. and {Dang}, L. and {Doyon}, R. and {Ehrenreich}, D. and {Flagg}, L. and {Johnstone}, D. and {Langeveld}, A.~B. and {Lim}, O. and {Piaulet-Ghorayeb}, C. and {Radica}, M. and {Rowe}, J. and {Taylor}, J. and {Turner}, J.~D.},
        title = "{Enriched volatiles and refractories but deficient titanium on the day-side atmosphere of WASP-121b revealed by JWST/NIRISS}",
      journal = {\aap},
         year = 2026,
        month = jan,
       volume = {706},
          eid = {A2},
        pages = {A2},
          doi = {10.1051/0004-6361/202556985},
archivePrefix = {arXiv},
       eprint = {2508.18341},
 primaryClass = {astro-ph.EP},
       adsurl = {https://ui.adsabs.harvard.edu/abs/2026A&A...706A...2P}
}

@ARTICLE{Pelletier2025_highres,
       author = {{Pelletier}, Stefan and {Benneke}, Bj{\"o}rn and {Chachan}, Yayaati and {Bazinet}, Luc and {Allart}, Romain and {Hoeijmakers}, H. Jens and {Lavail}, Alexis and {Prinoth}, Bibiana and {Coulombe}, Louis-Philippe and {Lothringer}, Joshua D. and {Parmentier}, Vivien and {Smith}, Peter and {Borsato}, Nicholas and {Thorsbro}, Brian},
        title = "{CRIRES$^{+}$ and ESPRESSO Reveal an Atmosphere Enriched in Volatiles Relative to Refractories on the Ultrahot Jupiter WASP-121b}",
      journal = {\aj},
         year = 2025,
        month = jan,
       volume = {169},
       number = {1},
          eid = {10},
        pages = {10},
          doi = {10.3847/1538-3881/ad8b28},
archivePrefix = {arXiv},
       eprint = {2410.18183},
 primaryClass = {astro-ph.EP},
       adsurl = {https://ui.adsabs.harvard.edu/abs/2025AJ....169...10P}
}

@ARTICLE{Gandhi2023_uhj,
       author = {{Gandhi}, Siddharth and {Kesseli}, Aurora and {Zhang}, Yapeng and {Louca}, Amy and {Snellen}, Ignas and {Brogi}, Matteo and {Miguel}, Yamila and {Casasayas-Barris}, N{\'u}ria and {Pelletier}, Stefan and {Landman}, Rico and {Maguire}, Cathal and {Gibson}, Neale P.},
        title = "{Retrieval Survey of Metals in Six Ultrahot Jupiters: Trends in Chemistry, Rain-out, Ionization, and Atmospheric Dynamics}",
      journal = {\aj},
         year = 2023,
        month = jun,
       volume = {165},
       number = {6},
          eid = {242},
        pages = {242},
          doi = {10.3847/1538-3881/accd65},
archivePrefix = {arXiv},
       eprint = {2305.17228},
 primaryClass = {astro-ph.EP},
       adsurl = {https://ui.adsabs.harvard.edu/abs/2023AJ....165..242G}
}

@ARTICLE{Tsai+17,
       author = {{Tsai}, Shang-Min and {Lyons}, James R. and {Grosheintz}, Luc and {Rimmer}, Paul B. and {Kitzmann}, Daniel and {Heng}, Kevin},
        title = "{VULCAN: An Open-source, Validated Chemical Kinetics Python Code for Exoplanetary Atmospheres}",
      journal = {\apjs},
         year = 2017,
        month = feb,
       volume = {228},
       number = {2},
          eid = {20},
        pages = {20},
          doi = {10.3847/1538-4365/228/2/20},
archivePrefix = {arXiv},
       eprint = {1607.00409},
 primaryClass = {astro-ph.EP},
       adsurl = {https://ui.adsabs.harvard.edu/abs/2017ApJS..228...20T}
}

@ARTICLE{Fu2024,
       author = {{Fu}, Guangwei and {Welbanks}, Luis and {Deming}, Drake and {Inglis}, Julie and {Zhang}, Michael and {Lothringer}, Joshua and {Ih}, Jegug and {Moses}, Julianne I. and {Schlawin}, Everett and {Knutson}, Heather A. and {Henry}, Gregory and {Greene}, Thomas and {Sing}, David K. and {Savel}, Arjun B. and {Kempton}, Eliza M. -R. and {Louie}, Dana R. and {Line}, Michael and {Nixon}, Matt},
        title = "{Hydrogen sulfide and metal-enriched atmosphere for a Jupiter-mass exoplanet}",
      journal = {Nature},
         year = 2024,
        month = aug,
       volume = {632},
       number = {8026},
        pages = {752-756},
          doi = {10.1038/s41586-024-07760-y},
archivePrefix = {arXiv},
       eprint = {2407.06163},
 primaryClass = {astro-ph.EP},
       adsurl = {https://ui.adsabs.harvard.edu/abs/2024Natur.632..752F}
}

@ARTICLE{Luck2017,
       author = {{Luck}, R. Earle},
        title = "{Abundances in the Local Region II: F, G, and K Dwarfs and Subgiants}",
      journal = {\aj},
         year = 2017,
        month = jan,
       volume = {153},
       number = {1},
          eid = {21},
        pages = {21},
          doi = {10.3847/1538-3881/153/1/21},
archivePrefix = {arXiv},
       eprint = {1611.02897},
 primaryClass = {astro-ph.SR},
       adsurl = {https://ui.adsabs.harvard.edu/abs/2017AJ....153...21L}
}

@ARTICLE{Kolecki2022,
       author = {{Kolecki}, Jared R. and {Wang}, Ji},
        title = "{Measuring Elemental Abundances of JWST Target Stars for Exoplanet Characterization. I. FGK Stars}",
      journal = {\aj},
         year = 2022,
        month = sep,
       volume = {164},
       number = {3},
          eid = {87},
        pages = {87},
          doi = {10.3847/1538-3881/ac7de3},
archivePrefix = {arXiv},
       eprint = {2112.02031},
 primaryClass = {astro-ph.EP},
       adsurl = {https://ui.adsabs.harvard.edu/abs/2022AJ....164...87K}
}

@ARTICLE{Thompson2023,
       author = {{Thompson}, William and {Marois}, Christian and {Do {\'O}}, Clarissa R. and {Konopacky}, Quinn and {Ruffio}, Jean-Baptiste and {Wang}, Jason and {Skemer}, Andy J. and {De Rosa}, Robert J. and {Macintosh}, Bruce},
        title = "{Deep Orbital Search for Additional Planets in the HR 8799 System}",
      journal = {\aj},
         year = 2023,
        month = jan,
       volume = {165},
       number = {1},
          eid = {29},
        pages = {29},
          doi = {10.3847/1538-3881/aca1af},
archivePrefix = {arXiv},
       eprint = {2210.14213},
 primaryClass = {astro-ph.EP},
       adsurl = {https://ui.adsabs.harvard.edu/abs/2023AJ....165...29T}
}

@ARTICLE{Lothringer2021,
       author = {{Lothringer}, Joshua D. and {Rustamkulov}, Zafar and {Sing}, David K. and {Gibson}, Neale P. and {Wilson}, Jamie and {Schlaufman}, Kevin C.},
        title = "{A New Window into Planet Formation and Migration: Refractory-to-Volatile Elemental Ratios in Ultra-hot Jupiters}",
      journal = {Astrophys. J.},
         year = 2021,
        month = jun,
       volume = {914},
       number = {1},
          eid = {12},
        pages = {12},
          doi = {10.3847/1538-4357/abf8a9},
archivePrefix = {arXiv},
       eprint = {2011.10626},
 primaryClass = {astro-ph.EP},
       adsurl = {https://ui.adsabs.harvard.edu/abs/2021ApJ...914...12L}
}

@ARTICLE{Ohno2023b,
       author = {{Ohno}, Kazumasa and {Fortney}, Jonathan J.},
        title = "{Nitrogen as a Tracer of Giant Planet Formation. II. Comprehensive Study of Nitrogen Photochemistry and Implications for Observing NH$_{3}$ and HCN in Transmission and Emission Spectra}",
      journal = {\apj},
         year = 2023,
        month = oct,
       volume = {956},
       number = {2},
          eid = {125},
        pages = {125},
          doi = {10.3847/1538-4357/ace531},
archivePrefix = {arXiv},
       eprint = {2211.16877},
 primaryClass = {astro-ph.EP},
       adsurl = {https://ui.adsabs.harvard.edu/abs/2023ApJ...956..125O}
}

@ARTICLE{Beiler2024b,
       author = {{Beiler}, Samuel A. and {Mukherjee}, Sagnick and {Cushing}, Michael C. and {Kirkpatrick}, J. Davy and {Schneider}, Adam C. and {Kothari}, Harshil and {Marley}, Mark S. and {Visscher}, Channon},
        title = "{A Tale of Two Molecules: The Underprediction of CO$_{2}$ and Overprediction of PH$_{3}$ in Late T and Y Dwarf Atmospheric Models}",
      journal = {\apj},
         year = 2024,
        month = sep,
       volume = {973},
       number = {1},
          eid = {60},
        pages = {60},
          doi = {10.3847/1538-4357/ad6759},
archivePrefix = {arXiv},
       eprint = {2407.15950},
 primaryClass = {astro-ph.EP},
       adsurl = {https://ui.adsabs.harvard.edu/abs/2024ApJ...973...60B}
}

@ARTICLE{Hsu2024_pds,
       author = {{Hsu}, Chih-Chun and {Wang}, Jason J. and {Blake}, Geoffrey A. and {Xuan}, Jerry W. and {Zhang}, Yapeng and {Ruffio}, Jean-Baptiste and {Horstman}, Katelyn and {Cronin}, Julianne and {Sappey}, Ben and {Xin}, Yinzi and {Finnerty}, Luke and {Echeverri}, Daniel and {Mawet}, Dimitri and {Jovanovic}, Nemanja and {Do {\'O}}, Clarissa R. and {Baker}, Ashley and {Bartos}, Randall and {Calvin}, Benjamin and {Cetre}, Sylvain and {Delorme}, Jacques-Robert and {Doppmann}, Gregory W. and {Fitzgerald}, Michael P. and {Liberman}, Joshua and {L{\'o}pez}, Ronald A. and {Morris}, Evan and {Pezzato-Rovner}, Jacklyn and {Schofield}, Tobias and {Skemer}, Andrew and {Wallace}, J. Kent and {Wang}, Ji},
        title = "{PDS 70b Shows Stellar-like Carbon-to-oxygen Ratio}",
      journal = {\apjl},
         year = 2024,
        month = dec,
       volume = {977},
       number = {2},
          eid = {L47},
        pages = {L47},
          doi = {10.3847/2041-8213/ad95e8},
archivePrefix = {arXiv},
       eprint = {2411.15117},
 primaryClass = {astro-ph.EP},
       adsurl = {https://ui.adsabs.harvard.edu/abs/2024ApJ...977L..47H}
}

@article{Xuan2022,
   author = {Jerry W Xuan and Jason Wang and Jean-Baptiste Ruffio and Heather Knutson and Dimitri Mawet and Paul Mollière and Jared Kolecki and Arthur Vigan and Sagnick Mukherjee and Nicole Wallack and Ji Wang and Ashley Baker and Randall Bartos and Geoffrey A Blake and Charlotte Z Bond and Marta Bryan and Benjamin Calvin and Sylvain Cetre and Mark Chun and Jacques-Robert Delorme and Greg Doppmann and Daniel Echeverri and Luke Finnerty and Michael P Fitzgerald and Katelyn Horstman and Julie Inglis and Nemanja Jovanovic and Ronald López and Emily C Martin and Evan Morris and Jacklyn Pezzato and Sam Ragland and Bin Ren and Garreth Ruane and Ben Sappey and Tobias Schofield and Andrew Skemer and Taylor Venenciano and J Kent Wallace and Peter Wizinowich},
   doi = {10.3847/1538-4357/ac8673},
   issn = {0004-637X},
   issue = {2},
   journal = {The Astrophysical Journal},
   pages = {54},
   title = {A Clear View of a Cloudy Brown Dwarf Companion from High-resolution Spectroscopy},
   volume = {937},
   url = {https://iopscience.iop.org/article/10.3847/1538-4357/ac8673/meta https://iopscience.iop.org/article/10.3847/1538-4357/ac8673/pdf https://iopscience.iop.org/article/10.3847/1538-4357/ac8673},
   year = {2022},
}

@article{Wang2023,
   author = {Ji Wang and Jason J Wang and Jean-Baptiste Ruffio and Geoffrey A Blake and Dimitri Mawet and Ashley Baker and Randall Bartos and Charlotte Z Bond and Benjamin Calvin and Sylvain Cetre and Jacques-Robert Delorme and Greg Doppmann and Daniel Echeverri and Luke Finnerty and Michael P Fitzgerald and Nemanja Jovanovic and Ronald Lopez and Emily C Martin and Evan Morris and Jacklyn Pezzato and Sam Ragland and Garreth Ruane and Ben Sappey and Tobias Schofield and Andrew Skemer and Taylor Venenciano and J Kent Wallace and Peter Wizinowich and Jerry W Xuan and Marta L Bryan and Arpita Roy and Nicole L Wallack},
   doi = {10.3847/1538-3881/ac9f19},
   issn = {0004-6256},
   journal = {The Astronomical Journal},
   pages = {4},
   title = {Retrieving C and O Abundance of HR 8799 c by Combining High- and Low-resolution Data},
   volume = {165},
   url = {https://ui.adsabs.harvard.edu/abs/2023AJ....165....4W https://ui.adsabs.harvard.edu/link_gateway/2023AJ....165....4W/ARTICLE},
   year = {2023},
}

@article{wang_dynamical_2018,
	title = {Dynamical {Constraints} on the {HR} 8799 {Planets} with {GPI}},
	volume = {156},
	issn = {0004-6256},
	url = {http://adsabs.harvard.edu/abs/2018AJ....156..192W},
	doi = {10.3847/1538-3881/aae150},
	urldate = {2019-04-11},
	journal = {\aj},
	author = {Wang, Jason J. and Graham, James R. and Dawson, Rebekah and Fabrycky, Daniel and De Rosa, Robert J. and Pueyo, Laurent and Konopacky, Quinn and Macintosh, Bruce and Marois, Christian and Chiang, Eugene and Ammons, S. Mark and Arriaga, Pauline and Bailey, Vanessa P. and Barman, Travis and Bulger, Joanna and Chilcote, Jeffrey and Cotten, Tara and Doyon, Rene and Duchene, Gaspard and Esposito, Thomas M. and Fitzgerald, Michael P. and Follette, Katherine B. and Gerard, Benjamin L. and Goodsell, Stephen J. and Greenbaum, Alexandra Z. and Hibon, Pascale and Hung, Li-Wei and Ingraham, Patrick and Kalas, Paul and Larkin, James E. and Maire, Jerome and Marchis, Franck and Marley, Mark S. and Metchev, Stanimir and Millar-Blanchaer, Maxwell A. and Nielsen, Eric L. and Oppenheimer, Rebecca and Palmer, David and Patience, Jennifer and Perrin, Marshall and Poyneer, Lisa and Rajan, Abhijith and Rameau, Julien and Rantakyro, Fredrik T. and Ruffio, Jean-Baptiste and Savransky, Dmitry and Schneider, Adam C. and Sivaramakrishnan, Anand and Song, Inseok and Soummer, Remi and Thomas, Sandrine and Wallace, J. Kent and Ward-Duong, Kimberly and Wiktorowicz, Sloane and Wolff, Schuyler},
	month = nov,
	year = {2018},
	pages = {192}
}

@article{konopacky_detection_2013,
	title = {Detection of {Carbon} {Monoxide} and {Water} {Absorption} {Lines} in an {Exoplanet} {Atmosphere}},
	volume = {339},
	issn = {0036-8075, 1095-9203},
	url = {http://science.sciencemag.org/content/339/6126/1398},
	doi = {10.1126/science.1232003},
	language = {en},
	number = {6126},
	urldate = {2019-04-05},
	journal = {Science},
	author = {Konopacky, Quinn M. and Barman, Travis S. and Macintosh, Bruce A. and Marois, Christian},
	month = mar,
	year = {2013},
	pmid = {23493423},
	pages = {1398--1401},
}

@ARTICLE{Kama2019,
       author = {{Kama}, Mihkel and {Shorttle}, Oliver and {Jermyn}, Adam S. and {Folsom}, Colin P. and {Furuya}, Kenji and {Bergin}, Edwin A. and {Walsh}, Catherine and {Keller}, Lindsay},
        title = "{Abundant Refractory Sulfur in Protoplanetary Disks}",
      journal = {Astrophys. J.},
         year = 2019,
        month = nov,
       volume = {885},
       number = {2},
          eid = {114},
        pages = {114},
          doi = {10.3847/1538-4357/ab45f8},
archivePrefix = {arXiv},
       eprint = {1908.05169},
 primaryClass = {astro-ph.EP},
       adsurl = {https://ui.adsabs.harvard.edu/abs/2019ApJ...885..114K}
}

@ARTICLE{Pacetti2022,
       author = {{Pacetti}, Elenia and {Turrini}, Diego and {Schisano}, Eugenio and {Molinari}, Sergio and {Fonte}, Sergio and {Politi}, Romolo and {Hennebelle}, Patrick and {Klessen}, Ralf and {Testi}, Leonardo and {Lebreuilly}, Ugo},
        title = "{Chemical Diversity in Protoplanetary Disks and Its Impact on the Formation History of Giant Planets}",
      journal = {Astrophys. J.},
         year = 2022,
        month = sep,
       volume = {937},
       number = {1},
          eid = {36},
        pages = {36},
          doi = {10.3847/1538-4357/ac8b11},
archivePrefix = {arXiv},
       eprint = {2206.14685},
 primaryClass = {astro-ph.EP},
       adsurl = {https://ui.adsabs.harvard.edu/abs/2022ApJ...937...36P}
}

@ARTICLE{Bowler2010,
       author = {{Bowler}, Brendan P. and {Liu}, Michael C. and {Dupuy}, Trent J. and {Cushing}, Michael C.},
        title = "{Near-infrared Spectroscopy of the Extrasolar Planet HR 8799 b}",
      journal = {\apj},
         year = 2010,
        month = nov,
       volume = {723},
       number = {1},
        pages = {850-868},
          doi = {10.1088/0004-637X/723/1/850},
archivePrefix = {arXiv},
       eprint = {1008.4582},
 primaryClass = {astro-ph.EP},
       adsurl = {https://ui.adsabs.harvard.edu/abs/2010ApJ...723..850B}
}

@ARTICLE{Janson2010,
       author = {{Janson}, M. and {Bergfors}, C. and {Goto}, M. and {Brandner}, W. and {Lafreni{\`e}re}, D.},
        title = "{Spatially Resolved Spectroscopy of the Exoplanet HR 8799 c}",
      journal = {\apjl},
         year = 2010,
        month = feb,
       volume = {710},
       number = {1},
        pages = {L35-L38},
          doi = {10.1088/2041-8205/710/1/L35},
archivePrefix = {arXiv},
       eprint = {1001.2017},
 primaryClass = {astro-ph.EP},
       adsurl = {https://ui.adsabs.harvard.edu/abs/2010ApJ...710L..35J}
}

@article{barman2015,
	title = {{SIMULTANEOUS} {DETECTION} {OF} {WATER}, {METHANE}, {AND} {CARBON} {MONOXIDE} {IN} {THE} {ATMOSPHERE} {OF} {EXOPLANET} {HR} 8799 b},
	volume = {804},
	issn = {0004-637X},
	url = {https://doi.org/10.1088%2F0004-637x%2F804%2F1%2F61},
	doi = {10.1088/0004-637X/804/1/61},
	language = {en},
	number = {1},
	urldate = {2019-04-05},
	journal = {\apj},
	author = {Barman, Travis S. and Konopacky, Quinn M. and Macintosh, Bruce and Marois, Christian},
	month = may,
	year = {2015},
	pages = {61},
}

@article{oberg_effects_2011,
	title = {The {Effects} of {Snowlines} on {C}/{O} in {Planetary} {Atmospheres}},
	volume = {743},
	issn = {2041-8205},
	url = {http://stacks.iop.org/2041-8205/743/i=1/a=L16},
	doi = {10.1088/2041-8205/743/1/L16},
	language = {en},
	number = {1},
	urldate = {2018-12-25},
	journal = {\apjl},
	author = {\"Oberg, Karin I. and Murray-Clay, Ruth and Bergin, Edwin A.},
	year = {2011},
	pages = {L16}
}

@ARTICLE{Schneider2021a,
       author = {{Schneider}, Aaron David and {Bitsch}, Bertram},
        title = "{How drifting and evaporating pebbles shape giant planets. I. Heavy element content and atmospheric C/O}",
      journal = {\aap},
         year = 2021,
        month = oct,
       volume = {654},
          eid = {A71},
        pages = {A71},
          doi = {10.1051/0004-6361/202039640},
archivePrefix = {arXiv},
       eprint = {2105.13267},
 primaryClass = {astro-ph.EP},
       adsurl = {https://ui.adsabs.harvard.edu/abs/2021A&A...654A..71S}
}

@article{oberg_bergin2021,
   author = {Karin I. Öberg and Edwin A. Bergin},
   doi = {10.1016/j.physrep.2020.09.004},
   issn = {03701573},
   journal = {Physics Reports},
   month = {1},
   pages = {1-48},
   publisher = {Elsevier B.V.},
   title = {Astrochemistry and compositions of planetary systems},
   volume = {893},
   year = {2021},
}

@ARTICLE{Schneider2021b,
       author = {{Schneider}, Aaron David and {Bitsch}, Bertram},
        title = "{How drifting and evaporating pebbles shape giant planets. II. Volatiles and refractories in atmospheres}",
      journal = {\aap},
         year = 2021,
        month = oct,
       volume = {654},
          eid = {A72},
        pages = {A72},
          doi = {10.1051/0004-6361/202141096},
archivePrefix = {arXiv},
       eprint = {2109.03589},
 primaryClass = {astro-ph.EP},
       adsurl = {https://ui.adsabs.harvard.edu/abs/2021A&A...654A..72S}
}

@article{smith_Estimation_1998,
  title = {Estimation of a {{Length Scale}} to {{Use}} with the {{Quench Level Approximation}} for {{Obtaining Chemical Abundances}}},
  author = {Smith, Michael D.},
  year = {1998},
  month = mar,
  journal = {Icarus},
  volume = {132},
  number = {1},
  pages = {176--184},
  issn = {0019-1035},
  doi = {10.1006/icar.1997.5886},
  langid = {english}
}

@ARTICLE{Zuckerman2011,
       author = {{Zuckerman}, B. and {Rhee}, Joseph H. and {Song}, Inseok and {Bessell}, M.~S.},
        title = "{The Tucana/Horologium, Columba, AB Doradus, and Argus Associations: New Members and Dusty Debris Disks}",
      journal = {\apj},
         year = 2011,
        month = may,
       volume = {732},
       number = {2},
          eid = {61},
        pages = {61},
          doi = {10.1088/0004-637X/732/2/61},
archivePrefix = {arXiv},
       eprint = {1104.0284},
 primaryClass = {astro-ph.SR},
       adsurl = {https://ui.adsabs.harvard.edu/abs/2011ApJ...732...61Z}
}

@ARTICLE{Gray2002,
       author = {{Gray}, R.~O. and {Corbally}, C.~J.},
        title = "{A Spectroscopic Search for {\ensuremath{\lambda}} Bootis and Other Peculiar A-Type Stars in Intermediate-Age Open Clusters}",
      journal = {\aj},
         year = 2002,
        month = aug,
       volume = {124},
       number = {2},
        pages = {989-1000},
          doi = {10.1086/341609},
       adsurl = {https://ui.adsabs.harvard.edu/abs/2002AJ....124..989G}
}

@ARTICLE{Currie2014_hr8799,
       author = {{Currie}, Thayne and {Burrows}, Adam and {Girard}, Julien H. and {Cloutier}, Ryan and {Fukagawa}, Misato and {Sorahana}, Satoko and {Kuchner}, Marc and {Kenyon}, Scott J. and {Madhusudhan}, Nikku and {Itoh}, Yoichi and {Jayawardhana}, Ray and {Matsumura}, Soko and {Pyo}, Tae-Soo},
        title = "{Deep Thermal Infrared Imaging of HR 8799 bcde: New Atmospheric Constraints and Limits on a Fifth Planet}",
      journal = {\apj},
         year = 2014,
        month = nov,
       volume = {795},
       number = {2},
          eid = {133},
        pages = {133},
          doi = {10.1088/0004-637X/795/2/133},
archivePrefix = {arXiv},
       eprint = {1409.5134},
 primaryClass = {astro-ph.EP},
       adsurl = {https://ui.adsabs.harvard.edu/abs/2014ApJ...795..133C}
}

@article{Phillips2020,
   author = {M. W. Phillips and P. Tremblin and I. Baraffe and G. Chabrier and N. F. Allard and F. Spiegelman and J. M. Goyal and B. Drummond and E. Hébrard},
   doi = {10.1051/0004-6361/201937381},
   issn = {14320746},
   journal = {Astronomy and Astrophysics},
   month = {5},
   publisher = {EDP Sciences},
   title = {A new set of atmosphere and evolution models for cool T-Y brown dwarfs and giant exoplanets},
   volume = {637},
   year = {2020},
}

@article{molliere_Interpreting_2022a,
  title = {Interpreting the {{Atmospheric Composition}} of {{Exoplanets}}: {{Sensitivity}} to {{Planet Formation Assumptions}}},
  shorttitle = {Interpreting the {{Atmospheric Composition}} of {{Exoplanets}}},
  author = {Molli{\`e}re, Paul and Molyarova, Tamara and Bitsch, Bertram and Henning, Thomas and Schneider, Aaron and Kreidberg, Laura and Eistrup, Christian and Burn, Remo and Nasedkin, Evert and Semenov, Dmitry and Mordasini, Christoph and Schlecker, Martin and Schwarz, Kamber R. and Lacour, Sylvestre and Nowak, Mathias and Schulik, Matth{\"a}us},
  year = {2022},
  month = jul,
  journal = {ApJ},
  volume = {934},
  pages = {74},
  issn = {0004-637X},
  doi = {10.3847/1538-4357/ac6a56}
}

@ARTICLE{Nasedkin2024,
       author = {{Nasedkin}, E. and {Molli{\`e}re}, P. and {Lacour}, S. and {Nowak}, M. and {Kreidberg}, L. and {Stolker}, T. and {Wang}, J.~J. and {Balmer}, W.~O. and {Kammerer}, J. and {Shangguan}, J. and {Abuter}, R. and {Amorim}, A. and {Asensio-Torres}, R. and {Benisty}, M. and {Berger}, J.-P. and {Beust}, H. and {Blunt}, S. and {Boccaletti}, A. and {Bonnefoy}, M. and {Bonnet}, H. and {Bordoni}, M.~S. and {Bourdarot}, G. and {Brandner}, W. and {Cantalloube}, F. and {Caselli}, P. and {Charnay}, B. and {Chauvin}, G. and {Chavez}, A. and {Choquet}, E. and {Christiaens}, V. and {Cl{\'e}net}, Y. and {Coud{\'e} Du Foresto}, V. and {Cridland}, A. and {Davies}, R. and {Dembet}, R. and {Dexter}, J. and {Drescher}, A. and {Duvert}, G. and {Eckart}, A. and {Eisenhauer}, F. and {F{\"o}rster Schreiber}, N.~M. and {Garcia}, P. and {Garcia Lopez}, R. and {Gendron}, E. and {Genzel}, R. and {Gillessen}, S. and {Girard}, J.~H. and {Grant}, S. and {Haubois}, X. and {Hei{\ss}el}, G. and {Henning}, Th. and {Hinkley}, S. and {Hippler}, S. and {Houll{\'e}}, M. and {Hubert}, Z. and {Jocou}, L. and {Keppler}, M. and {Kervella}, P. and {Kurtovic}, N.~T. and {Lagrange}, A.-M. and {Lapeyr{\`e}re}, V. and {Le Bouquin}, J.-B. and {Lutz}, D. and {Maire}, A.-L. and {Mang}, F. and {Marleau}, G.-D. and {M{\'e}rand}, A. and {Monnier}, J.~D. and {Mordasini}, C. and {Ott}, T. and {Otten}, G.~P.~P.~L. and {Paladini}, C. and {Paumard}, T. and {Perraut}, K. and {Perrin}, G. and {Pfuhl}, O. and {Pourr{\'e}}, N. and {Pueyo}, L. and {Ribeiro}, D.~C. and {Rickman}, E. and {Ruffio}, J.~B. and {Rustamkulov}, Z. and {Shimizu}, T. and {Sing}, D. and {Stadler}, J. and {Straub}, O. and {Straubmeier}, C. and {Sturm}, E. and {Tacconi}, L.~J. and {van Dishoeck}, E.~F. and {Vigan}, A. and {Vincent}, F. and {von Fellenberg}, S.~D. and {Widmann}, F. and {Winterhalder}, T.~O. and {Woillez}, J. and {Yazici}, {\c{S}}. and {Gravity Collaboration}},
        title = "{Four-of-a-kind? Comprehensive atmospheric characterisation of the HR 8799 planets with VLTI/GRAVITY}",
      journal = {\aap},
         year = 2024,
        month = jul,
       volume = {687},
          eid = {A298},
        pages = {A298},
          doi = {10.1051/0004-6361/202449328},
archivePrefix = {arXiv},
       eprint = {2404.03776},
 primaryClass = {astro-ph.EP},
       adsurl = {https://ui.adsabs.harvard.edu/abs/2024A&A...687A.298N}
}

@article{Saumon_2008,
	doi = {10.1086/592734},
	url = {https://doi.org/10.1086%2F592734},
	year = 2008,
	month = {dec},
	publisher = {{IOP} Publishing},
	volume = {689},
	number = {2},
	pages = {1327--1344},
	author = {D. Saumon and Mark S. Marley},
	title = {The Evolution of L and T Dwarfs in Color-Magnitude Diagrams},
	journal = {\apj},
}

@article{Biazzo2017,
   author = {K. Biazzo and A. Frasca and J. M. Alcalá and M. Zusi and E. Covino and S. Randich and M. Esposito and C. F. Manara and S. Antoniucci and B. Nisini and E. Rigliaco and F. Getman},
   doi = {10.1051/0004-6361/201730850},
   issn = {14320746},
   journal = {A\&A},
   month = {9},
   publisher = {EDP Sciences},
   title = {X-shooter spectroscopy of young stellar objects in Lupus: Lithium, iron, and barium elemental abundances},
   volume = {605},
   year = {2017},
}

@ARTICLE{Santos2008,
       author = {{Santos}, N.~C. and {Melo}, C. and {James}, D.~J. and {Gameiro}, J.~F. and {Bouvier}, J. and {Gomes}, J.~I.},
        title = "{Chemical abundances in six nearby star-forming regions. Implications for galactic evolution and planet searches around very young stars}",
      journal = {\aap},
         year = 2008,
        month = mar,
       volume = {480},
       number = {3},
        pages = {889-897},
          doi = {10.1051/0004-6361:20079083},
archivePrefix = {arXiv},
       eprint = {0801.2529},
 primaryClass = {astro-ph},
       adsurl = {https://ui.adsabs.harvard.edu/abs/2008A&A...480..889S}
}

@article{DOrazi2011,
   author = {V. D'Orazi and K. Biazzo and S. Randich},
   doi = {10.1051/0004-6361/201015616},
   issn = {00046361},
   issue = {13},
   journal = {A\&A},
   month = {2},
   title = {Chemical composition of the Taurus-Auriga association},
   volume = {526},
   year = {2011},
}

@article{Rothman2010,
   author = {L. S. Rothman and I. E. Gordon and R. J. Barber and H. Dothe and R. R. Gamache and A. Goldman and V. I. Perevalov and S. A. Tashkun and J. Tennyson},
   doi = {10.1016/j.jqsrt.2010.05.001},
   issn = {00224073},
   issue = {15},
   journal = {Journal of Quantitative Spectroscopy and Radiative Transfer},
   month = {10},
   pages = {2139-2150},
   title = {HITEMP, the high-temperature molecular spectroscopic database},
   volume = {111},
   year = {2010},
}

@article{Coles2019,
   author = {Phillip A. Coles and Sergei N. Yurchenko and Jonathan Tennyson},
   doi = {10.1093/mnras/stz2778},
   issn = {13652966},
   issue = {4},
   journal = {Monthly Notices of the Royal Astronomical Society},
   month = {12},
   pages = {4481-4488},
   publisher = {Oxford University Press},
   title = {ExoMol molecular line lists - XXXV. A rotation-vibration line list for hot ammonia},
   volume = {490},
   year = {2019},
}

@article{Polyansky2018,
   author = {Oleg L Polyansky and Aleksandra A Kyuberis and Nikolai F Zobov and Jonathan Tennyson and Sergei N Yurchenko and Lorenzo Lodi},
   doi = {10.1093/mnras/sty1877},
   issn = {0035-8711},
   issue = {2},
   journal = {Monthly Notices of the Royal Astronomical Society},
   pages = {2597-2608},
   title = {ExoMol molecular line lists XXX: a complete high-accuracy line list for water},
   volume = {480},
   url = {https://doi.org/10.1093/mnras/sty1877 https://academic.oup.com/mnras/article-pdf/480/2/2597/28250193/sty1877.pdf https://academic.oup.com/mnras/article/480/2/2597/5054049},
   year = {2018},
}

@article{Azzam2016,
   author = {Ala'a A.A. Azzam and Jonathan Tennyson and Sergei N. Yurchenko and Olga V. Naumenko},
   doi = {10.1093/mnras/stw1133},
   issn = {13652966},
   issue = {4},
   journal = {Monthly Notices of the Royal Astronomical Society},
   month = {8},
   pages = {4063-4074},
   publisher = {Oxford University Press},
   title = {ExoMol molecular line lists - XVI. The rotation-vibration spectrum of hot H2S},
   volume = {460},
   year = {2016},
}

@article{Bernath2020,
   author = {Peter F. Bernath},
   doi = {10.1016/j.jqsrt.2019.106687},
   issn = {00224073},
   journal = {Journal of Quantitative Spectroscopy and Radiative Transfer},
   month = {1},
   publisher = {Elsevier Ltd},
   title = {MoLLIST: Molecular Line Lists, Intensities and Spectra},
   volume = {240},
   year = {2020},
}

@article{ackerman_Precipitating_2001,
  title = {Precipitating {{Condensation Clouds}} in {{Substellar Atmospheres}}},
  author = {Ackerman, Andrew S. and Marley, Mark S.},
  year = {2001},
  month = aug,
  journal = {The Astrophysical Journal},
  volume = {556},
  pages = {872--884},
  issn = {0004-637X},
  doi = {10.1086/321540}
}

@article{asplund_Chemical_2009,
  title = {The {{Chemical Composition}} of the {{Sun}}},
  author = {Asplund, Martin and Grevesse, Nicolas and Sauval, A. Jacques and Scott, Pat},
  year = {2009},
  month = sep,
  journal = {Annual Review of Astronomy \&amp; Astrophysics, vol. 47, Issue 1, pp.481-522},
  volume = {47},
  number = {1},
  pages = {481},
  issn = {0066-4146},
  doi = {10.1146/annurev.astro.46.060407.145222},
  language = {en}
}

@article{brewer_SPECTRAL_2016,
  title = {{{SPECTRAL PROPERTIES OF COOL STARS}}: {{EXTENDED ABUNDANCE ANALYSIS OF}} 1,617 {{PLANET}}-{{SEARCH STARS}}},
  shorttitle = {{{SPECTRAL PROPERTIES OF COOL STARS}}},
  author = {Brewer, John M. and Fischer, Debra A. and Valenti, Jeff A. and Piskunov, Nikolai},
  year = {2016},
  month = aug,
  journal = {ApJS},
  volume = {225},
  number = {2},
  pages = {32},
  publisher = {{American Astronomical Society}},
  issn = {0067-0049},
  doi = {10.3847/0067-0049/225/2/32},
  language = {en}
}

@article{Zhang2023,
doi = {10.3847/1538-3881/acf768},
url = {https://dx.doi.org/10.3847/1538-3881/acf768},
year = {2023},
month = {oct},
publisher = {The American Astronomical Society},
volume = {166},
number = {5},
pages = {198},
author = {Zhoujian Zhang and Paul Mollière and Keith Hawkins and Catherine Manea and Jonathan J. Fortney and Caroline V. Morley and Andrew Skemer and Mark S. Marley and Brendan P. Bowler and Aarynn L. Carter and Kyle Franson and Zachary G. Maas and Christopher Sneden},
title = {ELemental abundances of Planets and brown dwarfs Imaged around Stars (ELPIS). I. Potential Metal Enrichment of the Exoplanet AF Lep b and a Novel Retrieval Approach for Cloudy Self-luminous Atmospheres},
journal = {AJ},
}

@ARTICLE{Voyer2025,
       author = {{Voyer}, Ma{\"e}l and {Changeat}, Quentin and {Lagage}, Pierre-Olivier and {Tremblin}, Pascal and {Waters}, Rens and {G{\"u}del}, Manuel and {Henning}, Thomas and {Absil}, Olivier and {Barrado}, David and {Boccaletti}, Anthony and {Bouwman}, Jeroen and {Coulais}, Alain and {Decin}, Leen and {Glauser}, Adrian M. and {Pye}, John and {Glasse}, Alistair and {Gastaud}, Ren{\'e} and {Kendrew}, Sarah and {Patapis}, Polychronis and {Rouan}, Daniel and {van Dishoeck}, Ewine F. and {{\"O}stlin}, G{\"o}ran and {Ray}, Tom P. and {Wright}, Gillian},
        title = "{MIRI-LRS Spectrum of a Cold Exoplanet around a White Dwarf: Water, Ammonia, and Methane Measurements}",
      journal = {\apjl},
         year = 2025,
        month = apr,
       volume = {982},
       number = {2},
          eid = {L38},
        pages = {L38},
          doi = {10.3847/2041-8213/adbd46},
archivePrefix = {arXiv},
       eprint = {2503.04531},
 primaryClass = {astro-ph.EP},
       adsurl = {https://ui.adsabs.harvard.edu/abs/2025ApJ...982L..38V}
}

@ARTICLE{Crossfield2023,
       author = {{Crossfield}, Ian J.~M.},
        title = "{Volatile-to-sulfur Ratios Can Recover a Gas Giant's Accretion History}",
      journal = {\apjl},
         year = 2023,
        month = jul,
       volume = {952},
       number = {1},
          eid = {L18},
        pages = {L18},
          doi = {10.3847/2041-8213/ace35f},
archivePrefix = {arXiv},
       eprint = {2303.17622},
 primaryClass = {astro-ph.EP},
       adsurl = {https://ui.adsabs.harvard.edu/abs/2023ApJ...952L..18C}
}

@article{hargreaves_Accurate_2020,
  title = {An {{Accurate}}, {{Extensive}}, and {{Practical Line List}} of {{Methane}} for the {{HITEMP Database}}},
  author = {Hargreaves, Robert J. and Gordon, Iouli E. and Rey, Michael and Nikitin, Andrei V. and Tyuterev, Vladimir G. and Kochanov, Roman V. and Rothman, Laurence S.},
  year = {2020},
  month = apr,
  journal = {The Astrophysical Journal Supplement Series},
  volume = {247},
  pages = {55},
  issn = {0067-0049},
  doi = {10.3847/1538-4365/ab7a1a}
}

@article{luna_Empirically_2021,
  title = {Empirically {{Determining Substellar Cloud Compositions}} in the {{Era}} of the {{James Webb Space Telescope}}},
  author = {Luna, Jessica L. and Morley, Caroline V.},
  year = {2021},
  month = oct,
  journal = {The Astrophysical Journal},
  volume = {920},
  pages = {146},
  issn = {0004-637X},
  doi = {10.3847/1538-4357/ac1865}
}

@article{molliere_petitRADTRANS_2019,
  title = {{{petitRADTRANS}}. {{A Python}} Radiative Transfer Package for Exoplanet Characterization and Retrieval},
  author = {Molli{\`e}re, P. and Wardenier, J. P. and {van Boekel}, R. and Henning, Th. and Molaverdikhani, K. and Snellen, I. A. G.},
  year = {2019},
  month = jul,
  journal = {Astronomy and Astrophysics},
  volume = {627},
  pages = {A67},
  issn = {0004-6361},
  doi = {10.1051/0004-6361/201935470}
}

@article{molliere_Retrieving_2020,
  title = {Retrieving Scattering Clouds and Disequilibrium Chemistry in the Atmosphere of {{HR}} 8799e},
  author = {Molli{\`e}re, P. and Stolker, T. and Lacour, S. and Otten, G. P. P. L. and Shangguan, J. and Charnay, B. and Molyarova, T. and Nowak, M. and Henning, Th. and Marleau, G.-D. and Semenov, D. A. and {van Dishoeck}, E. and Eisenhauer, F. and Garcia, P. and Garcia Lopez, R. and Girard, J. H. and Greenbaum, A. Z. and Hinkley, S. and Kervella, P. and Kreidberg, L. and Maire, A.-L. and Nasedkin, E. and Pueyo, L. and Snellen, I. A. G. and Vigan, A. and Wang, J. and {de Zeeuw}, P. T. and Zurlo, A.},
  year = {2020},
  month = aug,
  journal = {Astronomy and Astrophysics},
  volume = {640},
  pages = {A131},
  issn = {0004-6361},
  doi = {10.1051/0004-6361/202038325}
}

@article{Zahnle_methane_2014,
  title = {Methane, {{Carbon Monoxide}}, and {{Ammonia}} in {{Brown Dwarfs}} and {{Self}}-{{Luminous Giant Planets}}},
  author = {Zahnle, Kevin J. and Marley, Mark S.},
  year = {2014},
  month = dec,
  journal = {The Astrophysical Journal},
  volume = {797},
  pages = {41},
  issn = {0004-637X},
  doi = {10.1088/0004-637X/797/1/41}
}

@article{zhang_13COrich_2021,
  title = {The {{13CO}}-Rich Atmosphere of a Young Accreting Super-{{Jupiter}}},
  author = {Zhang, Yapeng and Snellen, Ignas A. G. and Bohn, Alexander J. and Molli{\`e}re, Paul and Ginski, Christian and Hoeijmakers, H. Jens and Kenworthy, Matthew A. and Mamajek, Eric E. and Meshkat, Tiffany and Reggiani, Maddalena and Snik, Frans},
  year = {2021},
  month = jul,
  journal = {Nature},
  volume = {595},
  number = {7867},
  pages = {370--372},
  publisher = {{Nature Publishing Group}},
  issn = {1476-4687},
  doi = {10.1038/s41586-021-03616-x},
  copyright = {2021 The Author(s), under exclusive licence to Springer Nature Limited},
  langid = {english}
}

@INCOLLECTION{Gierasch_convect1985,
       author = {{Gierasch}, P.~J. and {Conrath}, B.~J.},
        title = "{Energy conversion processes in the outer planets.}",
    booktitle = {Recent Advances in Planetary Meteorology},
         year = 1985,
       editor = {{Hunt}, G.~E.},
        pages = {121-146},
        publisher = {Cambridge University Press},
       adsurl = {https://ui.adsabs.harvard.edu/abs/1985rapm.book..121G},
}

@article{zhang_Globalmean_2018,
  title = {Global-Mean {{Vertical Tracer Mixing}} in {{Planetary Atmospheres}}. {{I}}. {{Theory}} and {{Fast-rotating Planets}}},
  author = {Zhang, Xi and Showman, Adam P.},
  year = {2018},
  month = oct,
  journal = {The Astrophysical Journal},
  volume = {866},
  number = {1},
  pages = {1},
  publisher = {{American Astronomical Society}},
  issn = {0004-637X},
  doi = {10.3847/1538-4357/aada85},
  langid = {english}
}

@article{ruffio_Deep_2021,
  title = {Deep {{Exploration}} of the {{Planets HR}} 8799 b, c, and d with {{Moderate-resolution Spectroscopy}}},
  author = {Ruffio, Jean-Baptiste and Konopacky, Quinn M. and Barman, Travis and Macintosh, Bruce and Hoch, Kielan K. W. and De Rosa, Robert J. and Wang, Jason J. and Czekala, Ian and Marois, Christian},
  year = {2021},
  month = dec,
  journal = {AJ},
  volume = {162},
  pages = {290},
  issn = {0004-6256},
  doi = {10.3847/1538-3881/ac273a}
}

@ARTICLE{Malin2025,
       author = {{M{\^a}lin}, Mathilde and {Boccaletti}, Anthony and {Perrot}, Cl{\'e}ment and {Baudoz}, Pierre and {Rouan}, Daniel and {Lagage}, Pierre-Olivier and {Waters}, Rens and {G{\"u}del}, Manuel and {Henning}, Thomas and {Vandenbussche}, Bart and {Absil}, Olivier and {Barrado}, David and {Charnay}, Benjamin and {Choquet}, Elodie and {Cossou}, Christophe and {Danielski}, Camilla and {Decin}, Leen and {Glauser}, Adrian M. and {Pye}, John and {Olofsson}, Goran and {Glasse}, Alistair and {Patapis}, Polychronis and {Royer}, Pierre and {Scheithauer}, Silvia and {Serabyn}, Eugene and {Tremblin}, Pascal and {Whiteford}, Niall and {van Dishoeck}, Ewine F. and {Ostlin}, G{\"o}ran and {Ray}, Tom P. and {Wright}, Gillian},
        title = "{First unambiguous detection of ammonia in the atmosphere of a planetary mass companion with JWST/MIRI coronagraphs}",
      journal = {\aap},
         year = 2025,
        month = jan,
       volume = {693},
          eid = {A315},
        pages = {A315},
          doi = {10.1051/0004-6361/202452695},
archivePrefix = {arXiv},
       eprint = {2501.00104},
 primaryClass = {astro-ph.EP},
       adsurl = {https://ui.adsabs.harvard.edu/abs/2025A&A...693A.315M}
}

@ARTICLE{Ruffio2024,
       author = {{Ruffio}, Jean-Baptiste and {Perrin}, Marshall D. and {Hoch}, Kielan K.~W. and {Kammerer}, Jens and {Konopacky}, Quinn M. and {Pueyo}, Laurent and {Madurowicz}, Alex and {Rickman}, Emily and {Theissen}, Christopher A. and {Agrawal}, Shubh and {Greenbaum}, Alexandra Z. and {Miles}, Brittany E. and {Barman}, Travis S. and {Balmer}, William O. and {Llop-Sayson}, Jorge and {Girard}, Julien H. and {Rebollido}, Isabel and {Soummer}, R{\'e}mi and {Allen}, Natalie H. and {Anderson}, Jay and {Beichman}, Charles A. and {Bellini}, Andrea and {Bryden}, Geoffrey and {Espinoza}, N{\'e}stor and {Glidden}, Ana and {Huang}, Jingcheng and {Lewis}, Nikole K. and {Libralato}, Mattia and {Louie}, Dana R. and {Sohn}, Sangmo Tony and {Seager}, Sara and {van der Marel}, Roeland P. and {Wakeford}, Hannah R. and {Watkins}, Laura L. and {Ygouf}, Marie and {Mountain}, C. Matt},        title = "{JWST-TST High Contrast: Achieving Direct Spectroscopy of Faint Substellar Companions Next to Bright Stars with the NIRSpec Integral Field Unit}",
      journal = {Astron. J.},
         year = 2024,
        month = aug,
       volume = {168},
       number = {2},
          eid = {73},
        pages = {73},
          doi = {10.3847/1538-3881/ad5281},
archivePrefix = {arXiv},
       eprint = {2310.09902},
 primaryClass = {astro-ph.EP},
       adsurl = {https://ui.adsabs.harvard.edu/abs/2024AJ....168...73R},
}

@article{RuffioXuan2026,
  author  = {Ruffio, Jean-Baptiste and Xuan, Jerry W. and Chachan, Yayaati and Kesseli, Aurora and Lee, Eve J. and Beichman, Charles and Hodapp, Klaus and Balmer, William O. and Konopacky, Quinn and Perrin, Marshall D. and Mawet, Dimitri and Knutson, Heather A. and Bryden, Geoffrey and Greene, Thomas P. and Johnstone, Doug and Leisenring, Jarron and Meyer, Michael and Ygouf, Marie},
  title   = {Jupiter-like uniform metal enrichment in a system of multiple giant exoplanets},
  journal = {Nature Astronomy},
  year    = {2026},
  date    = {2026-02-09},
  doi     = {10.1038/s41550-026-02783-z},
  url     = {https://doi.org/10.1038/s41550-026-02783-z},
  issn    = {2397-3366},
}

@ARTICLE{Zhang2024,
       author = {{Zhang}, Yapeng and {Xuan}, Jerry W. and {Mawet}, Dimitri and {Wang}, Jason J. and {Hsu}, Chih-Chun and {Ruffio}, Jean-Bapiste and {Knutson}, Heather A. and {Inglis}, Julie and {Blake}, Geoffrey A. and {Chachan}, Yayaati and {Horstman}, Katelyn and {Baker}, Ashley and {Bartos}, Randall and {Calvin}, Benjamin and {Cetre}, Sylvain and {Delorme}, Jacques-Robert and {Doppmann}, Greg and {Echeverri}, Daniel and {Finnerty}, Luke and {Fitzgerald}, Michael P. and {Jovanovic}, Nemanja and {Liberman}, Joshua and {L{\'o}pez}, Ronald A. and {Morris}, Evan and {Pezzato}, Jacklyn and {Sappey}, Ben and {Schofield}, Tobias and {Skemer}, Andrew and {Wallace}, J. Kent and {Wang}, Ji and {Do {\'O}}, Clarissa R.},
        title = "{Atmospheric Characterization of the Super-Jupiter HIP 99770 b with KPIC}",
      journal = {\aj},
         year = 2024,
        month = sep,
       volume = {168},
       number = {3},
          eid = {131},
        pages = {131},
          doi = {10.3847/1538-3881/ad6609},
archivePrefix = {arXiv},
       eprint = {2407.20952},
 primaryClass = {astro-ph.EP},
       adsurl = {https://ui.adsabs.harvard.edu/abs/2024AJ....168..131Z}
}

@ARTICLE{Zurlo2016,
       author = {{Zurlo}, A. and {Vigan}, A. and {Galicher}, R. and {Maire}, A. -L. and {Mesa}, D. and {Gratton}, R. and {Chauvin}, G. and {Kasper}, M. and {Moutou}, C. and {Bonnefoy}, M. and {Desidera}, S. and {Abe}, L. and {Apai}, D. and {Baruffolo}, A. and {Baudoz}, P. and {Baudrand}, J. and {Beuzit}, J. -L. and {Blancard}, P. and {Boccaletti}, A. and {Cantalloube}, F. and {Carle}, M. and {Cascone}, E. and {Charton}, J. and {Claudi}, R.~U. and {Costille}, A. and {de Caprio}, V. and {Dohlen}, K. and {Dominik}, C. and {Fantinel}, D. and {Feautrier}, P. and {Feldt}, M. and {Fusco}, T. and {Gigan}, P. and {Girard}, J.~H. and {Gisler}, D. and {Gluck}, L. and {Gry}, C. and {Henning}, T. and {Hugot}, E. and {Janson}, M. and {Jaquet}, M. and {Lagrange}, A. -M. and {Langlois}, M. and {Llored}, M. and {Madec}, F. and {Magnard}, Y. and {Martinez}, P. and {Maurel}, D. and {Mawet}, D. and {Meyer}, M.~R. and {Milli}, J. and {Moeller-Nilsson}, O. and {Mouillet}, D. and {Orign{\'e}}, A. and {Pavlov}, A. and {Petit}, C. and {Puget}, P. and {Quanz}, S.~P. and {Rabou}, P. and {Ramos}, J. and {Rousset}, G. and {Roux}, A. and {Salasnich}, B. and {Salter}, G. and {Sauvage}, J. -F. and {Schmid}, H.~M. and {Soenke}, C. and {Stadler}, E. and {Suarez}, M. and {Turatto}, M. and {Udry}, S. and {Vakili}, F. and {Wahhaj}, Z. and {Wildi}, F. and {Antichi}, J.},
        title = "{First light of the VLT planet finder SPHERE. III. New spectrophotometry and astrometry of the HR 8799 exoplanetary system}",
      journal = {Astron. Astrophys.},
         year = 2016,
        month = mar,
       volume = {587},
          eid = {A57},
        pages = {A57},
          doi = {10.1051/0004-6361/201526835},
archivePrefix = {arXiv},
       eprint = {1511.04083},
 primaryClass = {astro-ph.EP},
       adsurl = {https://ui.adsabs.harvard.edu/abs/2016A&A...587A..57Z}
}

@ARTICLE{Mukherjee2022,
       author = {{Mukherjee}, Sagnick and {Fortney}, Jonathan J. and {Batalha}, Natasha E. and {Karalidi}, Theodora and {Marley}, Mark S. and {Visscher}, Channon and {Miles}, Brittany E. and {Skemer}, Andrew J.~I.},
        title = "{Probing the Extent of Vertical Mixing in Brown Dwarf Atmospheres with Disequilibrium Chemistry}",
      journal = {\apj},
         year = 2022,
        month = oct,
       volume = {938},
       number = {2},
          eid = {107},
        pages = {107},
          doi = {10.3847/1538-4357/ac8dfb},
archivePrefix = {arXiv},
       eprint = {2208.14317},
 primaryClass = {astro-ph.EP},
       adsurl = {https://ui.adsabs.harvard.edu/abs/2022ApJ...938..107M}
}

@ARTICLE{Xuan2024d,
       author = {{Xuan}, Jerry W. and {Perrin}, Marshall D. and {Mawet}, Dimitri and {Knutson}, Heather A. and {Mukherjee}, Sagnick and {Zhang}, Yapeng and {Hoch}, Kielan K.~W. and {Wang}, Jason J. and {Inglis}, Julie and {Wallack}, Nicole L. and {Ruffio}, Jean-Baptiste},
        title = "{Atmospheric Abundances and Bulk Properties of the Binary Brown Dwarf Gliese 229Bab from JWST/MIRI Spectroscopy}",
      journal = {\apjl},
         year = 2024,
        month = dec,
       volume = {977},
       number = {2},
          eid = {L32},
        pages = {L32},
          doi = {10.3847/2041-8213/ad92f9},
archivePrefix = {arXiv},
       eprint = {2411.10571},
 primaryClass = {astro-ph.SR},
       adsurl = {https://ui.adsabs.harvard.edu/abs/2024ApJ...977L..32X}
}

@ARTICLE{Gozdziewski2014,
       author = {{Go{\'z}dziewski}, Krzysztof and {Migaszewski}, Cezary},
        title = "{Multiple mean motion resonances in the HR 8799 planetary system}",
      journal = {Mon. Not. R. Astron. Soc.},
         year = 2014,
        month = jun,
       volume = {440},
       number = {4},
        pages = {3140-3171},
          doi = {10.1093/mnras/stu455},
       adsurl = {https://ui.adsabs.harvard.edu/abs/2014MNRAS.440.3140G},
}

@ARTICLE{Bonnefoy2016,
       author = {{Bonnefoy}, M. and {Zurlo}, A. and {Baudino}, J.~L. and {Lucas}, P. and {Mesa}, D. and {Maire}, A. -L. and {Vigan}, A. and {Galicher}, R. and {Homeier}, D. and {Marocco}, F. and {Gratton}, R. and {Chauvin}, G. and {Allard}, F. and {Desidera}, S. and {Kasper}, M. and {Moutou}, C. and {Lagrange}, A. -M. and {Antichi}, J. and {Baruffolo}, A. and {Baudrand}, J. and {Beuzit}, J. -L. and {Boccaletti}, A. and {Cantalloube}, F. and {Carbillet}, M. and {Charton}, J. and {Claudi}, R.~U. and {Costille}, A. and {Dohlen}, K. and {Dominik}, C. and {Fantinel}, D. and {Feautrier}, P. and {Feldt}, M. and {Fusco}, T. and {Gigan}, P. and {Girard}, J.~H. and {Gluck}, L. and {Gry}, C. and {Henning}, T. and {Janson}, M. and {Langlois}, M. and {Madec}, F. and {Magnard}, Y. and {Maurel}, D. and {Mawet}, D. and {Meyer}, M.~R. and {Milli}, J. and {Moeller-Nilsson}, O. and {Mouillet}, D. and {Pavlov}, A. and {Perret}, D. and {Pujet}, P. and {Quanz}, S.~P. and {Rochat}, S. and {Rousset}, G. and {Roux}, A. and {Salasnich}, B. and {Salter}, G. and {Sauvage}, J. -F. and {Schmid}, H.~M. and {Sevin}, A. and {Soenke}, C. and {Stadler}, E. and {Turatto}, M. and {Udry}, S. and {Vakili}, F. and {Wahhaj}, Z. and {Wildi}, F.},
        title = "{First light of the VLT planet finder SPHERE. IV. Physical and chemical properties of the planets around HR8799}",
      journal = {\aap},
         year = 2016,
        month = mar,
       volume = {587},
          eid = {A58},
        pages = {A58},
          doi = {10.1051/0004-6361/201526906},
archivePrefix = {arXiv},
       eprint = {1511.04082},
 primaryClass = {astro-ph.EP},
       adsurl = {https://ui.adsabs.harvard.edu/abs/2016A&A...587A..58B}
}

@article{skemer_Directly_2014,
  title = {Directly {{Imaged L-T Transition Exoplanets}} in the {{Mid-infrared}}},
  author = {Skemer, Andrew J. and Marley, Mark S. and Hinz, Philip M. and Morzinski, Katie M. and Skrutskie, Michael F. and Leisenring, Jarron M. and Close, Laird M. and Saumon, Didier and Bailey, Vanessa P. and Briguglio, Runa and Defrere, Denis and Esposito, Simone and Follette, Katherine B. and Hill, John M. and Males, Jared R. and Puglisi, Alfio and Rodigas, Timothy J. and Xompero, Marco},
  year = {2014},
  month = sep,
  journal = {ApJ},
  volume = {792},
  pages = {17},
  issn = {0004-637X},
  doi = {10.1088/0004-637X/792/1/17}
}

@ARTICLE{Feroz2019,
       author = {{Feroz}, Farhan and {Hobson}, Michael P. and {Cameron}, Ewan and {Pettitt}, Anthony N.},
        title = "{Importance Nested Sampling and the MultiNest Algorithm}",
      journal = {The Open Journal of Astrophysics},
         year = 2019,
        month = nov,
       volume = {2},
       number = {1},
          eid = {10},
        pages = {10},
          doi = {10.21105/astro.1306.2144},
archivePrefix = {arXiv},
       eprint = {1306.2144},
 primaryClass = {astro-ph.IM},
       adsurl = {https://ui.adsabs.harvard.edu/abs/2019OJAp....2E..10F}
}

@ARTICLE{Feroz2009,
       author = {{Feroz}, F. and {Hobson}, M.~P. and {Bridges}, M.},
        title = "{MULTINEST: an efficient and robust Bayesian inference tool for cosmology and particle physics}",
      journal = {Mon. Not. R. Astron. Soc.},
         year = 2009,
        month = oct,
       volume = {398},
       number = {4},
        pages = {1601-1614},
          doi = {10.1111/j.1365-2966.2009.14548.x},
archivePrefix = {arXiv},
       eprint = {0809.3437},
 primaryClass = {astro-ph},
       adsurl = {https://ui.adsabs.harvard.edu/abs/2009MNRAS.398.1601F}
}

@ARTICLE{Buchner2014,
       author = {{Buchner}, J. and {Georgakakis}, A. and {Nandra}, K. and {Hsu}, L. and {Rangel}, C. and {Brightman}, M. and {Merloni}, A. and {Salvato}, M. and {Donley}, J. and {Kocevski}, D.},
        title = "{X-ray spectral modelling of the AGN obscuring region in the CDFS: Bayesian model selection and catalogue}",
      journal = {Astron. Astrophys.},
         year = 2014,
        month = apr,
       volume = {564},
          eid = {A125},
        pages = {A125},
          doi = {10.1051/0004-6361/201322971},
archivePrefix = {arXiv},
       eprint = {1402.0004},
 primaryClass = {astro-ph.HE},
       adsurl = {https://ui.adsabs.harvard.edu/abs/2014A&A...564A.125B}
}

@ARTICLE{Thorngren2019,
       author = {{Thorngren}, Daniel and {Fortney}, Jonathan J.},
        title = "{Connecting Giant Planet Atmosphere and Interior Modeling: Constraints on Atmospheric Metal Enrichment}",
      journal = {\apjl},
         year = 2019,
        month = apr,
       volume = {874},
       number = {2},
          eid = {L31},
        pages = {L31},
          doi = {10.3847/2041-8213/ab1137},
archivePrefix = {arXiv},
       eprint = {1811.11859},
 primaryClass = {astro-ph.EP},
       adsurl = {https://ui.adsabs.harvard.edu/abs/2019ApJ...874L..31T}
}

@ARTICLE{Sepulveda2022,
       author = {{Sepulveda}, Aldo G. and {Bowler}, Brendan P.},
        title = "{Dynamical Mass of the Exoplanet Host Star HR 8799}",
      journal = {\aj},
         year = 2022,
        month = feb,
       volume = {163},
       number = {2},
          eid = {52},
        pages = {52},
          doi = {10.3847/1538-3881/ac3bb5},
archivePrefix = {arXiv},
       eprint = {2111.12090},
 primaryClass = {astro-ph.EP},
       adsurl = {https://ui.adsabs.harvard.edu/abs/2022AJ....163...52S}
}

@ARTICLE{Zurlo2022,
       author = {{Zurlo}, A. and {Go{\'z}dziewski}, K. and {Lazzoni}, C. and {Mesa}, D. and {Nogueira}, P. and {Desidera}, S. and {Gratton}, R. and {Marzari}, F. and {Langlois}, M. and {Pinna}, E. and {Chauvin}, G. and {Delorme}, P. and {Girard}, J.~H. and {Hagelberg}, J. and {Henning}, Th. and {Janson}, M. and {Rickman}, E. and {Kervella}, P. and {Avenhaus}, H. and {Bhowmik}, T. and {Biller}, B. and {Boccaletti}, A. and {Bonaglia}, M. and {Bonavita}, M. and {Bonnefoy}, M. and {Cantalloube}, F. and {Cheetham}, A. and {Claudi}, R. and {D'Orazi}, V. and {Feldt}, M. and {Galicher}, R. and {Ghose}, E. and {Lagrange}, A. -M. and {le Coroller}, H. and {Ligi}, R. and {Kasper}, M. and {Maire}, A. -L. and {Medard}, F. and {Meyer}, M. and {Peretti}, S. and {Perrot}, C. and {Puglisi}, A.~T. and {Rossi}, F. and {Rothberg}, B. and {Schmidt}, T. and {Sissa}, E. and {Vigan}, A. and {Wahhaj}, Z.},
        title = "{Orbital and dynamical analysis of the system around HR 8799. New astrometric epochs from VLT/SPHERE and LBT/LUCI}",
      journal = {Astron. Astrophys.},
         year = 2022,
        month = oct,
       volume = {666},
          eid = {A133},
        pages = {A133},
          doi = {10.1051/0004-6361/202243862},
archivePrefix = {arXiv},
       eprint = {2207.10684},
 primaryClass = {astro-ph.EP},
       adsurl = {https://ui.adsabs.harvard.edu/abs/2022A&A...666A.133Z}
}

@ARTICLE{Godziewski2020,
       author = {{Go{\'z}dziewski}, Krzysztof and {Migaszewski}, Cezary},
        title = "{An Exact, Generalized Laplace Resonance in the HR 8799 Planetary System}",
      journal = {\apjl},
         year = 2020,
        month = oct,
       volume = {902},
       number = {2},
          eid = {L40},
        pages = {L40},
          doi = {10.3847/2041-8213/abb881},
archivePrefix = {arXiv},
       eprint = {2009.07006},
 primaryClass = {astro-ph.EP},
       adsurl = {https://ui.adsabs.harvard.edu/abs/2020ApJ...902L..40G}
}

@ARTICLE{Su2009,
       author = {{Su}, K.~Y.~L. and {Rieke}, G.~H. and {Stapelfeldt}, K.~R. and {Malhotra}, R. and {Bryden}, G. and {Smith}, P.~S. and {Misselt}, K.~A. and {Moro-Martin}, A. and {Williams}, J.~P.},
        title = "{The Debris Disk Around HR 8799}",
      journal = {\apj},
         year = 2009,
        month = nov,
       volume = {705},
       number = {1},
        pages = {314-327},
          doi = {10.1088/0004-637X/705/1/314},
archivePrefix = {arXiv},
       eprint = {0909.2687},
 primaryClass = {astro-ph.EP},
       adsurl = {https://ui.adsabs.harvard.edu/abs/2009ApJ...705..314S}
}

@ARTICLE{Oberg2016,
       author = {{{\"O}berg}, Karin I. and {Bergin}, Edwin A.},
        title = "{Excess C/O and C/H in Outer Protoplanetary Disk Gas}",
      journal = {\apjl},
         year = 2016,
        month = nov,
       volume = {831},
       number = {2},
          eid = {L19},
        pages = {L19},
          doi = {10.3847/2041-8205/831/2/L19},
archivePrefix = {arXiv},
       eprint = {1610.07859},
 primaryClass = {astro-ph.GA},
       adsurl = {https://ui.adsabs.harvard.edu/abs/2016ApJ...831L..19O}
}

@ARTICLE{Booth2017,
       author = {{Booth}, Richard A. and {Clarke}, Cathie J. and {Madhusudhan}, Nikku and {Ilee}, John D.},
        title = "{Chemical enrichment of giant planets and discs due to pebble drift}",
      journal = {\mnras},
         year = 2017,
        month = aug,
       volume = {469},
       number = {4},
        pages = {3994-4011},
          doi = {10.1093/mnras/stx1103},
archivePrefix = {arXiv},
       eprint = {1705.03305},
 primaryClass = {astro-ph.EP},
       adsurl = {https://ui.adsabs.harvard.edu/abs/2017MNRAS.469.3994B}
}

@ARTICLE{Booth2016,
       author = {{Booth}, Mark and {Jord{\'a}n}, Andr{\'e}s and {Casassus}, Simon and {Hales}, Antonio S. and {Dent}, William R.~F. and {Faramaz}, Virginie and {Matr{\`a}}, Luca and {Barkats}, Denis and {Brahm}, Rafael and {Cuadra}, Jorge},
        title = "{Resolving the planetesimal belt of HR 8799 with ALMA}",
      journal = {\mnras},
         year = 2016,
        month = jul,
       volume = {460},
       number = {1},
        pages = {L10-L14},
          doi = {10.1093/mnrasl/slw040},
archivePrefix = {arXiv},
       eprint = {1603.04853},
 primaryClass = {astro-ph.EP},
       adsurl = {https://ui.adsabs.harvard.edu/abs/2016MNRAS.460L..10B}
}

@ARTICLE{Boccaletti2024,
       author = {{Boccaletti}, Anthony and {M{\^a}lin}, Mathilde and {Baudoz}, Pierre and {Tremblin}, Pascal and {Perrot}, Cl{\'e}ment and {Rouan}, Daniel and {Lagage}, Pierre-Olivier and {Whiteford}, Niall and {Molli{\`e}re}, Paul and {Waters}, Rens and {Henning}, Thomas and {Decin}, Leen and {G{\"u}del}, Manuel and {Vandenbussche}, Bart and {Absil}, Olivier and {Argyriou}, Ioannis and {Bouwman}, Jeroen and {Cossou}, Christophe and {Coulais}, Alain and {Gastaud}, Ren{\'e} and {Glasse}, Alistair and {Glauser}, Adrian M. and {Kamp}, Inga and {Kendrew}, Sarah and {Krause}, Oliver and {Lahuis}, Fred and {Mueller}, Michael and {Olofsson}, Goran and {Patapis}, Polychronis and {Pye}, John and {Royer}, Pierre and {Serabyn}, Eugene and {Scheithauer}, Silvia and {Colina}, Luis and {van Dishoeck}, Ewine F. and {Ostlin}, G{\"o}ran and {Ray}, Tom P. and {Wright}, Gillian},
        title = "{Imaging detection of the inner dust belt and the four exoplanets in the HR 8799 system with JWST's MIRI coronagraph}",
      journal = {\aap},
         year = 2024,
        month = jun,
       volume = {686},
          eid = {A33},
        pages = {A33},
          doi = {10.1051/0004-6361/202347912},
archivePrefix = {arXiv},
       eprint = {2310.13414},
 primaryClass = {astro-ph.EP},
       adsurl = {https://ui.adsabs.harvard.edu/abs/2024A&A...686A..33B}
}

@ARTICLE{Wang2022,
       author = {{Wang}, Jason J. and {Gao}, Peter and {Chilcote}, Jeffrey and {Lozi}, Julien and {Guyon}, Olivier and {Marois}, Christian and {De Rosa}, Robert J. and {Sahoo}, Ananya and {Groff}, Tyler D. and {Vievard}, Sebastien and {Jovanovic}, Nemanja and {Greenbaum}, Alexandra Z. and {Macintosh}, Bruce},
        title = "{Atmospheric Monitoring and Precise Spectroscopy of the HR 8799 Planets with SCExAO/CHARIS}",
      journal = {\aj},
         year = 2022,
        month = oct,
       volume = {164},
       number = {4},
          eid = {143},
        pages = {143},
          doi = {10.3847/1538-3881/ac8984},
archivePrefix = {arXiv},
       eprint = {2208.05594},
 primaryClass = {astro-ph.EP},
       adsurl = {https://ui.adsabs.harvard.edu/abs/2022AJ....164..143W}
}

@ARTICLE{Bell2015,
       author = {{Bell}, Cameron P.~M. and {Mamajek}, Eric E. and {Naylor}, Tim},
        title = "{A self-consistent, absolute isochronal age scale for young moving groups in the solar neighbourhood}",
      journal = {\mnras},
         year = 2015,
        month = nov,
       volume = {454},
       number = {1},
        pages = {593-614},
          doi = {10.1093/mnras/stv1981},
archivePrefix = {arXiv},
       eprint = {1508.05955},
 primaryClass = {astro-ph.SR},
       adsurl = {https://ui.adsabs.harvard.edu/abs/2015MNRAS.454..593B}
}

@ARTICLE{Godziewski2018,
       author = {{Go{\'z}dziewski}, Krzysztof and {Migaszewski}, Cezary},
        title = "{The Orbital Architecture and Debris Disks of the HR 8799 Planetary System}",
      journal = {\apjs},
         year = 2018,
        month = sep,
       volume = {238},
       number = {1},
          eid = {6},
        pages = {6},
          doi = {10.3847/1538-4365/aad3d3},
archivePrefix = {arXiv},
       eprint = {1807.05898},
 primaryClass = {astro-ph.EP},
       adsurl = {https://ui.adsabs.harvard.edu/abs/2018ApJS..238....6G}
}

@ARTICLE{Konopacky2016,
       author = {{Konopacky}, Q.~M. and {Marois}, C. and {Macintosh}, B.~A. and {Galicher}, R. and {Barman}, T.~S. and {Metchev}, S.~A. and {Zuckerman}, B.},
        title = "{Astrometric Monitoring of the HR 8799 Planets: Orbit Constraints from Self-consistent Measurements}",
      journal = {\aj},
         year = 2016,
        month = aug,
       volume = {152},
       number = {2},
          eid = {28},
        pages = {28},
          doi = {10.3847/0004-6256/152/2/28},
archivePrefix = {arXiv},
       eprint = {1604.08157},
 primaryClass = {astro-ph.EP},
       adsurl = {https://ui.adsabs.harvard.edu/abs/2016AJ....152...28K}
}

@ARTICLE{Poblete2025,
       author = {{Poblete}, Pedro P. and {Pearce}, Tim D. and {Charalambous}, Carolina},
        title = "{The HR 8799 debris disk: Shaped by planetary migration and a possible fifth outermost planet}",
      journal = {\aap},
         year = 2025,
        month = aug,
       volume = {700},
          eid = {A148},
        pages = {A148},
          doi = {10.1051/0004-6361/202554802},
archivePrefix = {arXiv},
       eprint = {2507.01589},
 primaryClass = {astro-ph.EP},
       adsurl = {https://ui.adsabs.harvard.edu/abs/2025A&A...700A.148P}
}

@article{Faramaz2021,
   author = {Virginie Faramaz and Sebastian Marino and Mark Booth and Luca Matrà and Eric E. Mamajek and Geoffrey Bryden and Karl R. Stapelfeldt and Simon Casassus and Jorge Cuadra and Antonio S. Hales and Alice Zurlo},
   doi = {10.3847/1538-3881/abf4e0},
   issn = {0004-6256},
   issue = {6},
   journal = {The Astronomical Journal},
   month = {6},
   pages = {271},
   publisher = {American Astronomical Society},
   title = {A Detailed Characterization of HR 8799's Debris Disk with ALMA in Band 7},
   volume = {161},
   year = {2021},
}

@ARTICLE{Xuan2024,
       author = {{Xuan}, Jerry W. and {Wang}, Jason and {Finnerty}, Luke and {Horstman}, Katelyn and {Grimm}, Simon and {Peck}, Anne E. and {Nielsen}, Eric and {Knutson}, Heather A. and {Mawet}, Dimitri and {Isaacson}, Howard and {Howard}, Andrew W. and {Liu}, Michael C. and {Walker}, Sam and {Phillips}, Mark W. and {Blake}, Geoffrey A. and {Ruffio}, Jean-Baptiste and {Zhang}, Yapeng and {Inglis}, Julie and {Wallack}, Nicole L. and {Sanghi}, Aniket and {Gonzales}, Erica J. and {Dai}, Fei and {Baker}, Ashley and {Bartos}, Randall and {Bond}, Charlotte Z. and {Bryan}, Marta L. and {Calvin}, Benjamin and {Cetre}, Sylvain and {Delorme}, Jacques-Robert and {Doppmann}, Greg and {Echeverri}, Daniel and {Fitzgerald}, Michael P. and {Jovanovic}, Nemanja and {Liberman}, Joshua and {L{\'o}pez}, Ronald A. and {Martin}, Emily C. and {Morris}, Evan and {Pezzato}, Jacklyn and {Ruane}, Garreth and {Sappey}, Ben and {Schofield}, Tobias and {Skemer}, Andrew and {Venenciano}, Taylor and {Wallace}, J. Kent and {Wang}, Ji and {Wizinowich}, Peter and {Xin}, Yinzi and {Agrawal}, Shubh and {Do {\'O}}, Clarissa R. and {Hsu}, Chih-Chun and {Phillips}, Caprice L.},
        title = "{Validation of Elemental and Isotopic Abundances in Late-M Spectral Types with the Benchmark HIP 55507 AB System}",
      journal = {\apj},
         year = 2024,
        month = feb,
       volume = {962},
       number = {1},
          eid = {10},
        pages = {10},
          doi = {10.3847/1538-4357/ad1243},
archivePrefix = {arXiv},
       eprint = {2312.02297},
 primaryClass = {astro-ph.SR},
       adsurl = {https://ui.adsabs.harvard.edu/abs/2024ApJ...962...10X}
}

@ARTICLE{Xuan2024b,
       author = {{Xuan}, Jerry W. and {Hsu}, Chih-Chun and {Finnerty}, Luke and {Wang}, Jason and {Ruffio}, Jean-Baptiste and {Zhang}, Yapeng and {Knutson}, Heather A. and {Mawet}, Dimitri and {Mamajek}, Eric E. and {Inglis}, Julie and {Wallack}, Nicole L. and {Bryan}, Marta L. and {Blake}, Geoffrey A. and {Molli{\`e}re}, Paul and {Hejazi}, Neda and {Baker}, Ashley and {Bartos}, Randall and {Calvin}, Benjamin and {Cetre}, Sylvain and {Delorme}, Jacques-Robert and {Doppmann}, Greg and {Echeverri}, Daniel and {Fitzgerald}, Michael P. and {Jovanovic}, Nemanja and {Liberman}, Joshua and {L{\'o}pez}, Ronald A. and {Morris}, Evan and {Pezzato}, Jacklyn and {Sappey}, Ben and {Schofield}, Tobias and {Skemer}, Andrew and {Wallace}, J. Kent and {Wang}, Ji and {Agrawal}, Shubh and {Horstman}, Katelyn},
        title = "{Are These Planets or Brown Dwarfs? Broadly Solar Compositions from High-resolution Atmospheric Retrievals of {\ensuremath{\sim}}10{\textendash}30 M $_{Jup}$ Companions}",
      journal = {\apj},
         year = 2024,
        month = jul,
       volume = {970},
       number = {1},
          eid = {71},
        pages = {71},
          doi = {10.3847/1538-4357/ad4796},
archivePrefix = {arXiv},
       eprint = {2405.13128},
 primaryClass = {astro-ph.EP},
       adsurl = {https://ui.adsabs.harvard.edu/abs/2024ApJ...970...71X}
}

@article{allen81,
  author = {{Allen}, M. and {Yung}, Y.~L. and {Waters}, J.~W.},
  title = {Vertical transport and photochemistry in the terrestrial mesosphere and lower thermosphere (50-120 km)},
  journal = {\jgr},
  year = {1981},
  volume = {86},
  pages = {3617-3627},
  doi = {0.1029/JA086iA05p03617}
}

@article{moses11,
  author = {{Moses}, J.~I. and {Visscher}, C. and {Fortney}, J.~J. and {Showman}, A.~P. and
       {Lewis}, N.~K. and {Griffith}, C.~A. and {Klippenstein}, S.~J. and
       {Shabram}, M. and {Friedson}, A.~J. and {Marley}, M.~S. and
       {Freedman}, R.~S.},
  title = {Disequilibrium Carbon, Oxygen, and Nitrogen Chemistry in the Atmospheres of {HD} 189733b and {HD} 209458b},
  journal = {\apj},
  year = {2011},
  volume = {737},
  pages = {15},
  doi = {10.1088/0004-637X/737/1/15}
}

@article{moses13gj436,
  author = {{Moses}, J.~I. and {Line}, M.~R. and {Visscher}, C. and {Richardson}, M.~R. and
       {Nettelmann}, N. and {Fortney}, J.~J. and {Barman}, T.~S. and
       {Stevenson}, K.~B. and {Madhusudhan}, N.},
  title = {Compositional Diversity in the Atmospheres of Hot {Neptunes}, with Application to {GJ} 436b},
  journal = {\apj},
  year = {2013},
  volume = {777},
  pages = {34},
  doi = {10.1088/0004-637X/777/1/34}
}

@article{moses14,
  author    = {{Moses}, J.~I.},
  title     = {Chemical Kinetics on Extrasolar Planets},
  journal   = {Phil. Trans. R. Soc. A},
  year      = {2014},
  volume    = {372},
  pages     = {20130073},
  doi = {10.1098/rsta.2013.0073}
}

@article{tsai23jwst,
  author = {{Tsai}, Shang-Min and {Lee}, Elspeth K.~H. and {Powell}, Diana and {Gao}, Peter and {Zhang}, Xi
  and {Moses}, Julianne and {H{\'e}brard}, Eric and {Venot}, Olivia and {Parmentier}, Vivien and {Jordan},
  Sean and {Hu}, Renyu and {Alam}, Munazza K. and {Alderson}, Lili and {Batalha}, Natalie M. and {Bean},
  Jacob L. and {Benneke}, Bj{\"o}rn and {Bierson}, Carver J. and {Brady}, Ryan P. and {Carone}, Ludmila and
  {Carter}, Aarynn L. and {Chubb}, Katy L. and {Inglis}, Julie and {Leconte}, J{\'e}r{\'e}my and {Line}, Michael
  and {L{\'o}pez-Morales}, Mercedes and {Miguel}, Yamila and {Molaverdikhani}, Karan and {Rustamkulov}, Zafar
  and {Sing}, David K. and {Stevenson}, Kevin B. and {Wakeford}, Hannah R. and {Yang}, Jeehyun and {Aggarwal},
  Keshav and {Baeyens}, Robin and {Barat}, Saugata and {de Val-Borro}, Miguel and {Daylan}, Tansu and {Fortney},
  Jonathan J. and {France}, Kevin and {Goyal}, Jayesh M. and {Grant}, David and {Kirk}, James and {Kreidberg},
  Laura and {Louca}, Amy and {Moran}, Sarah E. and {Mukherjee}, Sagnick and {Nasedkin}, Evert and {Ohno}, Kazumasa
  and {Rackham}, Benjamin V. and {Redfield}, Seth and {Taylor}, Jake and {Tremblin}, Pascal and {Visscher}, Channon
  and {Wallack}, Nicole L. and {Welbanks}, Luis and {Youngblood}, Allison and {Ahrer}, Eva-Maria and {Batalha},
  Natasha E. and {Behr}, Patrick and {Berta-Thompson}, Zachory K. and {Blecic}, Jasmina and {Casewell}, S.~L. and
  {Crossfield}, Ian J.~M. and {Crouzet}, Nicolas and {Cubillos}, Patricio E. and {Decin}, Leen and {D{\'e}sert},
  Jean-Michel and {Feinstein}, Adina D. and {Gibson}, Neale P. and {Harrington}, Joseph and {Heng}, Kevin and
  {Henning}, Thomas and {Kempton}, Eliza M. -R. and {Krick}, Jessica and {Lagage}, Pierre-Olivier and {Lendl},
  Monika and {Lothringer}, Joshua D. and {Mansfield}, Megan and {Mayne}, N.~J. and {Mikal-Evans}, Thomas and
  {Palle}, Enric and {Schlawin}, Everett and {Shorttle}, Oliver and {Wheatley}, Peter J. and {Yurchenko}, Sergei N.},
  title = {{Photochemically produced SO$_{2}$ in the atmosphere of WASP-39b}},
  journal = {Nature},
  year = {2023},
  volume = {617},
  ages = {483-487},
  doi = {10.1038/s41586-023-05902-2}
}

@article{yung84,
   author    = {{Yung}, Y.~L. and {Allen}, M. and {Pinto}, J.~P.},
   title     = {Photochemistry of the atmosphere of {Titan}: Comparison
                  between model and observations},
   journal   = {\apjs},
   year      = {1984},
   volume    = {55},
   pages     = {465-506},
   doi = {10.1086/190963}
}

@article{zahnle16,
  author    = {{Zahnle}, K. and {Marley}, M.~S. and {Morley}, C.~V. and {Moses}, J.~I.},
  title     = {Photolytic hazes in the atmosphere of 51 {Eri} {b}},
  journal   = {\apj},
  year      = {2016},
  volume    = {824},
  pages     = {137},
  doi = {10.3847/0004-637X/824/2/137}
}

@ARTICLE{Wogan2025,
       author = {{Wogan}, Nicholas F. and {Mang}, James and {Batalha}, Natasha E. and {Zahnle}, Kevin and {Mukherjee}, Sagnick and {Visscher}, Channon and {Fortney}, Jonathan J. and {Marley}, Mark S. and {Morley}, Caroline V.},
        title = "{The Sonora Substellar Atmosphere Models. V. A Correction to the Disequilibrium Abundance of CO$_{2}$ for Sonora Elf Owl}",
      journal = {Research Notes of the American Astronomical Society},
         year = 2025,
        month = may,
       volume = {9},
       number = {5},
          eid = {108},
        pages = {108},
          doi = {10.3847/2515-5172/add407},
archivePrefix = {arXiv},
       eprint = {2505.03994},
 primaryClass = {astro-ph.EP},
       adsurl = {https://ui.adsabs.harvard.edu/abs/2025RNAAS...9..108W}
}

@ARTICLE{Ruffio2017,
       author = {{Ruffio}, Jean-Baptiste and {Macintosh}, Bruce and {Wang}, Jason J. and {Pueyo}, Laurent and {Nielsen}, Eric L. and {De Rosa}, Robert J. and {Czekala}, Ian and {Marley}, Mark S. and {Arriaga}, Pauline and {Bailey}, Vanessa P. and {Barman}, Travis and {Bulger}, Joanna and {Chilcote}, Jeffrey and {Cotten}, Tara and {Doyon}, Rene and {Duch{\^e}ne}, Gaspard and {Fitzgerald}, Michael P. and {Follette}, Katherine B. and {Gerard}, Benjamin L. and {Goodsell}, Stephen J. and {Graham}, James R. and {Greenbaum}, Alexandra Z. and {Hibon}, Pascale and {Hung}, Li-Wei and {Ingraham}, Patrick and {Kalas}, Paul and {Konopacky}, Quinn and {Larkin}, James E. and {Maire}, J{\'e}r{\^o}me and {Marchis}, Franck and {Marois}, Christian and {Metchev}, Stanimir and {Millar-Blanchaer}, Maxwell A. and {Morzinski}, Katie M. and {Oppenheimer}, Rebecca and {Palmer}, David and {Patience}, Jennifer and {Perrin}, Marshall and {Poyneer}, Lisa and {Rajan}, Abhijith and {Rameau}, Julien and {Rantakyr{\"o}}, Fredrik T. and {Savransky}, Dmitry and {Schneider}, Adam C. and {Sivaramakrishnan}, Anand and {Song}, Inseok and {Soummer}, Remi and {Thomas}, Sandrine and {Wallace}, J. Kent and {Ward-Duong}, Kimberly and {Wiktorowicz}, Sloane and {Wolff}, Schuyler},
        title = "{Improving and Assessing Planet Sensitivity of the GPI Exoplanet Survey with a Forward Model Matched Filter}",
      journal = {Astrophys. J.},
         year = 2017,
        month = jun,
       volume = {842},
       number = {1},
          eid = {14},
        pages = {14},
          doi = {10.3847/1538-4357/aa72dd},
archivePrefix = {arXiv},
       eprint = {1705.05477},
 primaryClass = {astro-ph.EP},
       adsurl = {https://ui.adsabs.harvard.edu/abs/2017ApJ...842...14R}
}

@ARTICLE{deRegt2024,
       author = {{de Regt}, S. and {Gandhi}, S. and {Snellen}, I.~A.~G. and {Zhang}, Y. and {Ginski}, C. and {Gonz{\'a}lez Picos}, D. and {Kesseli}, A.~Y. and {Landman}, R. and {Molli{\`e}re}, P. and {Nasedkin}, E. and {S{\'a}nchez-L{\'o}pez}, A. and {Stolker}, T.},
        title = "{The ESO SupJup Survey. I. Chemical and isotopic characterisation of the late L-dwarf DENIS J0255-4700 with CRIRES$^{+}$}",
      journal = {Astron. Astrophys.},
         year = 2024,
        month = aug,
       volume = {688},
          eid = {A116},
        pages = {A116},
          doi = {10.1051/0004-6361/202348508},
archivePrefix = {arXiv},
       eprint = {2405.10841},
 primaryClass = {astro-ph.EP},
       adsurl = {https://ui.adsabs.harvard.edu/abs/2024A&A...688A.116D}
}

@ARTICLE{Rowland2024,
       author = {{Rowland}, Melanie J. and {Morley}, Caroline V. and {Miles}, Brittany E. and {Suarez}, Genaro and {Faherty}, Jacqueline K. and {Skemer}, Andrew J. and {Beiler}, Samuel A. and {Line}, Michael R. and {Bjoraker}, Gordon L. and {Fortney}, Jonathan J. and {Vos}, Johanna M. and {Alejandro Merchan}, Sherelyn and {Marley}, Mark and {Burningham}, Ben and {Freedman}, Richard and {Gharib-Nezhad}, Ehsan and {Batalha}, Natasha and {Lupu}, Roxana and {Visscher}, Channon and {Schneider}, Adam C. and {Geballe}, T.~R. and {Carter}, Aarynn and {Allers}, Katelyn and {Mang}, James and {Apai}, D{\'a}niel and {Limbach}, Mary Anne and {Wilson}, Mikayla J.},
        title = "{Protosolar D-to-H Abundance and One Part per Billion PH$_{3}$ in the Coldest Brown Dwarf}",
      journal = {\apjl},
         year = 2024,
        month = dec,
       volume = {977},
       number = {2},
          eid = {L49},
        pages = {L49},
          doi = {10.3847/2041-8213/ad9744},
archivePrefix = {arXiv},
       eprint = {2411.14541},
 primaryClass = {astro-ph.SR},
       adsurl = {https://ui.adsabs.harvard.edu/abs/2024ApJ...977L..49R}
}

@ARTICLE{Moses2016,
       author = {{Moses}, J.~I. and {Marley}, M.~S. and {Zahnle}, K. and {Line}, M.~R. and {Fortney}, J.~J. and {Barman}, T.~S. and {Visscher}, C. and {Lewis}, N.~K. and {Wolff}, M.~J.},
        title = "{On the Composition of Young, Directly Imaged Giant Planets}",
      journal = {\apj},
         year = 2016,
        month = oct,
       volume = {829},
       number = {2},
          eid = {66},
        pages = {66},
          doi = {10.3847/0004-637X/829/2/66},
archivePrefix = {arXiv},
       eprint = {1608.08643},
 primaryClass = {astro-ph.EP},
       adsurl = {https://ui.adsabs.harvard.edu/abs/2016ApJ...829...66M}
}

@article{Cridland2020,
   author = {Alex J. Cridland and Ewine F. Van Dishoeck and Matthew Alessi and Ralph E. Pudritz},
   doi = {10.1051/0004-6361/202038767},
   issn = {14320746},
   journal = {Astronomy and Astrophysics},
   month = {10},
   publisher = {EDP Sciences},
   title = {Connecting planet formation and astrochemistry: C/Os and N/Os of warm giant planets and Jupiter analogues},
   volume = {642},
   year = {2020},
}

@article{Turrini2021,
   author = {D. Turrini and E. Schisano and S. Fonte and S. Molinari and R. Politi and D. Fedele and O. Panić and M. Kama and Q. Changeat and G. Tinetti},
   doi = {10.3847/1538-4357/abd6e5},
   issn = {0004-637X},
   issue = {1},
   journal = {The Astrophysical Journal},
   month = {3},
   pages = {40},
   publisher = {American Astronomical Society},
   title = {Tracing the Formation History of Giant Planets in Protoplanetary Disks with Carbon, Oxygen, Nitrogen, and Sulfur},
   volume = {909},
   year = {2021},
}

@ARTICLE{Reggiani2024,
       author = {{Reggiani}, Henrique and {Galarza}, Jhon Yana and {Schlaufman}, Kevin C. and {Sing}, David K. and {Healy}, Brian F. and {McWilliam}, Andrew and {Lothringer}, Joshua D. and {Pueyo}, Laurent},
        title = "{Insight into the Formation of {\ensuremath{\beta}} Pic b through the Composition of Its Parent Protoplanetary Disk as Revealed by the {\ensuremath{\beta}} Pic Moving Group Member HD 181327}",
      journal = {\aj},
         year = 2024,
        month = jan,
       volume = {167},
       number = {1},
          eid = {45},
        pages = {45},
          doi = {10.3847/1538-3881/ad0f93},
archivePrefix = {arXiv},
       eprint = {2311.12210},
 primaryClass = {astro-ph.SR},
       adsurl = {https://ui.adsabs.harvard.edu/abs/2024AJ....167...45R}
}

@article{Morley2019,
   author = {Caroline V. Morley and Andrew J. Skemer and Brittany E. Miles and Michael R. Line and Eric D. Lopez and Matteo Brogi and Richard S. Freedman and Mark S. Marley},
   doi = {10.3847/2041-8213/ab3c65},
   issn = {20418213},
   issue = {2},
   journal = {The Astrophysical Journal},
   month = {9},
   pages = {L29},
   publisher = {American Astronomical Society},
   title = {Measuring the D/H Ratios of Exoplanets and Brown Dwarfs},
   volume = {882},
   year = {2019},
}

@article{Ohno2023,
   author = {Kazumasa Ohno and Jonathan J. Fortney},
   doi = {10.3847/1538-4357/acafed},
   issn = {0004-637X},
   issue = {1},
   journal = {The Astrophysical Journal},
   month = {3},
   pages = {18},
   publisher = {American Astronomical Society},
   title = {Nitrogen as a Tracer of Giant Planet Formation. I. A Universal Deep Adiabatic Profile and Semianalytical Predictions of Disequilibrium Ammonia Abundances in Warm Exoplanetary Atmospheres},
   volume = {946},
   year = {2023},
}

@ARTICLE{Chachan2023,
       author = {{Chachan}, Yayaati and {Knutson}, Heather A. and {Lothringer}, Joshua and {Blake}, Geoffrey A.},
        title = "{Breaking Degeneracies in Formation Histories by Measuring Refractory Content in Gas Giants}",
      journal = {\apj},
         year = 2023,
        month = feb,
       volume = {943},
       number = {2},
          eid = {112},
        pages = {112},
          doi = {10.3847/1538-4357/aca614},
archivePrefix = {arXiv},
       eprint = {2211.09080},
}

@ARTICLE{Eistrup2018,
       author = {{Eistrup}, Christian and {Walsh}, Catherine and {van Dishoeck}, Ewine F.},
        title = "{Molecular abundances and C/O ratios in chemically evolving planet-forming disk midplanes}",
      journal = {\aap},
         year = 2018,
        month = may,
       volume = {613},
          eid = {A14},
        pages = {A14},
          doi = {10.1051/0004-6361/201731302},
archivePrefix = {arXiv},
       eprint = {1709.07863},
 primaryClass = {astro-ph.EP},
       adsurl = {https://ui.adsabs.harvard.edu/abs/2018A&A...613A..14E}
}

@ARTICLE{Eistrup2016,
       author = {{Eistrup}, Christian and {Walsh}, Catherine and {van Dishoeck}, Ewine F.},
        title = "{Setting the volatile composition of (exo)planet-building material. Does chemical evolution in disk midplanes matter?}",
      journal = {\aap},
         year = 2016,
        month = nov,
       volume = {595},
          eid = {A83},
        pages = {A83},
          doi = {10.1051/0004-6361/201628509},
archivePrefix = {arXiv},
       eprint = {1607.06710},
 primaryClass = {astro-ph.EP},
       adsurl = {https://ui.adsabs.harvard.edu/abs/2016A&A...595A..83E}
}

@ARTICLE{Andrews2018,
       author = {{Andrews}, Sean M. and {Terrell}, Marie and {Tripathi}, Anjali and {Ansdell}, Megan and {Williams}, Jonathan P. and {Wilner}, David J.},
        title = "{Scaling Relations Associated with Millimeter Continuum Sizes in Protoplanetary Disks}",
      journal = {\apj},
         year = 2018,
        month = oct,
       volume = {865},
       number = {2},
          eid = {157},
        pages = {157},
          doi = {10.3847/1538-4357/aadd9f},
archivePrefix = {arXiv},
       eprint = {1808.10510},
 primaryClass = {astro-ph.EP},
       adsurl = {https://ui.adsabs.harvard.edu/abs/2018ApJ...865..157A}
}

@ARTICLE{Hoch2023,
       author = {{Hoch}, Kielan K.~W. and {Konopacky}, Quinn M. and {Theissen}, Christopher A. and {Ruffio}, Jean-Baptiste and {Barman}, Travis S. and {Rickman}, Emily L. and {Perrin}, Marshall D. and {Macintosh}, Bruce and {Marois}, Christian},
        title = "{Assessing the C/O Ratio Formation Diagnostic: A Potential Trend with Companion Mass}",
      journal = {\aj},
         year = 2023,
        month = sep,
       volume = {166},
       number = {3},
          eid = {85},
        pages = {85},
          doi = {10.3847/1538-3881/ace442},
archivePrefix = {arXiv},
       eprint = {2212.04557},
 primaryClass = {astro-ph.EP},
       adsurl = {https://ui.adsabs.harvard.edu/abs/2023AJ....166...85H}
}

@article{mordasini_Imprint_2016,
  title = {The {{Imprint}} of {{Exoplanet Formation History}} on {{Observable Present-day Spectra}} of {{Hot Jupiters}}},
  author = {Mordasini, C. and {van Boekel}, R. and Molli{\`e}re, P. and Henning, Th. and Benneke, Bj{\"o}rn},
  year = {2016},
  month = nov,
  journal = {The Astrophysical Journal},
  volume = {832},
  pages = {41},
  issn = {0004-637X},
  doi = {10.3847/0004-637X/832/1/41}
}

@article{Balmer2025b,
doi = {10.3847/1538-3881/adb1c6},
url = {https://dx.doi.org/10.3847/1538-3881/adb1c6},
year = {2025},
month = {mar},
publisher = {The American Astronomical Society},
volume = {169},
number = {4},
pages = {209},
author = {Balmer, William O. and Kammerer, Jens and Pueyo, Laurent and Perrin, Marshall D. and Girard, Julien H. and Leisenring, Jarron M. and Lawson, Kellen and Dennen, Henry and van der Marel, Roeland P. and Beichman, Charles A. and Bryden, Geoffrey and Llop-Sayson, Jorge and Valenti, Jeff A. and Lothringer, Joshua D. and Lewis, Nikole K. and Mâlin, Mathilde and Rebollido, Isabel and Rickman, Emily and Hoch, Kielan K. W. and Soummer, Rémi and Clampin, Mark and Mountain, C. Matt},
title = {JWST-TST High Contrast: Living on the Wedge, or, NIRCam Bar Coronagraphy Reveals CO2 in the HR 8799 and 51 Eri Exoplanets’ Atmospheres},
journal = {Astron. J.}
}

@ARTICLE{Tobin20,
       author = {{Tobin}, John J. and {Sheehan}, Patrick D. and {Megeath}, S. Thomas and {D{\'\i}az-Rodr{\'\i}guez}, Ana Karla and {Offner}, Stella S.~R. and {Murillo}, Nadia M. and {van 't Hoff}, Merel L.~R. and {van Dishoeck}, Ewine F. and {Osorio}, Mayra and {Anglada}, Guillem and {Furlan}, Elise and {Stutz}, Amelia M. and {Reynolds}, Nickalas and {Karnath}, Nicole and {Fischer}, William J. and {Persson}, Magnus and {Looney}, Leslie W. and {Li}, Zhi-Yun and {Stephens}, Ian and {Chandler}, Claire J. and {Cox}, Erin and {Dunham}, Michael M. and {Tychoniec}, {\L}ukasz and {Kama}, Mihkel and {Kratter}, Kaitlin and {Kounkel}, Marina and {Mazur}, Brian and {Maud}, Luke and {Patel}, Lisa and {Perez}, Laura and {Sadavoy}, Sarah I. and {Segura-Cox}, Dominique and {Sharma}, Rajeeb and {Stephenson}, Brian and {Watson}, Dan M. and {Wyrowski}, Friedrich},
        title = "{The VLA/ALMA Nascent Disk and Multiplicity (VANDAM) Survey of Orion Protostars. II. A Statistical Characterization of Class 0 and Class I Protostellar Disks}",
      journal = {Astrophys. J.},
         year = 2020,
        month = feb,
       volume = {890},
       number = {2},
          eid = {130},
        pages = {130},
          doi = {10.3847/1538-4357/ab6f64},
       adsurl = {https://ui.adsabs.harvard.edu/abs/2020ApJ...890..130T}
}

@ARTICLE{McClure2023,
       author = {{McClure}, M.~K. and {Rocha}, W.~R.~M. and {Pontoppidan}, K.~M. and {Crouzet}, N. and {Chu}, L.~E.~U. and {Dartois}, E. and {Lamberts}, T. and {Noble}, J.~A. and {Pendleton}, Y.~J. and {Perotti}, G. and {Qasim}, D. and {Rachid}, M.~G. and {Smith}, Z.~L. and {Sun}, Fengwu and {Beck}, Tracy L. and {Boogert}, A.~C.~A. and {Brown}, W.~A. and {Caselli}, P. and {Charnley}, S.~B. and {Cuppen}, Herma M. and {Dickinson}, H. and {Drozdovskaya}, M.~N. and {Egami}, E. and {Erkal}, J. and {Fraser}, H. and {Garrod}, R.~T. and {Harsono}, D. and {Ioppolo}, S. and {Jim{\'e}nez-Serra}, I. and {Jin}, M. and {J{\o}rgensen}, J.~K. and {Kristensen}, L.~E. and {Lis}, D.~C. and {McCoustra}, M.~R.~S. and {McGuire}, Brett A. and {Melnick}, G.~J. and {{\"O}berg}, Karin I. and {Palumbo}, M.~E. and {Shimonishi}, T. and {Sturm}, J.~A. and {van Dishoeck}, E.~F. and {Linnartz}, H.},
        title = "{An Ice Age JWST inventory of dense molecular cloud ices}",
      journal = {Nature Astronomy},
         year = 2023,
        month = apr,
       volume = {7},
        pages = {431-443},
          doi = {10.1038/s41550-022-01875-w},
archivePrefix = {arXiv},
       eprint = {2301.09140},
 primaryClass = {astro-ph.GA},
       adsurl = {https://ui.adsabs.harvard.edu/abs/2023NatAs...7..431M}
}

@ARTICLE{Tychoniec2020,
       author = {{Tychoniec}, {\L}ukasz and {Manara}, Carlo F. and {Rosotti}, Giovanni P. and {van Dishoeck}, Ewine F. and {Cridland}, Alexander J. and {Hsieh}, Tien-Hao and {Murillo}, Nadia M. and {Segura-Cox}, Dominique and {van Terwisga}, Sierk E. and {Tobin}, John J.},
        title = "{Dust masses of young disks: constraining the initial solid reservoir for planet formation}",
      journal = {Astron. Astrophys.},
         year = 2020,
        month = aug,
       volume = {640},
          eid = {A19},
        pages = {A19},
          doi = {10.1051/0004-6361/202037851},
archivePrefix = {arXiv},
       eprint = {2006.02812},
 primaryClass = {astro-ph.EP},
       adsurl = {https://ui.adsabs.harvard.edu/abs/2020A&A...640A..19T}
}

@ARTICLE{Inglis2024b,
       author = {{Inglis}, Julie and {Batalha}, Natasha E. and {Lewis}, Nikole K. and {Kataria}, Tiffany and {Knutson}, Heather A. and {Kilpatrick}, Brian M. and {Gagnebin}, Anna and {Mukherjee}, Sagnick and {Pettyjohn}, Maria M. and {Crossfield}, Ian J.~M. and {Foote}, Trevor O. and {Grant}, David and {Henry}, Gregory W. and {Lally}, Maura and {McKemmish}, Laura K. and {Sing}, David K. and {Wakeford}, Hannah R. and {Zapata Trujillo}, Juan C. and {Zellem}, Robert T.},
        title = "{Quartz Clouds in the Dayside Atmosphere of the Quintessential Hot Jupiter HD 189733 b}",
      journal = {Astrophys. J. Lett.},
         year = 2024,
        month = oct,
       volume = {973},
       number = {2},
          eid = {L41},
        pages = {L41},
          doi = {10.3847/2041-8213/ad725e},
archivePrefix = {arXiv},
       eprint = {2409.11395},
 primaryClass = {astro-ph.EP},
       adsurl = {https://ui.adsabs.harvard.edu/abs/2024ApJ...973L..41I}
}

@ARTICLE{Fletcher2009,
       author = {{Fletcher}, L.~N. and {Orton}, G.~S. and {Teanby}, N.~A. and {Irwin}, P.~G.~J. and {Bjoraker}, G.~L.},
        title = "{Methane and its isotopologues on Saturn from Cassini/CIRS observations}",
      journal = {Icarus},
         year = 2009,
        month = feb,
       volume = {199},
       number = {2},
        pages = {351-367},
          doi = {10.1016/j.icarus.2008.09.019},
       adsurl = {https://ui.adsabs.harvard.edu/abs/2009Icar..199..351F}
}

@ARTICLE{Briggs1989,
       author = {{Briggs}, F.~H. and {Sackett}, P.~D.},
        title = "{Radio observations of Saturn as a probe of its atmosphere and cloud structure}",
      journal = {Icarus},
         year = 1989,
        month = jul,
       volume = {80},
       number = {1},
        pages = {77-103},
          doi = {10.1016/0019-1035(89)90162-0},
       adsurl = {https://ui.adsabs.harvard.edu/abs/1989Icar...80...77B}
}

@ARTICLE{Kamp2001,
       author = {{Kamp}, I. and {Iliev}, I. Kh. and {Paunzen}, E. and {Pintado}, O.~I. and {Solano}, E. and {Barzova}, I.~S.},
        title = "{Light element non-LTE abundances of lambda Bootis stars. II. Nitrogen and sulphur}",
      journal = {Astron. Astrophys.},
         year = 2001,
        month = sep,
       volume = {375},
        pages = {899-908},
          doi = {10.1051/0004-6361:20010886},
       adsurl = {https://ui.adsabs.harvard.edu/abs/2001A&A...375..899K}
}

@ARTICLE{Li2020,
       author = {{Li}, Cheng and {Ingersoll}, Andrew and {Bolton}, Scott and {Levin}, Steven and {Janssen}, Michael and {Atreya}, Sushil and {Lunine}, Jonathan and {Steffes}, Paul and {Brown}, Shannon and {Guillot}, Tristan and {Allison}, Michael and {Arballo}, John and {Bellotti}, Amadeo and {Adumitroaie}, Virgil and {Gulkis}, Samuel and {Hodges}, Amoree and {Li}, Liming and {Misra}, Sidharth and {Orton}, Glenn and {Oyafuso}, Fabiano and {Santos-Costa}, Daniel and {Waite}, Hunter and {Zhang}, Zhimeng},
        title = "{The water abundance in Jupiter's equatorial zone}",
      journal = {Nature Astronomy},
         year = 2020,
        month = feb,
       volume = {4},
        pages = {609-616},
          doi = {10.1038/s41550-020-1009-3},
archivePrefix = {arXiv},
       eprint = {2012.10305},
 primaryClass = {astro-ph.EP},
       adsurl = {https://ui.adsabs.harvard.edu/abs/2020NatAs...4..609L}
}

@inbook{Atreya2018, place={Cambridge}, series={Cambridge Planetary Science}, title={The Origin and Evolution of Saturn, with Exoplanet Perspective}, booktitle={Saturn in the 21st Century}, publisher={Cambridge University Press}, author={Atreya, Sushil K. and Crida, Aurélien and Guillot, Tristan and Lunine, Jonathan I. and Madhusudhan, Nikku and Mousis, Olivier}, editor={Baines, Kevin H. and Flasar, F. Michael and Krupp, Norbert and Stallard, TomEditors}, year={2018}, pages={5–43}, collection={Cambridge Planetary Science}}

@ARTICLE{Wong2004,
       author = {{Wong}, Michael H. and {Mahaffy}, Paul R. and {Atreya}, Sushil K. and {Niemann}, Hasso B. and {Owen}, Tobias C.},
        title = "{Updated Galileo probe mass spectrometer measurements of carbon, oxygen, nitrogen, and sulfur on Jupiter}",
      journal = {Icarus},
         year = 2004,
        month = sep,
       volume = {171},
       number = {1},
        pages = {153-170},
          doi = {10.1016/j.icarus.2004.04.010},
       adsurl = {https://ui.adsabs.harvard.edu/abs/2004Icar..171..153W}
}

@ARTICLE{Lei2025,
       author = {{Lei}, Elise and {Molli{\`e}re}, Paul},
        title = "{easyCHEM: A Python package for calculating chemical equilibrium abundances in exoplanet atmospheres}",
      journal = {The Journal of Open Source Software},
         year = 2025,
        month = aug,
       volume = {10},
       number = {112},
          eid = {7712},
        pages = {7712},
          doi = {10.21105/joss.07712},
archivePrefix = {arXiv},
       eprint = {2410.21364},
 primaryClass = {astro-ph.IM},
       adsurl = {https://ui.adsabs.harvard.edu/abs/2025JOSS...10.7712L}
}

@article{Oberg2019,
  title = {Jupiter's {{Composition Suggests}} Its {{Core Assembled Exterior}} to the {{N}}{\textsubscript{2}} {{Snowline}}},
  author = {{\"O}berg, Karin I and Wordsworth, Robin},
  year = {2019},
  month = nov,
  journal = {Astron. J.},
  volume = {158},
  number = {5},
  pages = {194},
  issn = {0004-6256, 1538-3881},
  doi = {10.3847/1538-3881/ab46a8},
  urldate = {2025-02-11},
  langid = {english}
}

@ARTICLE{Allard2019,
       author = {{Allard}, N.~F. and {Spiegelman}, F. and {Leininger}, T. and {Molliere}, P.},
        title = "{New study of the line profiles of sodium perturbed by H$_{2}$}",
      journal = {Astron. Astrophys.},
         year = 2019,
        month = aug,
       volume = {628},
          eid = {A120},
        pages = {A120},
          doi = {10.1051/0004-6361/201935593},
archivePrefix = {arXiv},
       eprint = {1908.01989},
 primaryClass = {astro-ph.SR},
       adsurl = {https://ui.adsabs.harvard.edu/abs/2019A&A...628A.120A}
}

@ARTICLE{Barber2014,
       author = {{Barber}, R.~J. and {Strange}, J.~K. and {Hill}, C. and {Polyansky}, O.~L. and {Mellau}, G. Ch. and {Yurchenko}, S.~N. and {Tennyson}, Jonathan},
        title = "{ExoMol line lists - III. An improved hot rotation-vibration line list for HCN and HNC}",
      journal = {Mon. Not. R. Astron. Soc.},
         year = 2014,
        month = jan,
       volume = {437},
       number = {2},
        pages = {1828-1835},
          doi = {10.1093/mnras/stt2011},
archivePrefix = {arXiv},
       eprint = {1311.1328},
 primaryClass = {astro-ph.SR},
       adsurl = {https://ui.adsabs.harvard.edu/abs/2014MNRAS.437.1828B}
}

@ARTICLE{Voronin2010,
       author = {{Voronin}, B.~A. and {Tennyson}, J. and {Tolchenov}, R.~N. and {Lugovskoy}, A.~A. and {Yurchenko}, S.~N.},
        title = "{A high accuracy computed line list for the HDO molecule}",
      journal = {\mnras},
         year = 2010,
        month = feb,
       volume = {402},
       number = {1},
        pages = {492-496},
          doi = {10.1111/j.1365-2966.2009.15904.x},
       adsurl = {https://ui.adsabs.harvard.edu/abs/2010MNRAS.402..492V}
}

@ARTICLE{Kama2015,
       author = {{Kama}, M. and {Folsom}, C.~P. and {Pinilla}, P.},
        title = "{Fingerprints of giant planets in the photospheres of Herbig stars}",
      journal = {Astron. Astrophys.},
         year = 2015,
        month = oct,
       volume = {582},
          eid = {L10},
        pages = {L10},
          doi = {10.1051/0004-6361/201527094},
archivePrefix = {arXiv},
       eprint = {1509.02741},
 primaryClass = {astro-ph.SR},
       adsurl = {https://ui.adsabs.harvard.edu/abs/2015A&A...582L..10K}
}

@ARTICLE{Alacoria2022,
       author = {{Alacoria}, J. and {Saffe}, C. and {Jaque Arancibia}, M. and {Angeloni}, R. and {Miquelarena}, P. and {Flores}, M. and {Veramendi}, M.~E. and {Collado}, A.},
        title = "{Testing the accretion scenario of {\ensuremath{\lambda}} Boo stars}",
      journal = {Astron. Astrophys.},
         year = 2022,
        month = apr,
       volume = {660},
          eid = {A98},
        pages = {A98},
          doi = {10.1051/0004-6361/202243058},
archivePrefix = {arXiv},
       eprint = {2202.05066},
 primaryClass = {astro-ph.SR},
       adsurl = {https://ui.adsabs.harvard.edu/abs/2022A&A...660A..98A}
}

@article{Jura2015,
doi = {10.1088/0004-6256/150/6/166},
url = {https://dx.doi.org/10.1088/0004-6256/150/6/166},
year = {2015},
month = {nov},
publisher = {The American Astronomical Society},
volume = {150},
number = {6},
pages = {166},
author = {Jura, M.},
title = {LAMBDA BOO ABUNDANCE PATTERNS: ACCRETION FROM ORBITING SOURCES},
journal = {Astron. J.}
}

@article{Vogt2014,
doi = {10.1086/676120},
url = {https://dx.doi.org/10.1086/676120},
year = {2014},
month = {apr},
publisher = {University of Chicago Press},
volume = {126},
number = {938},
pages = {359},
author = {Steven S. Vogt and Matthew Radovan and Robert Kibrick and R. Paul Butler and Barry Alcott and Steve Allen and Pamela Arriagada and Mike Bolte and Jennifer Burt and Jerry Cabak and Kostas Chloros and David Cowley and William Deich and Brian Dupraw and Wayne Earthman and Harland Epps and Sandra Faber and Debra Fischer and Elinor Gates and David Hilyard and Brad Holden and Ken Johnston and Sandy Keiser and Dick Kanto and Myra Katsuki and Lee Laiterman and Kyle Lanclos and Greg Laughlin and Jeff Lewis and Chris Lockwood and Paul Lynam and Geoffrey Marcy and Maureen McLean and Joe Miller and Tony Misch and Michael Peck and Terry Pfister and Andrew Phillips and Eugenio Rivera and Dale Sandford and Mike Saylor and Richard Stover and Matthew Thompson and Bernie Walp and James Ward and John Wareham and Mingzhi Wei and Chris Wright},
title = {APF—The Lick Observatory Automated Planet Finder},
journal = {Publications of the Astronomical Society of the Pacific},
}

@ARTICLE{Jermyn2018,
       author = {{Jermyn}, Adam S. and {Kama}, Mihkel},
        title = "{Stellar photospheric abundances as a probe of discs and planets}",
      journal = {Mon. Not. R. Astron. Soc.},
         year = 2018,
        month = jun,
       volume = {476},
       number = {4},
        pages = {4418-4434},
          doi = {10.1093/mnras/sty429},
archivePrefix = {arXiv},
       eprint = {1804.06414},
 primaryClass = {astro-ph.SR},
       adsurl = {https://ui.adsabs.harvard.edu/abs/2018MNRAS.476.4418J}
}

@ARTICLE{Hejazi2025,
       author = {{Hejazi}, Neda and {Xuan}, Jerry W. and {Coria}, David R. and {Sawczynec}, Erica and {Crossfield}, Ian J.~M. and {Cristofari}, Paul I. and {Zhang}, Zhoujian and {Rhem}, Maleah},
        title = "{Chemical Links between a Young M-type T Tauri Star and Its Substellar Companion: Spectral Analysis and C/O Measurement of DH Tau A}",
      journal = {\apj},
         year = 2025,
        month = jan,
       volume = {978},
       number = {1},
          eid = {42},
        pages = {42},
          doi = {10.3847/1538-4357/ad968c},
archivePrefix = {arXiv},
       eprint = {2411.15591},
 primaryClass = {astro-ph.SR},
       adsurl = {https://ui.adsabs.harvard.edu/abs/2025ApJ...978...42H}
}

@misc{MOCAPY,
  author       = {Gagné, Jonathan},
  title        = {MOCAPY: Bayesian classification of young stars},
  howpublished = {\url{https://github.com/jgagneastro/mocapy}},
  note         = {Accessed 2025-08-07},
  year         = {2025}
}

@ARTICLE{Gagne2018,
       author = {{Gagn{\'e}}, Jonathan and {Mamajek}, Eric E. and {Malo}, Lison and {Riedel}, Adric and {Rodriguez}, David and {Lafreni{\`e}re}, David and {Faherty}, Jacqueline K. and {Roy-Loubier}, Olivier and {Pueyo}, Laurent and {Robin}, Annie C. and {Doyon}, Ren{\'e}},
        title = "{BANYAN. XI. The BANYAN {\ensuremath{\Sigma}} Multivariate Bayesian Algorithm to Identify Members of Young Associations with 150 pc}",
      journal = {\apj},
         year = 2018,
        month = mar,
       volume = {856},
       number = {1},
          eid = {23},
        pages = {23},
          doi = {10.3847/1538-4357/aaae09},
archivePrefix = {arXiv},
       eprint = {1801.09051},
 primaryClass = {astro-ph.SR},
       adsurl = {https://ui.adsabs.harvard.edu/abs/2018ApJ...856...23G}
}

@ARTICLE{Gagne2024,
       author = {{Gagn{\'e}}, Jonathan},
        title = "{A Quick Guide to Nearby Young Associations}",
      journal = {\pasp},
         year = 2024,
        month = jun,
       volume = {136},
       number = {6},
          eid = {063001},
        pages = {063001},
          doi = {10.1088/1538-3873/ad4e6a},
archivePrefix = {arXiv},
       eprint = {2405.12860},
 primaryClass = {astro-ph.SR},
       adsurl = {https://ui.adsabs.harvard.edu/abs/2024PASP..136f3001G}
}

@article{Baburaj2025,
  title = {A {{High-Resolution Spectroscopic Survey}} of {{Directly Imaged Companion Hosts}}: {{I}}. {{Determination}} of Diagnostic Stellar Abundances for Planet Formation and Composition},
  shorttitle = {A {{High-Resolution Spectroscopic Survey}} of {{Directly Imaged Companion Hosts}}},
  author = {Baburaj, Aneesh and Konopacky, Quinn M. and Theissen, Christopher A. and Peacock, Sarah and Huseby, Lori and Fulton, Benjamin and Gerasimov, Roman and Barman, Travis S. and Hoch, Kielan K. W.},
  year = {2025},
  month = feb,
  journal = {Astron. J.},
  volume = {169},
  number = {2},
  eprint = {2409.14239},
  primaryclass = {astro-ph},
  pages = {55},
  issn = {0004-6256, 1538-3881},
  doi = {10.3847/1538-3881/ad8dfc},
  urldate = {2025-02-11},
  archiveprefix = {arXiv},
  langid = {english}
}

@ARTICLE{Apogee2022,
       author = {{Abdurro'uf} and {Accetta}, Katherine and {Aerts}, Conny and {Silva Aguirre}, V{\'\i}ctor and {Ahumada}, Romina and {Ajgaonkar}, Nikhil and {Filiz Ak}, N. and {Alam}, Shadab and {Allende Prieto}, Carlos and {Almeida}, Andr{\'e}s and {Anders}, Friedrich and {Anderson}, Scott F. and {Andrews}, Brett H. and {Anguiano}, Borja and {Aquino-Ort{\'\i}z}, Erik and {Arag{\'o}n-Salamanca}, Alfonso and {Argudo-Fern{\'a}ndez}, Maria and {Ata}, Metin and {Aubert}, Marie and {Avila-Reese}, Vladimir and {Badenes}, Carles and {Barb{\'a}}, Rodolfo H. and {Barger}, Kat and {Barrera-Ballesteros}, Jorge K. and {Beaton}, Rachael L. and {Beers}, Timothy C. and {Belfiore}, Francesco and {Bender}, Chad F. and {Bernardi}, Mariangela and {Bershady}, Matthew A. and {Beutler}, Florian and {Bidin}, Christian Moni and {Bird}, Jonathan C. and {Bizyaev}, Dmitry and {Blanc}, Guillermo A. and {Blanton}, Michael R. and {Boardman}, Nicholas Fraser and {Bolton}, Adam S. and {Boquien}, M{\'e}d{\'e}ric and {Borissova}, Jura and {Bovy}, Jo and {Brandt}, W.~N. and {Brown}, Jordan and {Brownstein}, Joel R. and {Brusa}, Marcella and {Buchner}, Johannes and {Bundy}, Kevin and {Burchett}, Joseph N. and {Bureau}, Martin and {Burgasser}, Adam and {Cabang}, Tuesday K. and {Campbell}, Stephanie and {Cappellari}, Michele and {Carlberg}, Joleen K. and {Wanderley}, F{\'a}bio Carneiro and {Carrera}, Ricardo and {Cash}, Jennifer and {Chen}, Yan-Ping and {Chen}, Wei-Huai and {Cherinka}, Brian and {Chiappini}, Cristina and {Choi}, Peter Doohyun and {Chojnowski}, S. Drew and {Chung}, Haeun and {Clerc}, Nicolas and {Cohen}, Roger E. and {Comerford}, Julia M. and {Comparat}, Johan and {da Costa}, Luiz and {Covey}, Kevin and {Crane}, Jeffrey D. and {Cruz-Gonzalez}, Irene and {Culhane}, Connor and {Cunha}, Katia and {Dai}, Y. Sophia and {Damke}, Guillermo and {Darling}, Jeremy and {Davidson}, Jr., James W. and {Davies}, Roger and {Dawson}, Kyle and {De Lee}, Nathan and {Diamond-Stanic}, Aleksandar M. and {Cano-D{\'\i}az}, Mariana and {S{\'a}nchez}, Helena Dom{\'\i}nguez and {Donor}, John and {Duckworth}, Chris and {Dwelly}, Tom and {Eisenstein}, Daniel J. and {Elsworth}, Yvonne P. and {Emsellem}, Eric and {Eracleous}, Mike and {Escoffier}, Stephanie and {Fan}, Xiaohui and {Farr}, Emily and {Feng}, Shuai and {Fern{\'a}ndez-Trincado}, Jos{\'e} G. and {Feuillet}, Diane and {Filipp}, Andreas and {Fillingham}, Sean P. and {Frinchaboy}, Peter M. and {Fromenteau}, Sebastien and {Galbany}, Llu{\'\i}s and {Garc{\'\i}a}, Rafael A. and {Garc{\'\i}a-Hern{\'a}ndez}, D.~A. and {Ge}, Junqiang and {Geisler}, Doug and {Gelfand}, Joseph and {G{\'e}ron}, Tobias and {Gibson}, Benjamin J. and {Goddy}, Julian and {Godoy-Rivera}, Diego and {Grabowski}, Kathleen and {Green}, Paul J. and {Greener}, Michael and {Grier}, Catherine J. and {Griffith}, Emily and {Guo}, Hong and {Guy}, Julien and {Hadjara}, Massinissa and {Harding}, Paul and {Hasselquist}, Sten and {Hayes}, Christian R. and {Hearty}, Fred and {Hern{\'a}ndez}, Jes{\'u}s and {Hill}, Lewis and {Hogg}, David W. and {Holtzman}, Jon A. and {Horta}, Danny and {Hsieh}, Bau-Ching and {Hsu}, Chin-Hao and {Hsu}, Yun-Hsin and {Huber}, Daniel and {Huertas-Company}, Marc and {Hutchinson}, Brian and {Hwang}, Ho Seong and {Ibarra-Medel}, H{\'e}ctor J. and {Chitham}, Jacob Ider and {Ilha}, Gabriele S. and {Imig}, Julie and {Jaekle}, Will and {Jayasinghe}, Tharindu and {Ji}, Xihan and {Johnson}, Jennifer A. and {Jones}, Amy and {J{\"o}nsson}, Henrik and {Katkov}, Ivan and {Khalatyan}, Dr., Arman and {Kinemuchi}, Karen and {Kisku}, Shobhit and {Knapen}, Johan H. and {Kneib}, Jean-Paul and {Kollmeier}, Juna A. and {Kong}, Miranda and {Kounkel}, Marina and {Kreckel}, Kathryn and {Krishnarao}, Dhanesh and {Lacerna}, Ivan and {Lane}, Richard R. and {Langgin}, Rachel and {Lavender}, Ramon and {Law}, David R. and {Lazarz}, Daniel and {Leung}, Henry W. and {Leung}, Ho-Hin and {Lewis}, Hannah M. and {Li}, Cheng and {Li}, Ran and {Lian}, Jianhui and {Liang}, Fu-Heng and {Lin}, Lihwai and {Lin}, Yen-Ting and {Lin}, Sicheng and {Lintott}, Chris and {Long}, Dan and {Longa-Pe{\~n}a}, Pen{\'e}lope and {L{\'o}pez-Cob{\'a}}, Carlos and {Lu}, Shengdong and {Lundgren}, Britt F. and {Luo}, Yuanze and {Mackereth}, J. Ted and {de la Macorra}, Axel and {Mahadevan}, Suvrath and {Majewski}, Steven R. and {Manchado}, Arturo and {Mandeville}, Travis and {Maraston}, Claudia and {Margalef-Bentabol}, Berta and {Masseron}, Thomas and {Masters}, Karen L. and {Mathur}, Savita and {McDermid}, Richard M. and {Mckay}, Myles and {Merloni}, Andrea and {Merrifield}, Michael and {Meszaros}, Szabolcs and {Miglio}, Andrea and {Di Mille}, Francesco and {Minniti}, Dante and {Minsley}, Rebecca and {Monachesi}, Antonela},
        title = "{The Seventeenth Data Release of the Sloan Digital Sky Surveys: Complete Release of MaNGA, MaStar, and APOGEE-2 Data}",
      journal = {\apjs},
         year = 2022,
        month = apr,
       volume = {259},
       number = {2},
          eid = {35},
        pages = {35},
          doi = {10.3847/1538-4365/ac4414},
archivePrefix = {arXiv},
       eprint = {2112.02026},
 primaryClass = {astro-ph.GA},
       adsurl = {https://ui.adsabs.harvard.edu/abs/2022ApJS..259...35A}
}

@ARTICLE{Borthakur2025,
       author = {{Borthakur}, Sandipan P.~D. and {Kama}, Mihkel and {Fossati}, Luca and {Kral}, Quentin and {Folsom}, Colin P. and {Teske}, Johanna and {Aret}, Anna},
        title = "{Abundance analysis of stars hosting gas-rich debris discs}",
      journal = {\aap},
         year = 2025,
        month = may,
       volume = {697},
          eid = {A59},
        pages = {A59},
          doi = {10.1051/0004-6361/202452840},
archivePrefix = {arXiv},
       eprint = {2503.03614},
 primaryClass = {astro-ph.SR},
       adsurl = {https://ui.adsabs.harvard.edu/abs/2025A&A...697A..59B}
}

@ARTICLE{Molliere2025,
       author = {{Molli{\`e}re}, P. and {K{\"u}hnle}, H. and {Matthews}, E.~C. and {Henning}, Th. and {Min}, M. and {Patapis}, P. and {Lagage}, P.-O. and {Waters}, L.~B.~F.~M. and {G{\"u}del}, M. and {J{\"a}ger}, C. and {Zhang}, Z. and {Decin}, L. and {Biller}, B.~A. and {Absil}, O. and {Argyriou}, I. and {Barrado}, D. and {Cossou}, C. and {Glasse}, A. and {Olofsson}, G. and {Pye}, J.~P. and {Rouan}, D. and {Samland}, M. and {Scheithauer}, S. and {Tremblin}, P. and {Whiteford}, N. and {van Dishoeck}, E.~F. and {{\"O}stlin}, G. and {Ray}, T.},
        title = "{Evidence for SiO cloud nucleation in the rogue planet PSO J318}",
      journal = {\aap},
         year = 2025,
        month = nov,
       volume = {703},
          eid = {A79},
        pages = {A79},
          doi = {10.1051/0004-6361/202555732},
archivePrefix = {arXiv},
       eprint = {2507.18691},
 primaryClass = {astro-ph.EP},
       adsurl = {https://ui.adsabs.harvard.edu/abs/2025A&A...703A..79M}
}

@ARTICLE{Marois2008science,
       author = {{Marois}, Christian and {Macintosh}, Bruce and {Barman}, Travis and
         {Zuckerman}, B. and {Song}, Inseok and {Patience}, Jennifer and
         {Lafreni{\`e}re}, David and {Doyon}, Ren{\'e}},
        title = "{Direct Imaging of Multiple Planets Orbiting the Star HR 8799}",
      journal = {Science},
         year = "2008",
        month = "Nov",
       volume = {322},
       number = {5906},
        pages = {1348},
          doi = {10.1126/science.1166585},
archivePrefix = {arXiv},
       eprint = {0811.2606},
 primaryClass = {astro-ph},
       adsurl = {https://ui.adsabs.harvard.edu/abs/2008Sci...322.1348M}
}
\bibliographystyle{aasjournalv7}

\end{document}